\documentclass[12pt,english]{article}
\usepackage{amsmath}
\usepackage{amssymb}
\usepackage{amsthm}
\usepackage{mathptmx}
\usepackage{newtxmath}
\usepackage[T1]{fontenc}
\usepackage[utf8]{inputenc}
\usepackage{geometry}
\usepackage{color}
\usepackage{babel}
\usepackage{float}
\usepackage{placeins}
\usepackage{url}
\usepackage{graphicx}
\usepackage{pdflscape}
\usepackage{array}
\usepackage{booktabs}
\usepackage{longtable}
\usepackage{setspace}
\usepackage[authoryear,round]{natbib}
\let\citepbase\citep
\let\citetbase\citet
\let\citebase\cite
\let\citealtbase\citealt
\let\citealpbase\citealp
\renewcommand{\citep}{\citepbase*}
\renewcommand{\citet}{\citetbase*}
\renewcommand{\cite}{\citebase*}
\renewcommand{\citealt}{\citealtbase*}
\renewcommand{\citealp}{\citealpbase*}
\usepackage[unicode=true,pdfusetitle,
 bookmarks=true,bookmarksnumbered=false,bookmarksopen=false,
 breaklinks=false,pdfborder={0 0 0},pdfborderstyle={},backref=false,colorlinks=true]
 {hyperref}
\hypersetup{
 linkcolor=black,urlcolor=black,citecolor=black}

\makeatletter
\providecommand{\tabularnewline}{\\}
\usepackage[bottom]{footmisc}
\usepackage{indentfirst}
\usepackage{pdflscape}
\usepackage{xcolor}
\usepackage{caption}
\usepackage{subcaption}
\makeatother

\newcolumntype{P}[1]{>{\raggedright\arraybackslash}p{#1}}
\newcounter{anch}
\renewcommand{\thetable}{\arabic{table}}

\graphicspath{{figures/}{./}}

\usepackage{atbegshi}
\makeatletter
\newsavebox{\lsf@box}
\newcount\lsf@abspage
\newtoks\lsf@savedattr
\newif\iflsf@turned
\newcommand*{\lsf@theabspage}{\the\lsf@abspage}
\newcommand*{\lsf@begin}[1]{%
  \@float{#1}[p]%
  \begin{lrbox}{\lsf@box}\begin{minipage}{\textheight}}
\newcommand*{\lsf@end}{%
  \end{minipage}\end{lrbox}%
  \noindent\rotatebox{90}{\usebox{\lsf@box}}%
  \protected@write\@auxout{\let\lsf@theabspage\relax}%
    {\string\newlabel{lsfloat@page@\lsf@theabspage}{{}{\thepage}}}%
  \end@float}

\AtBeginShipout{%
  \global\advance\lsf@abspage\@ne
  \ifdefined\pdfpageattr
    \iflsf@turned
      \edef\lsf@now{\the\pdfpageattr}%
      \ifx\lsf@now\lsf@set \global\pdfpageattr\expandafter{\the\lsf@savedattr}\fi
      \global\lsf@turnedfalse
    \fi
    \@ifundefined{r@lsfloat@page@\the\lsf@abspage}{}{%
      \global\lsf@savedattr\expandafter{\the\pdfpageattr}%
      \global\pdfpageattr\expandafter{\the\pdfpageattr/Rotate 90}%
      \xdef\lsf@set{\the\pdfpageattr}%
      \global\lsf@turnedtrue}%
  \fi}
\makeatother

\makeatletter
\let\ftype@figure\ftype@table
\newcount\tpt@part

\makeatother

\title{\textbf{\Large The Rise of Agents and the Labor Market}}
\author{Nicola Borri, Anton Cheremukhin, Yukun Liu, and Aleh Tsyvinski\thanks{Nicola Borri is with Luiss University. Anton Cheremukhin is with the Federal Reserve Bank of Dallas. Yukun Liu is with the University of Rochester, Simon Business School. Aleh Tsyvinski is with Yale University and NBER. The views expressed in this paper are those of the authors and do not necessarily reflect the views of the Reserve Bank of Dallas or the Federal Reserve System. Data provided by OpenRouter, Inc. used under license. The analysis is based only on aggregated and anonymized derivative data constructed by the authors. The original OpenRouter data are not redistributed or disclosed. We are grateful to OpenRouter for providing access to the data, especially to Justin Summerville for his assistance throughout the data-access process, Ron Borzekowski and the Yale Data-Intensive Social Science Center for assistance in onboarding the data. Data on job postings are from Emsi Lightcast.}}
\date{\today}

\begin{document}
\pagecolor{white}

\hypersetup{pageanchor=false}
\maketitle
\thispagestyle{plain}

\begin{abstract} \singlespacing \noindent We provide the first economy-wide and cross-AI model evidence on the effects of agentic AI on labor demand. We combine proprietary OpenRouter data covering about two quadrillion tokens of realized AI consumption (2 percent of global AI token consumption) across more than 400 AI models, including 1.1 quadrillion agentic tokens, with the full text of 34.8 million U.S. job postings for public firms from Lightcast. We compute our measure of firm-level exposure to aggregate agentic AI consumption (agentic beta) using the comovement of U.S. public firms’ stock returns with realized weekly aggregate agentic AI token consumption. We find that a one-standard-deviation increase in a firm's exposure to agentic AI predicts 2.11 percent lower growth in job postings over the next six months and 3.20 percent over nine months. We then conduct an event study based on the release of Claude Code 2.0, which significantly expanded the capabilities of AI agents, and show that a one-standard-deviation higher pre-release exposure to agentic AI predicts 4.73 percent lower posting growth over six months post the release. Furthermore, we show that the decrease in job postings is concentrated among non-senior jobs and is statistically insignificant for senior jobs. In contrast, we find no statistically significant relation between broad AI realized consumption exposure and subsequent job posting growth once agentic exposure is controlled for. Finally, we validate our market-implied agentic exposure measure using independent evidence from firms’ mentions of AI deployment in securities filings and earnings calls.

\medskip
\noindent\textbf{JEL codes:} J23, J24, O33.
\end{abstract}

\newpage
\hypersetup{pageanchor=true}

\section{Introduction}\label{sec:intro}

The rise of AI agents is considered to be the most important development in AI since the introduction of LLMs (see, e.g., \citealp{Altman2025Reflections,Amodei2024Machines}). Agentic AI can carry out multi-step workflows that involve planning, calling external tools, taking actions, and coordinating with other agents, and can substitute for many tasks that are performed by humans. The rise of agents may lead to significant changes in human work and to disruptions in the labor market that are distinct from the advances in LLMs. The evidence on the effect of agents on the labor market so far is limited to reports based on the data from individual AI labs: Perplexity \citep{YangEtAl2025Agents,YangEtAl2026KnowledgeWork}, OpenAI's Codex (\citealp{JohnstonEtAl2026Codex}), and Anthropic's Claude Code (\citealp{AnthropicResearch2026Expertise}). Notably, the report ``How Employment Is Changing at Anthropic'' \citep{WilmersCatiponMassenkoffMcCrory2026} gives early evidence of how LLMs and agentic AI are significantly changing human work and the organization of work within one organization.

Our paper provides the first economy-wide and cross-model evidence of the effects of agentic AI on the labor market, and specifically on job postings. We combine and match two datasets. The first is the proprietary data from OpenRouter that contains granular information on about two quadrillion tokens of realized AI consumption (1.1 quadrillion tokens of agentic realized AI consumption) across more than 400 models and covers millions of users---about two percent of global realized token consumption, the largest dataset of cross-model AI token consumption. The second is data from Lightcast that gives the full text of 34.8 million U.S. job postings.\footnote{The full dataset is nearly 194 million job postings, of which 34.8 million are in our matched sample of public firms.} From the OpenRouter data, we document the rise of agentic realized token consumption, construct the  aggregate agentic consumption index, and calculate the stock-level beta exposure of public companies to this aggregate index of agentic token consumption. We validate our market-implied agentic exposure measure using independent evidence from firms' discussions of AI deployment in securities filings and earnings calls.

Our main finding is that greater market-implied exposure of firms to the aggregate agentic index (stock-level agentic AI beta) predicts slower subsequent growth in firms' job postings. Specifically, a one-standard-deviation increase in agentic exposure corresponds to a 2.11 percent lower job posting growth over the next six months and 3.20 percent over the next nine months. We then show that the decrease in job postings is concentrated among non-senior jobs and is statistically insignificant for senior jobs. In contrast, we find no statistically significant relation between overall realized AI consumption exposure and subsequent job postings once agentic exposure is controlled for. 

We measure agentic AI consumption as the tokens associated with responses that end by calling such tools. Our constructed agentic factor is the aggregate weekly log growth of this token consumption. We then compute the stock-level firm exposure (agentic betas) with respect to aggregate agentic AI consumption using the comovement of U.S. public firms' stock returns with our agentic factor. The resulting agentic betas capture public firms' market-implied sensitivity to growth in realized aggregate agentic AI token consumption. 

Our measure of exposure draws on the classic literature linking stock returns to aggregate economic variables (\citealp{Merton1973}, \citealp{CoxIngersollRoss1985}, and \citealp{ChenRollRoss1986}). More recently,  \citet{KoganPapanikolaou2014}, \citet{BakerBloomDavis2016}, and \citet{MeeuwisPapanikolaouRothbaumSchmidtForthcoming} study how individual stock exposures to aggregate variables---investment-specific technology shock, political uncertainty, the risk premium---predict firms' future real outcomes. Similarly, agentic betas reflect how stock markets aggregate dispersed information and beliefs on the technological disruptions that affect firms today and in the future, and specifically how those affect firms' job postings.

We then validate our market-based agentic betas using firms' subsequent disclosures and earnings call. We show that firms with higher agentic betas subsequently report more about AI agents and about the deployment of AI in their disclosure filings and management's statements during earnings calls. Specifically, a one-standard-deviation increase in the agentic beta is associated with a 1.39 percentage point higher probability of mentioning AI agents in SEC filings over the following six months and a 1.46 percentage point higher probability of management mentioning AI agents in earnings calls over the same horizon. Both estimates are significant at the 1 percent level. The agentic beta remains a significant predictor of whether firms mention AI agents or AI deployment after controlling for broad AI exposure. In these joint regressions, the coefficients on the broad AI beta are never positive and statistically significant.

Having computed the agentic exposure, we then link firms' agentic betas to the firm-specific labor demand using online job postings from Lightcast. Lightcast (formerly Burning Glass Technologies) collects advertisements from company career pages and job boards, providing detailed information on employers and job requirements. We match postings from firms and their subsidiaries to listed parent companies and count new postings each month. Our matched sample contains 1,784 listed parent companies linked to 3,689 Lightcast employers and covers 34.8 million U.S. job postings over the sample period, or approximately 18 percent of the Lightcast postings. From Lightcast we also obtain detailed descriptions of job characteristics, which we use to classify postings by occupation, seniority, required experience, and expected use of AI tools, as well as the full text of each job posting. Similarly, \citet{AcemogluAutorHazellRestrepo2022} use Lightcast data to study automation and labor demand. 

The paper has three main results. First, a firm's greater exposure to our aggregate agentic AI token consumption index predicts slower subsequent labor demand growth, measured using firms' total job postings. A one-standard-deviation increase in agentic beta is associated with $2.11$ percent lower posting growth over six months and $3.20$ percent lower posting growth over nine months, after controlling for industry and time fixed effects and firm characteristics. Both estimates are significant at the 1 percent level. In contrast, we find that the coefficients on broad AI exposure are statistically insignificant at the three- and six-month horizons. For broad AI token consumption exposure we use AI betas from our paper on the AI premium and the asset pricing implications of AI exposures \citep{BorriLiuTsyvinski2026}. When we include both the agentic beta and the broad AI beta in the same regression, only the agentic beta predicts lower growth in job postings: $2.47$ percent at six months and $2.96$ percent at nine months, significant at the 1 and 5 percent levels, respectively.

Second, we complement the panel evidence with an event study in order to account for possible changes in firms' growth opportunities or anticipated labor adjustments which may affect both stock prices and job postings. We use the September 2025 release of Claude Code 2.0, which significantly expanded the capabilities of AI agents \citep{Anthropic2025Code20,Anthropic2025Sonnet45}.\footnote{We also document that the release was followed by a large increase in agentic consumption. From June through September 2025, the agentic share of total tokens had ranged between 8.2 and 8.7 percent. Since October 2025, the agentic share increased every month, reaching 21.5 percent in December 2025 and 53.9 percent in April 2026.} We measure firms' exposure using their agentic betas as of September 2025, just before the release, and hold these exposures fixed. We assume that absent the release of Claude Code 2.0, firms with higher and lower agentic exposure would have experienced similar posting growth, after accounting for industry and firm characteristics. We find no statistically significant relation between agentic exposure and job posting growth before the release. After the release, we instead find lower posting growth for firms with higher agentic betas in every month from October 2025. The event-study estimates are larger than those from the panel specification. With industry fixed effects and controls for firm characteristics, a one-standard-deviation higher pre-release agentic beta is associated with $4.73$ percent lower posting growth over the next six months and $4.33$ percent lower growth over nine months. Both estimates are significant at the 1 percent level. When the broad AI betas are also included, the point estimates on agentic betas are quantitatively similar at $4.15$ and $3.88$, both significant at the 5 percent level. Furthermore, for within occupation variations we find that a one-standard-deviation higher agentic beta is associated with $3.28$ percent lower posting growth over the next six months and $3.24$ percent lower growth over nine months, both significant at the 5 percent level.

Third, extracting the granular characteristics of job postings from the text of job postings in Lightcast data, we examine how the relation between agentic AI exposure and labor demand varies across job types and characteristics. We find that the negative relation is stronger for non-senior positions and jobs requiring less experience. Over the six months following the release of Claude Code 2.0, the estimates for senior positions and jobs requiring at least six years of experience are statistically insignificant. For non-senior positions and jobs requiring less than six years of experience, a one-standard-deviation higher agentic beta predicts $4.54$ and $4.87$ percent lower posting growth, respectively. Both estimates are significant at the 5 percent level, as are the differences between senior and non-senior postings and between the two experience groups. We also find little response among postings that expect workers to use AI tools. In contrast, postings without such AI tool requirements show about $5.23$ percent lower growth, significant at the 1 percent level. 

Moreover, we show that the negative relationship between agentic exposure and the decline in labor demand is prevalent across industries and regions. The six-month panel estimates are negative in all five Fama-French industries, and the Health industry estimate is the largest in magnitude. When agentic beta is included alone, a one-standard-deviation higher agentic beta is associated with $7.50$ percent lower posting growth in the Health industry. Across the four Census regions, the estimates on agentic betas are similar in magnitude and are mostly significant at the 5 percent level.

Importantly, agentic exposure predicts lower posting growth after controlling for occupational AI exposure measures from \citet{Eloundou2024}, \citet{webb2020impact}, and \citet{MassenkoffMcCrory2026}. We then show that controlling for firms' characteristics, these occupational AI exposure measures have limited predictive power for changes in firms' job postings. 

Additionally, despite lower posting growth, we find no decline in wages advertised in Lightcast job postings. For investment, we find that a one-standard-deviation increase in the broad AI beta is associated with about $9.15$ percent higher capital-expenditure growth, significant at the 1 percent level. In contrast, we find no statistically significant relation between agentic exposure and firms' investment once broad AI exposure is controlled for.

Finally, we present a broad set of robustness tests. First, we find that our main posting-growth results hold across alternative beta-estimation windows from 20 to 34 weeks. The posting-growth results also hold when we estimate the agentic betas using alternative factor models, including the three-factor model of \citet{FamaFrench1993}, the four-factor model of \citet{Carhart1997}, and the five-factor model of \citet{FamaFrench2015}. We then estimate quantile regressions to assess whether this relation is concentrated in a particular part of the posting-growth distribution. We find that pre-release agentic exposure predicts lower posting growth at the bottom, middle, and top of the conditional distribution. Furthermore, we find that there is limited predictive power for changes in job postings to the exposure of aggregate AI agentic token consumption of open-weight models, and thus changes in job postings are mainly associated with exposures to aggregate AI agentic consumption of closed-weight frontier models.

We now briefly discuss the related literature. Our paper contributes to two strands of the literature. The first literature studies how technological change reallocates tasks and affects labor demand. \citet{AutorLevyMurnane2003} show that computerization substitutes for workers in some activities and complements workers in others.  \citet{AcemogluRestrepo2018} show how automation reallocates tasks from labor to capital and how new tasks can counteract displacement. The second strand examines AI, firm behavior, and labor-market adjustment. \citet{AcemogluAutorHazellRestrepo2022} document reduced hiring in non-AI positions at more AI-exposed establishments. \citet{HampolePapanikolaouSchmidtSeegmiller2026} find that AI reduces demand for exposed tasks, but that task reallocation and growth at adopting firms offset much of the effect on employment. \citet{EisfeldtSchubertZhangTaska2026} show that firms with greater task-based workforce exposure to generative AI gained in value after ChatGPT's release and reduced postings for exposed occupations. \citet{BrynjolfssonChandarChen2026}, using administrative payroll records,  document weaker employment growth among young workers in AI-exposed occupations, driven primarily by reduced hiring. \citet{HosseiniLichtinger2026} document a relative decline in junior employment at firms that start recruiting for generative AI integration roles. \citet{KoganPapanikolaouSeruStoffman2017} use stock returns around patent grants to estimate the private economic value of patents. \citet{GreenLiPapanikolaou2026} find that firms whose products are easier to deliver digitally underperformed around the Cowork Plugins launch. 

The rest of the paper is organized as follows. Section~\ref{sec:data} describes the data and measures. Section~\ref{sec:methodology} presents the empirical specifications and identification strategy. Section~\ref{sec:results} reports the main empirical results. Section~\ref{sec:additional_results} presents the additional results and robustness checks.

\section{Data}\label{sec:data} 

Our analysis combines granular AI consumption data from OpenRouter and job postings data from Lightcast. We also use data on US stock prices and firm characteristics.

\subsection{Agentic AI Factor}\label{sec:openrouter_data}

We measure AI consumption using data covering about two quadrillion tokens from OpenRouter, Inc., an AI inference provider that connects millions of users to hundreds of large language models through a single platform. The data are recorded at the anonymized user-day-model level. Users submit a prompt and select a model, and OpenRouter routes the request to a provider that hosts it. Available models on OpenRouter range from closed-weight models such as OpenAI's GPT, Anthropic's Claude, and Google's Gemini to open-weight models such as Meta's Llama, DeepSeek, and Qwen. Importantly, the data measure realized paid consumption from actual requests rather than self-reported consumption.

We define the agentic factor as the weekly log growth of aggregate agentic tokens.\footnote{We define a week to be from Monday to Sunday.} For each model response, OpenRouter records whether the model ends its turn by calling an external software tool. Tool calls allow models to request actions such as querying a database, running code, or reading or writing a file. We define \emph{agentic tokens} as the tokens associated with responses that end in a tool call. The agentic factor has a weekly standard deviation of about 36 percent. Additionally, in order to measure broad AI token consumption as opposed to agentic AI-specific token consumption, we consider the AI factor of \citet{BorriLiuTsyvinski2026}, defined as the first principal component of standardized weekly log growth in total tokens, dollar expenditure, and the weekly average number of daily active users.

Table~\ref{tab:tokens} reports the monthly number of tokens processed through OpenRouter and the share of agentic tokens. Agentic token usage is first recorded in our data on 18 January 2025. Agentic tokens account for 0.3 percent of the 2.2 trillion tokens processed that month. The table shows two main stylized facts. First, the agentic share is relatively stable in the months immediately preceding the 29 September 2025 release of Claude Code 2.0, ranging from 8.2 to 8.7 percent between June and September 2025. Second, the share rises every month after the release, reaching 21.5 percent in December 2025 and 76.8 percent in September 2026. By the summer of 2026, agentic consumption accounts for the majority of tokens processed through OpenRouter.

\begin{table}[!htbp]
\stepcounter{anch}
{\small\caption{{\small\textbf{Agentic Share of OpenRouter Aggregate Token Consumption by Month\label{tab:tokens}}}}
\medskip{}
}{\small\par}

{\footnotesize This table reports the monthly volume of aggregate realized token consumption through OpenRouter and the share of realized agentic token consumption. The last month covers its first 27 days.\par}

{\small\medskip{}
}{\small\par}
\centering{}{\small{}%
\renewcommand{\arraystretch}{1.10}%
\begin{tabular}{lccc}
\hline 
 & {\small (1)} & {\small (2)} & {\small (3)}\tabularnewline
{\small Month} & {\small Total tokens (trillions)} & {\small Agentic tokens (trillions)} & {\small Agentic share (\%)}\tabularnewline
\hline 
 &  &  & \tabularnewline
{\small January 2025} & {\small $2.23$} & {\small $0.01$} & {\small $0.3$}\tabularnewline
{\small February 2025} & {\small $3.50$} & {\small $0.04$} & {\small $1.0$}\tabularnewline
{\small March 2025} & {\small $6.31$} & {\small $0.22$} & {\small $3.5$}\tabularnewline
{\small April 2025} & {\small $8.29$} & {\small $0.60$} & {\small $7.2$}\tabularnewline
{\small May 2025} & {\small $9.86$} & {\small $1.51$} & {\small $15.4$}\tabularnewline
{\small June 2025} & {\small $9.68$} & {\small $0.82$} & {\small $8.5$}\tabularnewline
{\small July 2025} & {\small $11.76$} & {\small $0.96$} & {\small $8.2$}\tabularnewline
{\small August 2025} & {\small $14.97$} & {\small $1.30$} & {\small $8.7$}\tabularnewline
{\small September 2025} & {\small $21.42$} & {\small $1.82$} & {\small $8.5$}\tabularnewline
{\small October 2025} & {\small $22.35$} & {\small $2.29$} & {\small $10.3$}\tabularnewline
{\small November 2025} & {\small $26.56$} & {\small $3.15$} & {\small $11.8$}\tabularnewline
{\small December 2025} & {\small $26.01$} & {\small $5.60$} & {\small $21.5$}\tabularnewline
{\small January 2026} & {\small $30.74$} & {\small $9.07$} & {\small $29.5$}\tabularnewline
{\small February 2026} & {\small $47.91$} & {\small $21.47$} & {\small $44.8$}\tabularnewline
{\small March 2026} & {\small $76.89$} & {\small $38.90$} & {\small $50.6$}\tabularnewline
{\small April 2026} & {\small $91.23$} & {\small $49.19$} & {\small $53.9$}\tabularnewline
{\small May 2026} & {\small $119.91$} & {\small $67.45$} & {\small $56.3$}\tabularnewline
{\small June 2026} & {\small $188.34$} & {\small $120.16$} & {\small $63.8$}\tabularnewline
{\small July 2026} & {\small $248.40$} & {\small $167.21$} & {\small $67.3$}\tabularnewline
{\small August 2026} & {\small $379.83$} & {\small $268.85$} & {\small $70.8$}\tabularnewline
{\small September 2026} & {\small $483.26$} & {\small $371.10$} & {\small $76.8$}\tabularnewline
 &  &  & \tabularnewline
\hline
\end{tabular}
}{\footnotesize\par}
\end{table}

We then describe the composition of OpenRouter requests by task type in Appendix Table~\ref{tab:tasks}. OpenRouter runs an agentic classifier over a sample of about 0.1 percent of the requests from its users, available from 14 June to 11 October 2026. Agent tasks, such as workflow execution and multi-step planning, account for 33.3 percent of the sampled tokens, and coding tasks, such as implementation and debugging, for another 33.2 percent. Data extraction and transformation account for 6.4 percent, and the remaining labels, led by classification and tagging, for 27.1 percent. Agent and coding requests tend to consume more tokens than other requests. An agent request consumes 48,282 tokens on average and a coding request 61,007, against 9,477 for the remaining tasks. 

We further distinguish agentic consumption on open-weight and closed-weight models. Open-weight models have publicly downloadable weights, whereas closed-weight models do not. For each group, we construct a separate agentic factor as the weekly log growth of its agentic token consumption. Models we cannot classify account for 6 percent of agentic tokens over the sample. We exclude them from both series. The two factors have a correlation of 0.04. %

\subsection{Agentic Betas}\label{sec:betas}

We infer firm-level agentic exposure from the comovement of a firm's weekly log excess return with the agentic AI factor, controlling for the log excess market return. Similarly, we also estimate firm-level exposure to the broad AI factor. Specifically, for firm $i$ in month $t$, we estimate the following regression separately for the AI and agentic factors:
\begin{equation}\label{eq:beta}
r^{e}_{i,\tau} = \alpha^{k}_{i,t} + \beta^{k}_{i,t} \times F^{k}_{\tau} + \beta^{rm,k}_{i,t} \times r^{e}_{m,\tau} + \epsilon^{k}_{i,\tau}, \quad k \in \{AI,Agentic\},
\end{equation}
where $F^{AI}_{\tau}$ and $F^{Agentic}_{\tau}$ are the AI and agentic factors, respectively, and $r^{e}_{i,\tau}$ and $r^{e}_{m,\tau}$ are the weekly log excess returns of firm $i$ and the market. The coefficient $\beta^{k}_{i,t}$ measures the firm's exposure to factor $k$ in month $t$.

Each regression uses the weeks $\tau \in \mathcal{W}_t$, where $\mathcal{W}_t$ is the 26-week window ending in the last week of month $t$. We update the estimates at the end of every month and require at least 18 weekly observations in each window. To limit the influence of extreme observations, we winsorize the firm's excess return and the factor at their 1st and 99th percentiles.

In the subsequent regressions, we standardize each beta to have a mean of zero and a standard deviation of one across firms within each month. Therefore, the resulting coefficient estimates measure the difference in the outcome associated with a one-standard-deviation increase in exposure. The mean cross-sectional correlation between the AI and agentic betas is about 0.6 across months.

\begin{figure}[!htbp]
\caption{{\small\textbf{Agentic and AI Betas by Industry\label{fig:IND}}}}
\medskip{}

This figure shows the market-value-weighted average of the firm-level agentic beta (blue) and AI beta (gray) of firms for each of the ten Fama-French industries. Each average is multiplied by 100 and by the factor's standard deviation over the weeks that are used to calculate the agentic betas. The industries are sorted by their agentic betas.

\medskip{}
\centering{}\includegraphics{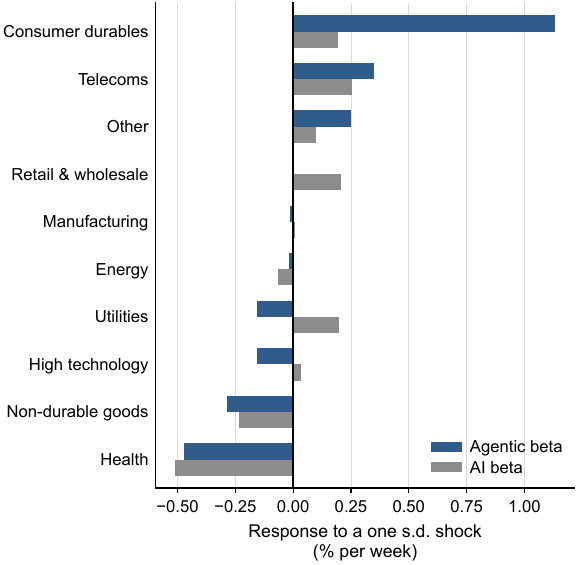}
\end{figure}

To illustrate how these exposures vary across industries, Figure~\ref{fig:IND} plots the market-value-weighted average agentic and AI betas for the ten Fama-French industries. We multiply each industry average by the weekly standard deviation of the corresponding factor and express the result in percent. Therefore, the plotted values measure the weekly return associated with a one-standard-deviation increase in the factor. Consumer durables has the largest agentic beta, followed by telecommunications and the Fama-French ``other'' industry. Health has the most negative exposure to both factors, at $-0.47$ for the agentic beta and $-0.51$ for the AI beta.

\subsection{Lightcast Job Postings}\label{sec:lightcast_data}

We use the data on online job postings from Lightcast (formerly Burning Glass Technologies) to measure firms' labor demand. Lightcast collects job advertisements from company career pages and job boards and removes duplicate advertisements of the same opening. The data identify individual postings and the day each posting first appeared. Our data cover nearly 194 million U.S. job postings from January 2022 to July 2026. Each record includes the employer, location, detailed occupation in the Standard Occupational Classification (SOC), education and experience requirements, advertised salary when available, remote or hybrid status, and the full text of the advertisement.

We link Lightcast employers to publicly listed companies, which we refer to as their parent companies, using employer names and subsidiary records. We first match standardized employer names to company names in Compustat, CRSP, or the list of registrants of the Securities and Exchange Commission (SEC). We also match brands, acronyms, and spelling variants of company names. We accept these matches automatically when the employer and parent have the same industry classification and review the remaining candidates manually. Branches and units that carry the parent's name are assigned to that parent. For subsidiaries with different names, we use the parent's latest annual report (Exhibit 21 of Form 10-K) or the Wharton Research Data Services (WRDS) company subsidiary data to establish the link.

We sum postings across all employers matched to a parent company including subsidiaries, as in \citet{LiuWu2026}. We fix the set of matched employers as of the end of 2024 and use the same set in every month, so that posting growth compares the same employers over time.\footnote{Franchisees are separate employers that often post through the brand's career site, so a brand's postings can include jobs at independently owned outlets. This affects a handful of restaurant, hotel, fitness, hair-salon, and car-rental brands.}

This procedure links 5,558 Lightcast employers to 3,370 listed parent companies with an AI beta in December 2024. We further restrict the regression sample to parents with at least eight postings per month on average in 2024, the last full year before agentic tokens appear in our data, to focus on companies for which Lightcast has adequate coverage. The resulting 1,784 parents account for 34.8 million postings, about 18 percent of all U.S. Lightcast postings over our sample period.

We aggregate postings to the firm-month level and assign zero to months in which a firm has no postings in Lightcast. We measure posting growth over the $h$ months following month $t$ as
\begin{equation}\label{eq:growth}
g_{i,t,h} = 100 \times \left[\ln\left(1 + \bar{N}_{i,t+1,t+h}\right) - \ln\left(1 + \bar{N}_{i,t-2,t}\right)\right],
\end{equation}
where $\bar{N}_{i,a,b}$ is firm $i$'s average number of postings per month from month $a$ to month $b$. The measure compares average monthly postings over the $h$ months after $t$ with the average over the three months ending in $t$. Averaging over this three-month base window smooths month-to-month noise in a firm's postings.\footnote{The results are qualitatively similar without this smoothing.} We multiply the log change by 100 to approximate percentage growth.

We then examine the composition of firms' job postings by seniority, experience requirements, expected AI-tool use, remote or hybrid status, and occupational exposure to AI. Table~\ref{tab:cuts} defines these posting types and reports their summary statistics. For junior and senior positions, we use a classifier trained on Lightcast's seniority labels. Based on job titles and other posting characteristics, Lightcast assigns each posting a probability of being junior and a probability of being senior. Summing these probabilities within a firm-month gives the probability-weighted number of junior and senior postings, respectively. Among postings that state a minimum experience requirement, we distinguish those asking for at most one year (0--1 year) and those asking for six or more years (6+ years). We identify postings that expect AI-tool use through phrases such as ``AI tools,'' ``leverage AI,'' or ``AI-assisted'' in the advertisement, and use Lightcast's labels to identify remote or hybrid positions.

\begin{table}[!htbp]
\stepcounter{anch}
{\footnotesize\caption{{\small\textbf{Job Postings by Type\label{tab:cuts}}}}
\medskip{}
}{\footnotesize\par}

{\footnotesize This table summarizes the job postings by types. For each job posting type, type ``X'' is the job postings by this type and ``not X'' is the firm's other postings. ``Mean X'' shows the average number of type-X postings per firm-month, including firm-months with no type-X postings. For junior and senior postings, counts are probability-weighted. Each posting is weighted by the classifier's estimated probability of belonging to the corresponding category, and these weights are summed within each firm-month. ``Share'' is the ratio of the number of job postings of type ``X'' to the sum of the numbers of job postings of type ``X'' and ``not X'', in percent.\par}

{\footnotesize\medskip{}
}{\footnotesize\par}
\centering{}{\footnotesize{}%
\renewcommand{\arraystretch}{1.10}%
\begin{tabular}{P{2.9cm}P{4.3cm}P{2.7cm}cc}
\hline 
{\footnotesize Cut} & {\footnotesize Type X} & {\footnotesize Not X} & {\footnotesize Mean X} & {\footnotesize Share (\%)}\tabularnewline
\hline 
 &  &  &  & \tabularnewline
{\footnotesize All postings} & {\footnotesize Every posting of the firm} & {\footnotesize } & {\footnotesize 334.4} & {\footnotesize 100.0}\tabularnewline
 &  &  &  & \tabularnewline
{\footnotesize Junior titles} & {\footnotesize Junior titles (probability-weighted)} & {\footnotesize Other scored postings} & {\footnotesize 32.7} & {\footnotesize 9.8}\tabularnewline
{\footnotesize Senior titles} & {\footnotesize Senior titles (probability-weighted)} & {\footnotesize Other scored postings} & {\footnotesize 36.4} & {\footnotesize 10.9}\tabularnewline
{\footnotesize Asks 0--1 years} & {\footnotesize Asks 0 to 1 years of experience} & {\footnotesize Asks 2 or more years} & {\footnotesize 50.7} & {\footnotesize 29.3}\tabularnewline
{\footnotesize Asks 6+ years} & {\footnotesize Asks 6 or more years of experience} & {\footnotesize Asks 0 to 5 years} & {\footnotesize 38.2} & {\footnotesize 22.1}\tabularnewline
{\footnotesize Expects AI-tool use} & {\footnotesize Text expects use of AI tools} & {\footnotesize All other postings} & {\footnotesize 7.9} & {\footnotesize 2.3}\tabularnewline
{\footnotesize Remote or hybrid} & {\footnotesize Remote or hybrid} & {\footnotesize On site} & {\footnotesize 31.5} & {\footnotesize 9.4}\tabularnewline
{\footnotesize Top half, Eloundou} & {\footnotesize Top half, \citeauthor{Eloundou2024} GPT exposure} & {\footnotesize Other scored occupations} & {\footnotesize 157.2} & {\footnotesize 50.1}\tabularnewline
{\footnotesize Top half, Webb AI} & {\footnotesize Top half, Webb AI exposure} & {\footnotesize Other scored occupations} & {\footnotesize 135.6} & {\footnotesize 52.8}\tabularnewline
{\footnotesize Top half, Anthropic exposure} & {\footnotesize Top half, Anthropic Economic Index observed exposure} & {\footnotesize Other scored occupations} & {\footnotesize 153.6} & {\footnotesize 50.4}\tabularnewline
 &  &  &  & \tabularnewline
\hline
\end{tabular}
}{\footnotesize\par}
\end{table}

For occupational AI exposure, we classify occupations using three measures from the literature. The first, from \citet{Eloundou2024}, measures the share of an occupation's tasks for which access to a language model cuts the time required to complete the task by at least half. Tasks receive half weight when that saving requires software built on top of the model. The second is the AI exposure percentile of \citet{webb2020impact}. The third, from the Anthropic Economic Index \citep{MassenkoffMcCrory2026}, measures the share of an occupation's work time spent on tasks that a language model could perform and that see substantial work-related Claude use.\footnote{The Anthropic measure is constructed after the release of Claude Code 2.0, so that we only used it as a retrospective, ex post benchmark. Section~\ref{sec:occupation_exposure} provides further discussions.} For each measure, we rank occupations by the exposure and select the most exposed occupations until they account for half of the sample firms' 2024 postings. We classify these occupations as having substantial overlap with AI under that measure.

For each characteristic of job postings described above, we apply equation~(\ref{eq:growth}) separately to postings with that characteristic, denoted ``X,'' and to the corresponding comparison job posting group without that characteristic, denoted ``Not X,'' to calculate labor demand growth for type ``X'' jobs and for type ``Not X'' jobs, respectively. %
Table \ref{tab:cuts} provides summary statistics of the job posting data based on these job posting characteristics.

Finally, we construct two additional sets of firm-level measures from job postings. First, for each of the three occupational exposure measures, a firm's \emph{AI overlap} is the posting-weighted average of its occupations' scores, calculated over occupations with an available score. Second, a firm's \emph{posted wage} in a given month is the mean log annualized salary across its postings that report salary information.

\subsection{Additional Data}\label{sec:other_data}

We use CRSP stock returns and Compustat financial data to construct the firm characteristics included as controls in our regressions. Compustat also provides our measures of investment. For the panel analysis, we use information available at the end of each month for these controls. For the event study, we use values from 2024, the year before agentic consumption first appears in our data. Appendix Table~\ref{tab:variable_definitions} reports the variable definitions and data sources.

\subsection{Validation of the Agentic Betas}\label{sec:validation}

The AI and agentic betas are inferred from the comovement of firms' stock returns with the growth of AI consumption. In this subsection, we validate them using firms' subsequent disclosure filings and earnings call. If the betas capture firms' exposure to agentic AI, we expect firms with higher betas to report more about agentic AI in their subsequent filings. We test this prediction using the annual and quarterly reports (Forms 10-K and 10-Q) of the firms in our sample. In each report, we use the business section (Item 1 of Form 10-K) and management's discussion and analysis (Item 7 of Form 10-K and Part I, Item 2 of Form 10-Q), where a firm describes its business and operations. We identify the sentences that contain an AI term, such as ``artificial intelligence,'' ``machine learning,'' ``generative AI,'' or ``large language model.'' We construct two measures of AI mentions in firms' disclosure filings. First, an \emph{agentic mention} is a sentence that contains a term for AI agents or agentic AI, such as ``agentic,'' ``AI agent,'' or ``autonomous agent.'' Second, an \emph{AI deployment mention} is a sentence in which a word of use or adoption, such as ``use'', ``deploy'', ``implement,'', or ``adopt,'' appears within 80 characters of an AI term. We use the two measures as proxies for a firm's adoption and deployment of agentic AI.

Then, for each firm and month $t$, we count the mentions in the reports that the firm files over the following six months. For each measure, we use three dependent variables: an indicator for at least one mention, the log of one plus the number of mentions, and the change in this log count from the same six months one year earlier. On average, 5.5 percent of the firm-month observations have at least one agentic mention over the following six months, and 26.9 percent have at least one AI deployment mention. We regress each dependent variable on the AI beta, the agentic beta, or both betas, measured at the end of month $t$ and standardized across firms within the month. All regressions include time fixed effects, and we cluster standard errors by firm.

Table~\ref{tab:mentions} reports the results. Panels A and B report coefficient estimates with the AI beta only and the agentic beta only, respectively. Panel C reports coefficient estimates from a specification that includes both betas. Columns (1) to (3) use agentic mentions, and columns (4) to (6) use AI deployment mentions.

\begin{table}[!htbp]
\stepcounter{anch}
{\footnotesize\caption{{\small\textbf{Future AI Mentions in SEC Filings on the AI and Agentic Betas\label{tab:mentions}}}}
\medskip{}
}{\footnotesize\par}

{\footnotesize This table reports panel regression results of future mentions of AI in firms' SEC filings on the trailing AI beta (Panel A), the trailing agentic beta (Panel B), and both betas in the same regression (Panel C). Mentions are sentences in the business section (Item 1 of Form 10-K) and in management's discussion and analysis (Item 7 of Form 10-K and Part I, Item 2 of Form 10-Q) of the annual and quarterly reports that a firm files in the six months after month $t$. An agentic mention (columns (1) to (3)) is a sentence that contains a term for AI agents or agentic AI. An AI deployment mention (columns (4) to (6)) is a sentence in which a word of use or adoption appears within 80 characters of an AI term. Both measures exclude sentences in risk-factor sections and forward-looking statements, sentences with risk language, and boilerplate text. AI deployment mentions also exclude sentences that mention regulation or competitors. The dependent variable is 100 times an indicator for at least one mention (Any mention), 100 times the log of one plus the number of mentions (Log count), or the change in Log count from the same six months one year earlier ($\Delta$ Log count). The beta is standardized within each month, so the coefficient estimates measure the change in the dependent variable per cross-sectional standard deviation of the beta. All regressions include time fixed effects. The number of observations and the adjusted $R^{2}$ are reported. t-statistics are reported in parentheses. Standard errors are clustered by firm. {*}, {*}{*}, and {*}{*}{*} denote statistical significance at the 10\%, 5\%, and 1\% levels, respectively.\par}

{\footnotesize\medskip{}
}{\footnotesize\par}
\centering{}{\footnotesize{}%
\renewcommand{\arraystretch}{1.10}%
\begin{tabular}{lcccccc}
\hline 
 & {\footnotesize (1)} & {\footnotesize (2)} & {\footnotesize (3)} & {\footnotesize (4)} & {\footnotesize (5)} & {\footnotesize (6)}\tabularnewline
 & \multicolumn{3}{c}{{\footnotesize Agentic mentions}} & \multicolumn{3}{c}{{\footnotesize AI deployment mentions}}\tabularnewline
{\footnotesize Dependent variable} & {\footnotesize Any mention} & {\footnotesize Log count} & {\footnotesize $\Delta$ Log count} & {\footnotesize Any mention} & {\footnotesize Log count} & {\footnotesize $\Delta$ Log count}\tabularnewline
\hline 
\multicolumn{7}{l}{{\footnotesize \textbf{Panel A: $\beta^{AI}$}}}\tabularnewline
 &  &  &  &  &  & \tabularnewline
{\footnotesize $\beta^{AI}$} & {\footnotesize $0.301$} & {\footnotesize $0.127$} & {\footnotesize $0.380$} & {\footnotesize $1.269${*}{*}} & {\footnotesize $1.434$} & {\footnotesize $-0.076$}\tabularnewline
 & {\footnotesize $(0.995)$} & {\footnotesize $(0.244)$} & {\footnotesize $(0.929)$} & {\footnotesize $(2.162)$} & {\footnotesize $(1.383)$} & {\footnotesize $(-0.126)$}\tabularnewline
 &  &  &  &  &  & \tabularnewline
{\footnotesize Adj $R^{2}$} & {\footnotesize $0.005$} & {\footnotesize $0.004$} & {\footnotesize $0.004$} & {\footnotesize $0.044$} & {\footnotesize $0.033$} & {\footnotesize $0.009$}\tabularnewline
{\footnotesize Obs} & {\footnotesize 16,471} & {\footnotesize 16,471} & {\footnotesize 16,427} & {\footnotesize 16,471} & {\footnotesize 16,471} & {\footnotesize 16,427}\tabularnewline
 &  &  &  &  &  & \tabularnewline
\hline 
\multicolumn{7}{l}{{\footnotesize \textbf{Panel B: $\beta^{Agentic}$}}}\tabularnewline
 &  &  &  &  &  & \tabularnewline
{\footnotesize $\beta^{Agentic}$} & {\footnotesize $1.394${*}{*}{*}} & {\footnotesize $2.684${*}{*}{*}} & {\footnotesize $2.024${*}{*}{*}} & {\footnotesize $2.149${*}{*}{*}} & {\footnotesize $4.383${*}{*}{*}} & {\footnotesize $1.450${*}{*}}\tabularnewline
 & {\footnotesize $(4.446)$} & {\footnotesize $(4.814)$} & {\footnotesize $(4.585)$} & {\footnotesize $(3.824)$} & {\footnotesize $(4.329)$} & {\footnotesize $(2.533)$}\tabularnewline
 &  &  &  &  &  & \tabularnewline
{\footnotesize Adj $R^{2}$} & {\footnotesize $0.008$} & {\footnotesize $0.008$} & {\footnotesize $0.008$} & {\footnotesize $0.045$} & {\footnotesize $0.036$} & {\footnotesize $0.010$}\tabularnewline
{\footnotesize Obs} & {\footnotesize 16,471} & {\footnotesize 16,471} & {\footnotesize 16,427} & {\footnotesize 16,471} & {\footnotesize 16,471} & {\footnotesize 16,427}\tabularnewline
 &  &  &  &  &  & \tabularnewline
\hline 
\multicolumn{7}{l}{{\footnotesize \textbf{Panel C: $\beta^{Agentic}$ and $\beta^{AI}$ in the same regression}}}\tabularnewline
 &  &  &  &  &  & \tabularnewline
{\footnotesize $\beta^{Agentic}$} & {\footnotesize $1.852${*}{*}{*}} & {\footnotesize $3.968${*}{*}{*}} & {\footnotesize $2.742${*}{*}{*}} & {\footnotesize $2.139${*}{*}{*}} & {\footnotesize $5.389${*}{*}{*}} & {\footnotesize $2.274${*}{*}{*}}\tabularnewline
 & {\footnotesize $(4.560)$} & {\footnotesize $(4.780)$} & {\footnotesize $(4.690)$} & {\footnotesize $(2.966)$} & {\footnotesize $(3.975)$} & {\footnotesize $(3.096)$}\tabularnewline
{\footnotesize $\beta^{AI}$} & {\footnotesize $-0.782${*}{*}} & {\footnotesize $-2.194${*}{*}{*}} & {\footnotesize $-1.226${*}{*}} & {\footnotesize $0.018$} & {\footnotesize $-1.718$} & {\footnotesize $-1.407${*}}\tabularnewline
 & {\footnotesize $(-1.991)$} & {\footnotesize $(-2.773)$} & {\footnotesize $(-2.247)$} & {\footnotesize $(0.023)$} & {\footnotesize $(-1.242)$} & {\footnotesize $(-1.835)$}\tabularnewline
 &  &  &  &  &  & \tabularnewline
{\footnotesize Adj $R^{2}$} & {\footnotesize $0.009$} & {\footnotesize $0.010$} & {\footnotesize $0.009$} & {\footnotesize $0.045$} & {\footnotesize $0.036$} & {\footnotesize $0.011$}\tabularnewline
{\footnotesize Obs} & {\footnotesize 16,471} & {\footnotesize 16,471} & {\footnotesize 16,427} & {\footnotesize 16,471} & {\footnotesize 16,471} & {\footnotesize 16,427}\tabularnewline
 &  &  &  &  &  & \tabularnewline
\hline
\end{tabular}
}{\footnotesize\par}
\end{table}

In Panel A, the coefficient estimates for AI exposure tend to be positive but are significant in only one of the six specifications, at the 5 percent level. In contrast, Panel B shows that firms with higher agentic betas subsequently report significantly more about AI agents. A one-standard-deviation increase in the agentic beta is associated with a $1.39$ percentage point higher likelihood of at least one agentic mention over the following six months, about 25 percent of the sample mean. The coefficients for the log count and for its change are $2.68$ and $2.02$, respectively. All three estimates are significant at the 1 percent level. The agentic beta also predicts AI deployment mentions. The corresponding coefficients in columns (4) to (6) are $2.15$, $4.38$, and $1.45$, significant at least at the 5 percent level. When we include both AI and agentic betas in Panel C, the coefficients on the agentic beta are positive and significant in all specifications.\footnote{Appendix Table~\ref{tab:mentions_controls} shows that we reach qualitatively similar conclusions controlling for industry fixed effects and the firm characteristics used in the baseline panel regression specification, which we later discuss.} The coefficients on the AI beta, on the other hand, are never positive and significant. Appendix Table~\ref{tab:mentions_filing} further shows that the agentic beta predicts both types of mentions separately in annual reports and agentic mentions in quarterly reports. %

Additionally, we repeat the test using firms' earnings conference calls. We apply the same two measures---agentic AI mentions and AI deployment mentions---to the sentences that the firm's management speaks in the prepared remarks and in the answers to analysts' questions. For each firm and month, we count the mentions in the earnings calls that the firm holds over the following six months and estimate the same regression specifications as in Table~\ref{tab:mentions}. Appendix Table~\ref{tab:mentions_calls} reports the results. In Panel A, the coefficient on the AI beta is significant only in one of the six specifications, at the 10 percent level. In contrast, Panel B shows that a one-standard-deviation increase in the agentic beta is associated with a $1.46$ percentage point higher likelihood of at least one agentic mention by management over the following six months, which is about 7 percent of the sample mean. The coefficients for the log count and for its change are $4.31$ and $2.07$, respectively. The three estimates are all significant at least at the 5 percent level. For AI deployment mentions, the coefficients for the indicator and for the log count are $1.70$ and $5.32$, significant at the 5 and 1 percent levels, respectively. The coefficient for the change in the log count is $1.54$ but is not statistically significant. When we include both betas in Panel C, the coefficients on the agentic beta are positive and largely significant in all six specifications. The coefficients on the AI beta are never positive and significant.

In sum, we find that firms with higher agentic betas subsequently mention AI agents and AI deployment more often in their SEC filings and earnings calls.

\section{Methodology}\label{sec:methodology}

We use a panel analysis framework to test whether firms' AI and agentic betas predict subsequent posting growth. This follows the classic methodology of \citet{ChenRollRoss1986}, \citet{CoxIngersollRoss1985} and \citet{Merton1973}, and more recently of  \citet{KoganPapanikolaou2014}, \citet{BakerBloomDavis2016}, and \citet{MeeuwisPapanikolaouRothbaumSchmidtForthcoming} of studying the effect of stock-level betas on firm-level real outcomes.  We complement the panel analysis with an event study based on the September 2025 release of Claude Code 2.0, comparing firms with different exposure to agentic AI measured before the release, to address the concern that changes in firms' growth opportunities or anticipated workforce adjustments may affect both stock returns and subsequent postings.

We pool firm-month observations and regress firm $i$'s posting growth over the $h \in \{3,6,9\}$ months after month $t$ on its AI exposure:
\begin{equation}\label{eq:posting_panel}
g_{i,t,h} = \alpha_{t,h} + b_h \widetilde{\beta}_{i,t} + \delta_{j(i),h} + X_{i,t}'\gamma_h + \varepsilon_{i,t,h},
\end{equation}
where $g_{i,t,h}$ is the posting growth defined in equation~(\ref{eq:growth}), and $\widetilde{\beta}_{i,t}$ is either the AI beta or the agentic beta, standardized to have a mean of zero and a standard deviation of one across firms in month $t$. We estimate a separate regression for each beta and horizon. The coefficient $b_h$ estimates the difference in subsequent posting growth associated with a one-standard-deviation increase in exposure.

All specifications include month fixed effects $\alpha_{t,h}$ to absorb changes in economic conditions common to firms, such as fluctuations in aggregate demand or commodity prices. In an alternative specification, we add fixed effects $\delta_{j(i),h}$ for firm $i$'s Fama-French 49 industry $j(i)$ to account for persistent differences in posting growth across industries. We also include the firm characteristics $X_{i,t}$ defined in Appendix Table~\ref{tab:variable_definitions} to control for differences in size, financial condition, and recent performance. These characteristics use information available at the end of month $t$, so the controls do not incorporate information from the subsequent posting window. We cluster standard errors by firm to allow regression residuals to be correlated across months, including when posting windows overlap.

For the event study, we focus on the 29 September 2025 release of Claude Code 2.0 alongside Claude Sonnet 4.5 \citep{Anthropic2025Code20,Anthropic2025Sonnet45}. The release combined a substantial improvement in autonomous model performance with tools for delegating and monitoring longer tasks. The release was followed by the sustained increase in the agentic share of AI consumption documented in Section~\ref{sec:openrouter_data}.

We fix each firm's exposure at its agentic beta dated 26 September 2025, the last estimate before the release, so that the measure does not incorporate subsequent stock-return responses. Let $\widetilde{\beta}^{Agentic}_{i,\mathrm{pre}}$ denote this pre-release beta, standardized across firms, and let $t_0$ denote September 2025. For each horizon from one to nine months, we estimate
\begin{equation}\label{eq:posting_shock}
g_{i,t_0,h} = a_h + \theta_h \widetilde{\beta}^{Agentic}_{i,\mathrm{pre}} + \delta_{j(i),h} + X_i'\gamma_h + u_{i,h},
\end{equation}
where $g_{i,t_0,h}$ compares average monthly postings from October 2025 through the end of horizon $h$ with the July--September 2025 base window. We estimate this regression with and without industry fixed effects and the 2024 firm characteristics $X_i$. The coefficient $\theta_h$ measures the difference in posting growth associated with a one-standard-deviation increase in the pre-release agentic beta. An alternative specification also includes the pre-release AI beta to account for broad AI exposure.

Our strategy is broadly similar to that of \citet{HershbeinKahn2018}, who compare changes in vacancy requirements after the 2008 financial crisis across metropolitan areas with different exposure to the crisis. In our setting, identification requires that, conditional on industry and firm characteristics, firms with different pre-release agentic betas would have experienced parallel changes in postings in the absence of the release. Under this assumption, $\theta_h$ identifies the differential labor demand response to the release associated with a one-standard-deviation increase in pre-release agentic exposure.

To assess this assumption and show when the response occurs, we further estimate regression~(\ref{eq:posting_shock}) separately for each calendar month from January 2024 to June 2026. Specifically, we replace the future-window average with that month's postings and keep the July--September 2025 base window and pre-release beta fixed. The pre-release coefficients show whether exposure was already associated with differences in posting growth, and the post-release coefficients show how the relation evolves after the release.

We also examine posting growth within occupations to account for differences in the types of workers firms recruit. Let $g_{i,o,t,h}$ denote posting growth for firm $i$ in occupation $o$, applying equation~(\ref{eq:growth}) to that firm's postings in the occupation. The panel specification is
\begin{equation}\label{eq:occupation_panel}
g_{i,o,t,h} = \alpha_{o,t,h} + b_h \widetilde{\beta}^{Agentic}_{i,t} + \delta_{j(i),h} + X_{i,t}'\gamma_h + \varepsilon_{i,o,t,h},
\end{equation}
where $\alpha_{o,t,h}$ are occupation-by-time fixed effects, so the regression compares firms' labor demand for the same occupation in the same month. The horizons are three, six, and nine months, as in the firm-level panel regression.

For the event study, the corresponding specification is
\begin{equation}\label{eq:occupation_shock}
g_{i,o,t_0,h} = a_{o,h} + \theta_h \widetilde{\beta}^{Agentic}_{i,\mathrm{pre}} + \delta_{j(i),h} + X_i'\gamma_h + u_{i,o,h},
\end{equation}
where $a_{o,h}$ are occupation fixed effects. We estimate a separate cross-sectional regression at each horizon from one to nine months, comparing posting growth for the same occupation across firms with different pre-release agentic betas. Both occupation-level specifications are estimated with and without industry fixed effects and firm characteristics.

For these firm-occupation level specifications, each firm-occupation observation is weighted by the occupation's share of the firm's postings in the corresponding three-month base window to account for firms' differential base labor demand for different occupations. For the panel regression, weights vary by month and give each firm-month a total weight of one. For the event-study specification, weights are fixed using July--September 2025 postings. We cluster standard errors by firm in both occupation-level specifications to allow for correlation across occupations and, in the panel, across months.

\section{Empirical Results}\label{sec:results}

We first show that higher agentic exposure predicts lower subsequent posting growth, whereas higher broad AI exposure generally does not. We then examine the response to the release of Claude Code 2.0, comparing firms with different pre-release agentic betas. Finally, we examine how the decline in posting growth varies across job types, industries, and regions.

\subsection{AI Exposure and Future Posting Growth}\label{sec:panel_results}

We compare broad AI and agentic exposure as predictors of firms' subsequent labor demand. Specifically, we estimate the panel regressions in specification~(\ref{eq:posting_panel}) and compare the coefficients on the AI and agentic betas over horizons of three, six, and nine months. The results are summarized in Table~\ref{tab:P1}. Panels A and B report estimates with the AI beta only and agentic beta only, respectively. Panel C reports estimates from a specification that includes both betas. Columns (1) to (3) include time fixed effects, and columns (4) to (6) further add industry fixed effects and firm characteristic controls.

\begin{table}[!htbp]
\stepcounter{anch}
{\footnotesize\caption{{\small\textbf{Future Posting Growth on the AI and Agentic Betas\label{tab:P1}}}}
\medskip{}
}{\footnotesize\par}

{\footnotesize This table reports panel regression results of future posting growth on the trailing AI beta (Panel A), the trailing agentic beta (Panel B), and both betas in the same regression (Panel C). The dependent variable is the growth of monthly job postings from the three months ending in month $t$ to the following 3, 6, or 9 months, in percent. The beta is standardized within each month, so the coefficient estimates measure approximate percent growth per cross-sectional standard deviation of the beta. Columns (1) to (3) include time fixed effects only. Columns (4) to (6) include Fama-French 49 industry fixed effects, time fixed effects, and control for nine firm characteristics measured at the end of month $t$: the market beta, momentum over the twelve months ending one month before month $t$, return volatility of the 52 weeks ending in month $t$, log market equity at its end, and sales growth, R\&D intensity, book-to-market, return on assets, and a loss dummy. The number of observations and the adjusted $R^{2}$ are reported. t-statistics are reported in parentheses. Standard errors are clustered by firm. {*}, {*}{*}, and {*}{*}{*} denote statistical significance at the 10\%, 5\%, and 1\% levels, respectively.\par}

{\footnotesize\medskip{}
}{\footnotesize\par}
\centering{}{\footnotesize{}%
\renewcommand{\arraystretch}{1.10}%
\begin{tabular}{lcccccc}
\hline 
 & {\footnotesize (1)} & {\footnotesize (2)} & {\footnotesize (3)} & {\footnotesize (4)} & {\footnotesize (5)} & {\footnotesize (6)}\tabularnewline
 & \multicolumn{3}{c}{{\footnotesize Time fixed effects}} & \multicolumn{3}{c}{{\footnotesize With controls}}\tabularnewline
{\footnotesize Horizon (months)} & {\footnotesize 3} & {\footnotesize 6} & {\footnotesize 9} & {\footnotesize 3} & {\footnotesize 6} & {\footnotesize 9}\tabularnewline
\hline 
\multicolumn{7}{l}{{\footnotesize \textbf{Panel A: $\beta^{AI}$}}}\tabularnewline
 &  &  &  &  &  & \tabularnewline
{\footnotesize $\beta^{AI}$} & {\footnotesize $-0.329$} & {\footnotesize $-0.861$} & {\footnotesize $-2.180${*}{*}} & {\footnotesize $-0.045$} & {\footnotesize $-0.414$} & {\footnotesize $-2.139${*}{*}}\tabularnewline
 & {\footnotesize $(-0.656)$} & {\footnotesize $(-1.205)$} & {\footnotesize $(-2.364)$} & {\footnotesize $(-0.089)$} & {\footnotesize $(-0.571)$} & {\footnotesize $(-2.119)$}\tabularnewline
 &  &  &  &  &  & \tabularnewline
{\footnotesize Adj $R^{2}$} & {\footnotesize $0.058$} & {\footnotesize $0.062$} & {\footnotesize $0.011$} & {\footnotesize $0.069$} & {\footnotesize $0.094$} & {\footnotesize $0.063$}\tabularnewline
{\footnotesize Obs} & {\footnotesize 17,090} & {\footnotesize 12,038} & {\footnotesize 6,920} & {\footnotesize 16,819} & {\footnotesize 11,847} & {\footnotesize 6,813}\tabularnewline
 &  &  &  &  &  & \tabularnewline
\hline 
\multicolumn{7}{l}{{\footnotesize \textbf{Panel B: $\beta^{Agentic}$}}}\tabularnewline
 &  &  &  &  &  & \tabularnewline
{\footnotesize $\beta^{Agentic}$} & {\footnotesize $-1.556${*}{*}{*}} & {\footnotesize $-3.033${*}{*}{*}} & {\footnotesize $-4.222${*}{*}{*}} & {\footnotesize $-0.869$} & {\footnotesize $-2.114${*}{*}{*}} & {\footnotesize $-3.198${*}{*}{*}}\tabularnewline
 & {\footnotesize $(-2.779)$} & {\footnotesize $(-4.031)$} & {\footnotesize $(-3.960)$} & {\footnotesize $(-1.506)$} & {\footnotesize $(-2.916)$} & {\footnotesize $(-3.023)$}\tabularnewline
 &  &  &  &  &  & \tabularnewline
{\footnotesize Adj $R^{2}$} & {\footnotesize $0.059$} & {\footnotesize $0.066$} & {\footnotesize $0.017$} & {\footnotesize $0.069$} & {\footnotesize $0.096$} & {\footnotesize $0.065$}\tabularnewline
{\footnotesize Obs} & {\footnotesize 17,090} & {\footnotesize 12,038} & {\footnotesize 6,920} & {\footnotesize 16,819} & {\footnotesize 11,847} & {\footnotesize 6,813}\tabularnewline
 &  &  &  &  &  & \tabularnewline
\hline 
\multicolumn{7}{l}{{\footnotesize \textbf{Panel C: $\beta^{Agentic}$ and $\beta^{AI}$ in the same regression}}}\tabularnewline
 &  &  &  &  &  & \tabularnewline
{\footnotesize $\beta^{Agentic}$} & {\footnotesize $-2.068${*}{*}{*}} & {\footnotesize $-3.384${*}{*}{*}} & {\footnotesize $-4.310${*}{*}{*}} & {\footnotesize $-1.281${*}} & {\footnotesize $-2.474${*}{*}{*}} & {\footnotesize $-2.960${*}{*}}\tabularnewline
 & {\footnotesize $(-2.920)$} & {\footnotesize $(-4.008)$} & {\footnotesize $(-3.325)$} & {\footnotesize $(-1.774)$} & {\footnotesize $(-3.027)$} & {\footnotesize $(-2.263)$}\tabularnewline
{\footnotesize $\beta^{AI}$} & {\footnotesize $0.877$} & {\footnotesize $0.741$} & {\footnotesize $0.162$} & {\footnotesize $0.707$} & {\footnotesize $0.776$} & {\footnotesize $-0.464$}\tabularnewline
 & {\footnotesize $(1.388)$} & {\footnotesize $(0.943)$} & {\footnotesize $(0.148)$} & {\footnotesize $(1.111)$} & {\footnotesize $(0.960)$} & {\footnotesize $(-0.373)$}\tabularnewline
 &  &  &  &  &  & \tabularnewline
{\footnotesize Adj $R^{2}$} & {\footnotesize $0.059$} & {\footnotesize $0.066$} & {\footnotesize $0.016$} & {\footnotesize $0.069$} & {\footnotesize $0.096$} & {\footnotesize $0.065$}\tabularnewline
{\footnotesize Obs} & {\footnotesize 17,090} & {\footnotesize 12,038} & {\footnotesize 6,920} & {\footnotesize 16,819} & {\footnotesize 11,847} & {\footnotesize 6,813}\tabularnewline
 &  &  &  &  &  & \tabularnewline
\hline
\end{tabular}
}{\footnotesize\par}
\end{table}

Panel A shows that the point estimates for the broad AI beta are negative. With time fixed effects, a one-standard-deviation increase in the AI beta is associated with about $0.33$, $0.86$, and $2.18$ percent lower posting growth over three, six, and nine months, respectively. With industry fixed effects and firm characteristic controls, the point estimates are $0.05$, $0.41$, and $2.14$ percent lower growth. Only the nine-month estimates are statistically significant, at the 5 percent level.

Panel B shows one of the paper's main results: a negative and statistically significant relation between a firm's agentic exposure and subsequent posting growth. When we include time fixed effects only, we find that a one-standard-deviation increase in the agentic beta is associated with about $1.56$ percent lower job posting growth over three months, $3.03$ percent over six months, and $4.22$ percent over nine months. These estimates are all significant at the 1 percent level. With industry fixed effects and firm characteristic controls, the point estimates remain negative and mostly significant at the 1 percent level. A one-standard-deviation increase in the agentic beta is associated with about $0.87$, $2.11$, and $3.20$ percent lower growth at the three-, six-, and nine-month horizons, respectively. The six- and nine-month estimates are significant at the 1 percent level.

Panel C reports the results of a specification that includes both the AI and agentic betas. With time fixed effects, a one-standard-deviation increase in the agentic beta is associated with about $2.07$, $3.38$, and $4.31$ percent lower job posting growth over three, six, and nine months, respectively. These coefficient estimates are all significant at the 1 percent level. In contrast, the coefficients on the AI beta are $0.88$, $0.74$, and $0.16$, and none is statistically significant. With industry fixed effects and firm characteristic controls, the agentic-beta estimates imply $1.28$, $2.47$, and $2.96$ percent lower posting growth. These estimates are significant at the 10, 1, and 5 percent levels, respectively. The AI-beta coefficients are about $0.71$, $0.78$, and $-0.46$, none of which is statistically significant. We find that the agentic beta predicts lower posting growth even after controlling for broad AI exposure.

We also examine whether the relation persists within occupations. We use the panel specification~(\ref{eq:occupation_panel}), with occupation-by-time fixed effects, to compare firms' labor demand for the same occupation. Appendix Table~\ref{tab:P2} shows that, with industry fixed effects and firm characteristic controls, a one-standard-deviation increase in the agentic beta is associated with about $0.74$, $1.67$, and $2.44$ percent lower posting growth over three, six, and nine months, respectively. The six- and nine-month estimates are significant at the 1 percent level.

\subsection{Claude Code 2.0 Release and Labor Demand Decline}\label{sec:release_results}

The panel regression results show that, with both betas in the regression, only firms' agentic exposure is negatively and significantly associated with subsequent labor demand growth. A potential endogeneity concern is that unobserved changes in firms' growth opportunities or anticipated workforce adjustments may affect both their agentic betas and subsequent labor demand. To address this concern, we estimate an event-study specification.

Specifically, we use the September 2025 release of Claude Code 2.0, which significantly expanded AI's agentic capabilities. We compare labor demand growth after the release across firms with different pre-release agentic betas. Measuring exposure before the release keeps subsequent stock-return responses out of the exposure measure. Under the parallel-trends assumption discussed in Section~\ref{sec:methodology}, comparing firms with different pre-release betas can help identify their differential labor demand response to the release.

We report the estimates of specification~(\ref{eq:posting_shock}) over one to nine months in Table~\ref{tab:S1}. We find that firms with higher pre-release agentic betas have lower labor demand growth after the release. In Panel A, with industry fixed effects and firm characteristic controls, a one-standard-deviation increase in the agentic beta is associated with about $2.17$ percent lower growth over the first month, $4.52$ percent over four months, $4.73$ percent over six months, and $4.33$ percent over the full nine months. The estimates are significant at the 5 percent level at every horizon from four months onward and at the 1 percent level from five through nine months.

\begin{table}[!p]
\stepcounter{anch}
{\footnotesize\caption{{\small\textbf{Posting Growth after the Claude Code 2.0 Release on the Agentic and AI Betas\label{tab:S1}}}}
\smallskip{}
}{\footnotesize\par}

{\footnotesize This table reports the regression results of posting growth after the 29 September 2025 release of Claude Code 2.0 on the agentic and AI betas dated 26 September 2025. The dependent variable is the growth of monthly postings from the three months before the shock to the months after it, in percent, one regression per horizon of 1 to 9 months. The betas are standardized across firms. Panel A includes the agentic beta alone. Column (1) has no controls. Column (2) adds Fama-French 49 industry fixed effects and nine firm characteristics: the market beta, momentum, return volatility, sales growth, R\&D intensity, book-to-market, log market equity, return on assets, and a loss dummy. Panel B includes both the agentic beta and the AI beta in the same regression, and each pair of columns reports the two coefficients of one regression: columns (1) and (2) have no controls, and columns (3) and (4) add the industry fixed effects and the nine characteristic controls. The number of observations and the adjusted $R^{2}$ of each regression are reported. t-statistics are reported in parentheses. {*}, {*}{*}, and {*}{*}{*} denote statistical significance at the 10\%, 5\%, and 1\% levels, respectively.\par}

{\footnotesize\smallskip{}
}{\footnotesize\par}

\centering
{\footnotesize
\renewcommand{\arraystretch}{0.87}
\begin{tabular}{lcccccc}
\hline
\multicolumn{7}{l}{{\footnotesize \textbf{Panel A: $\beta^{Agentic}$ alone}}}\tabularnewline
 & \multicolumn{2}{c}{{\footnotesize (1)}} & \multicolumn{2}{c}{{\footnotesize (2)}} & \multicolumn{2}{c}{{\footnotesize Adj $R^{2}$}}\tabularnewline
{\footnotesize Months after the shock} & \multicolumn{2}{c}{{\footnotesize No controls}} & \multicolumn{2}{c}{{\footnotesize With controls}} & {\footnotesize (1)} & {\footnotesize (2)}\tabularnewline
\hline
{\footnotesize 1} & \multicolumn{2}{c}{{\footnotesize $-0.870$}} & \multicolumn{2}{c}{{\footnotesize $-2.167$}} & {\footnotesize $0.000$} & {\footnotesize $0.002$}\tabularnewline
 & \multicolumn{2}{c}{{\footnotesize $(-0.535)$}} & \multicolumn{2}{c}{{\footnotesize $(-1.011)$}} &  & \tabularnewline
{\footnotesize 2} & \multicolumn{2}{c}{{\footnotesize $-2.697${*}}} & \multicolumn{2}{c}{{\footnotesize $-3.836${*}{*}}} & {\footnotesize $0.002$} & {\footnotesize $0.008$}\tabularnewline
 & \multicolumn{2}{c}{{\footnotesize $(-1.796)$}} & \multicolumn{2}{c}{{\footnotesize $(-2.052)$}} &  & \tabularnewline
{\footnotesize 3} & \multicolumn{2}{c}{{\footnotesize $-2.689${*}}} & \multicolumn{2}{c}{{\footnotesize $-3.383${*}}} & {\footnotesize $0.002$} & {\footnotesize $0.009$}\tabularnewline
 & \multicolumn{2}{c}{{\footnotesize $(-1.736)$}} & \multicolumn{2}{c}{{\footnotesize $(-1.772)$}} &  & \tabularnewline
{\footnotesize 4} & \multicolumn{2}{c}{{\footnotesize $-3.884${*}{*}}} & \multicolumn{2}{c}{{\footnotesize $-4.521${*}{*}}} & {\footnotesize $0.006$} & {\footnotesize $0.017$}\tabularnewline
 & \multicolumn{2}{c}{{\footnotesize $(-2.560)$}} & \multicolumn{2}{c}{{\footnotesize $(-2.531)$}} &  & \tabularnewline
{\footnotesize 5} & \multicolumn{2}{c}{{\footnotesize $-4.502${*}{*}{*}}} & \multicolumn{2}{c}{{\footnotesize $-4.863${*}{*}{*}}} & {\footnotesize $0.008$} & {\footnotesize $0.033$}\tabularnewline
 & \multicolumn{2}{c}{{\footnotesize $(-3.125)$}} & \multicolumn{2}{c}{{\footnotesize $(-3.051)$}} &  & \tabularnewline
{\footnotesize 6} & \multicolumn{2}{c}{{\footnotesize $-4.785${*}{*}{*}}} & \multicolumn{2}{c}{{\footnotesize $-4.731${*}{*}{*}}} & {\footnotesize $0.009$} & {\footnotesize $0.043$}\tabularnewline
 & \multicolumn{2}{c}{{\footnotesize $(-3.332)$}} & \multicolumn{2}{c}{{\footnotesize $(-2.979)$}} &  & \tabularnewline
{\footnotesize 7} & \multicolumn{2}{c}{{\footnotesize $-4.890${*}{*}{*}}} & \multicolumn{2}{c}{{\footnotesize $-4.345${*}{*}{*}}} & {\footnotesize $0.010$} & {\footnotesize $0.057$}\tabularnewline
 & \multicolumn{2}{c}{{\footnotesize $(-3.353)$}} & \multicolumn{2}{c}{{\footnotesize $(-2.716)$}} &  & \tabularnewline
{\footnotesize 8} & \multicolumn{2}{c}{{\footnotesize $-4.610${*}{*}{*}}} & \multicolumn{2}{c}{{\footnotesize $-4.196${*}{*}{*}}} & {\footnotesize $0.008$} & {\footnotesize $0.064$}\tabularnewline
 & \multicolumn{2}{c}{{\footnotesize $(-3.103)$}} & \multicolumn{2}{c}{{\footnotesize $(-2.615)$}} &  & \tabularnewline
{\footnotesize 9} & \multicolumn{2}{c}{{\footnotesize $-4.772${*}{*}{*}}} & \multicolumn{2}{c}{{\footnotesize $-4.326${*}{*}{*}}} & {\footnotesize $0.009$} & {\footnotesize $0.070$}\tabularnewline
 & \multicolumn{2}{c}{{\footnotesize $(-3.199)$}} & \multicolumn{2}{c}{{\footnotesize $(-2.732)$}} &  & \tabularnewline
{\footnotesize Obs} & \multicolumn{2}{c}{{\footnotesize 1,725}} & \multicolumn{2}{c}{{\footnotesize 1,394}} &  & \tabularnewline
\hline
\noalign{\vskip 6pt}
\multicolumn{7}{l}{{\footnotesize \textbf{Panel B: $\beta^{Agentic}$ and $\beta^{AI}$ in the same regression}}}\tabularnewline
 & {\footnotesize (1)} & {\footnotesize (2)} & {\footnotesize (3)} & {\footnotesize (4)} & \multicolumn{2}{c}{{\footnotesize Adj $R^{2}$}}\tabularnewline
 & \multicolumn{2}{c}{{\footnotesize No controls}} & \multicolumn{2}{c}{{\footnotesize With controls}} &  & \tabularnewline
{\footnotesize Months after the shock} & {\footnotesize $\beta^{Agentic}$} & {\footnotesize $\beta^{AI}$} & {\footnotesize $\beta^{Agentic}$} & {\footnotesize $\beta^{AI}$} & {\footnotesize (1)--(2)} & {\footnotesize (3)--(4)}\tabularnewline
\hline
{\footnotesize 1} & {\footnotesize $-0.600$} & {\footnotesize $-0.504$} & {\footnotesize $-0.683$} & {\footnotesize $-2.664$} & {\footnotesize $-0.001$} & {\footnotesize $0.003$}\tabularnewline
 & {\footnotesize $(-0.308)$} & {\footnotesize $(-0.325)$} & {\footnotesize $(-0.281)$} & {\footnotesize $(-1.304)$} &  & \tabularnewline
{\footnotesize 2} & {\footnotesize $-3.280${*}} & {\footnotesize $1.090$} & {\footnotesize $-3.342$} & {\footnotesize $-0.887$} & {\footnotesize $0.002$} & {\footnotesize $0.007$}\tabularnewline
 & {\footnotesize $(-1.753)$} & {\footnotesize $(0.743)$} & {\footnotesize $(-1.497)$} & {\footnotesize $(-0.485)$} &  & \tabularnewline
{\footnotesize 3} & {\footnotesize $-2.615$} & {\footnotesize $-0.138$} & {\footnotesize $-2.374$} & {\footnotesize $-1.811$} & {\footnotesize $0.002$} & {\footnotesize $0.009$}\tabularnewline
 & {\footnotesize $(-1.369)$} & {\footnotesize $(-0.093)$} & {\footnotesize $(-1.058)$} & {\footnotesize $(-1.003)$} &  & \tabularnewline
{\footnotesize 4} & {\footnotesize $-4.042${*}{*}} & {\footnotesize $0.295$} & {\footnotesize $-3.297$} & {\footnotesize $-2.198$} & {\footnotesize $0.005$} & {\footnotesize $0.017$}\tabularnewline
 & {\footnotesize $(-2.190)$} & {\footnotesize $(0.206)$} & {\footnotesize $(-1.612)$} & {\footnotesize $(-1.282)$} &  & \tabularnewline
{\footnotesize 5} & {\footnotesize $-4.753${*}{*}{*}} & {\footnotesize $0.470$} & {\footnotesize $-4.118${*}{*}} & {\footnotesize $-1.338$} & {\footnotesize $0.008$} & {\footnotesize $0.032$}\tabularnewline
 & {\footnotesize $(-2.721)$} & {\footnotesize $(0.331)$} & {\footnotesize $(-2.205)$} & {\footnotesize $(-0.782)$} &  & \tabularnewline
{\footnotesize 6} & {\footnotesize $-5.020${*}{*}{*}} & {\footnotesize $0.440$} & {\footnotesize $-4.151${*}{*}} & {\footnotesize $-1.040$} & {\footnotesize $0.009$} & {\footnotesize $0.043$}\tabularnewline
 & {\footnotesize $(-2.852)$} & {\footnotesize $(0.308)$} & {\footnotesize $(-2.230)$} & {\footnotesize $(-0.616)$} &  & \tabularnewline
{\footnotesize 7} & {\footnotesize $-5.103${*}{*}{*}} & {\footnotesize $0.399$} & {\footnotesize $-4.018${*}{*}} & {\footnotesize $-0.587$} & {\footnotesize $0.009$} & {\footnotesize $0.056$}\tabularnewline
 & {\footnotesize $(-2.867)$} & {\footnotesize $(0.276)$} & {\footnotesize $(-2.163)$} & {\footnotesize $(-0.347)$} &  & \tabularnewline
{\footnotesize 8} & {\footnotesize $-4.695${*}{*}} & {\footnotesize $0.159$} & {\footnotesize $-3.824${*}{*}} & {\footnotesize $-0.667$} & {\footnotesize $0.008$} & {\footnotesize $0.063$}\tabularnewline
 & {\footnotesize $(-2.570)$} & {\footnotesize $(0.108)$} & {\footnotesize $(-2.053)$} & {\footnotesize $(-0.396)$} &  & \tabularnewline
{\footnotesize 9} & {\footnotesize $-4.763${*}{*}{*}} & {\footnotesize $-0.017$} & {\footnotesize $-3.877${*}{*}} & {\footnotesize $-0.806$} & {\footnotesize $0.008$} & {\footnotesize $0.069$}\tabularnewline
 & {\footnotesize $(-2.598)$} & {\footnotesize $(-0.012)$} & {\footnotesize $(-2.092)$} & {\footnotesize $(-0.479)$} &  & \tabularnewline
{\footnotesize Obs} & {\footnotesize 1,725} & {\footnotesize 1,725} & {\footnotesize 1,394} & {\footnotesize 1,394} &  & \tabularnewline
\hline
\end{tabular}
\par}
\end{table}

To distinguish agentic exposure from broad AI exposure, we include both the AI and agentic betas in the same regression in Panel B. Firms with higher agentic betas continue to have lower posting growth. With industry fixed effects and firm characteristic controls, the agentic coefficients are $-4.15$ at the six-month horizon and $-3.88$ at the nine-month horizon. They are significant at the 5 percent level at every horizon from five through nine months. The corresponding AI-beta estimates are $-1.04$ and $-0.81$, respectively, and the coefficient on the AI beta is statistically insignificant at every horizon. At the sample mean of 334.4 monthly postings, the six-month agentic estimate in Panel B implies about 82 fewer postings out of 2,006. This corresponds to 4.1 percent lower labor demand. At nine months, the point estimate implies about 114 fewer postings out of 3,010, corresponding to 3.8 percent lower labor demand.

We also estimate a non-agentic beta using specification~(\ref{eq:beta}), with the weekly log growth of tokens associated with responses that do not end in a tool call as the factor. Appendix Table~\ref{tab:nonagentic} shows that the coefficient estimates on the non-agentic beta are generally insignificant, both alone and alongside the agentic beta, while those on the agentic beta remain close to the baseline.

To complement the conditional estimates, we sort firms into quartiles of their pre-release agentic betas and compare the average posting growth within each quartile (Appendix Table~\ref{tab:agentic_beta_quartiles}). Growth is measured as $100$ times the log change in one plus average monthly postings over the first $h$ post-release months relative to July--September 2025. The table shows that the average growth is negative in every quartile over the first three months. At six months, the average is $1.39$ in the lowest-exposure quartile and $-6.31$ in the highest, a difference of about $-7.70$ percent. At nine months, the corresponding averages are $7.10$ and $-1.33$, with the latter near zero, and the difference is about $-8.43$ percent. Both differences are significant at the 5 percent level.

Moreover, we examine whether the relation between agentic exposure and labor demand persists within occupations. We estimate equation~(\ref{eq:occupation_shock}) separately at each horizon, including occupation fixed effects and weighting observations by their base-window posting shares. With industry fixed effects and firm characteristic controls, we find that a one-standard-deviation increase in the agentic beta is associated with about $3.28$ percent lower posting growth at the six-month horizon and about $3.24$ percent lower posting growth at the nine-month horizon in Table~\ref{tab:S7}. Both estimates are significant at the 5 percent level.

\begin{table}[!htbp]
\stepcounter{anch}
{\footnotesize\caption{{\small\textbf{Posting Growth after the Claude Code 2.0 Release within Occupation on the Agentic Beta\label{tab:S7}}}}
\medskip{}
}{\footnotesize\par}

{\footnotesize This table repeats the regression results of Panel A of Table~\ref{tab:S1} within occupation. Observations are at the occupation-firm level. The dependent variable is the growth of the occupation's monthly postings from the three months before the release to the months after the release, in percent, one regression per horizon of 1 to 9 months. The independent variable is the agentic beta dated 26 September 2025, standardized across firms. Each occupation is weighted by its share of the firm's postings in the three months before the release. Both columns include occupation fixed effects. Column (2) adds Fama-French 49 industry fixed effects and the nine firm characteristic controls. The number of observations and the adjusted $R^{2}$ of each regression are reported. t-statistics are reported in parentheses. Standard errors are clustered by firm. {*}, {*}{*}, and {*}{*}{*} denote statistical significance at the 10\%, 5\%, and 1\% levels, respectively.\par}

{\footnotesize\medskip{}
}{\footnotesize\par}
\centering{}{\footnotesize{}%
\renewcommand{\arraystretch}{1.10}%
\begin{tabular}{lcccc}
\hline 
 & {\footnotesize (1)} & {\footnotesize (2)} & \multicolumn{2}{c}{{\footnotesize Adj $R^{2}$}}\tabularnewline
{\footnotesize Months after the shock} & {\footnotesize No controls} & {\footnotesize With controls} & {\footnotesize (1)} & {\footnotesize (2)}\tabularnewline
\hline 
 &  &  &  & \tabularnewline
{\footnotesize 1} & {\footnotesize $-1.231$} & {\footnotesize $-2.209$} & {\footnotesize $0.037$} & {\footnotesize $0.060$}\tabularnewline
 & {\footnotesize $(-0.975)$} & {\footnotesize $(-1.198)$} &  & \tabularnewline
{\footnotesize 2} & {\footnotesize $-2.072${*}} & {\footnotesize $-2.520$} & {\footnotesize $0.035$} & {\footnotesize $0.056$}\tabularnewline
 & {\footnotesize $(-1.844)$} & {\footnotesize $(-1.514)$} &  & \tabularnewline
{\footnotesize 3} & {\footnotesize $-2.012${*}} & {\footnotesize $-2.551$} & {\footnotesize $0.038$} & {\footnotesize $0.059$}\tabularnewline
 & {\footnotesize $(-1.833)$} & {\footnotesize $(-1.543)$} &  & \tabularnewline
{\footnotesize 4} & {\footnotesize $-2.040${*}{*}} & {\footnotesize $-2.974${*}} & {\footnotesize $0.041$} & {\footnotesize $0.062$}\tabularnewline
 & {\footnotesize $(-1.980)$} & {\footnotesize $(-1.953)$} &  & \tabularnewline
{\footnotesize 5} & {\footnotesize $-2.307${*}{*}} & {\footnotesize $-3.331${*}{*}} & {\footnotesize $0.044$} & {\footnotesize $0.067$}\tabularnewline
 & {\footnotesize $(-2.214)$} & {\footnotesize $(-2.156)$} &  & \tabularnewline
{\footnotesize 6} & {\footnotesize $-2.353${*}{*}} & {\footnotesize $-3.284${*}{*}} & {\footnotesize $0.047$} & {\footnotesize $0.070$}\tabularnewline
 & {\footnotesize $(-2.241)$} & {\footnotesize $(-2.118)$} &  & \tabularnewline
{\footnotesize 7} & {\footnotesize $-2.282${*}{*}} & {\footnotesize $-3.110${*}{*}} & {\footnotesize $0.050$} & {\footnotesize $0.074$}\tabularnewline
 & {\footnotesize $(-2.172)$} & {\footnotesize $(-2.011)$} &  & \tabularnewline
{\footnotesize 8} & {\footnotesize $-2.222${*}{*}} & {\footnotesize $-3.036${*}} & {\footnotesize $0.052$} & {\footnotesize $0.077$}\tabularnewline
 & {\footnotesize $(-2.107)$} & {\footnotesize $(-1.956)$} &  & \tabularnewline
{\footnotesize 9} & {\footnotesize $-2.375${*}{*}} & {\footnotesize $-3.236${*}{*}} & {\footnotesize $0.053$} & {\footnotesize $0.080$}\tabularnewline
 & {\footnotesize $(-2.266)$} & {\footnotesize $(-2.110)$} &  & \tabularnewline
 &  &  &  & \tabularnewline
{\footnotesize Obs} & {\footnotesize 92,102} & {\footnotesize 77,133} &  & \tabularnewline
 &  &  &  & \tabularnewline
\hline
\end{tabular}
}{\footnotesize\par}
\end{table}

\begin{figure}[!htbp]
\caption{{\small\textbf{Postings by Month around the Release -- Agentic Beta\label{fig:S1}}}}
\medskip{}

This figure shows the monthly point estimates on the agentic beta for growth in total firm postings (Panel A) and in postings within occupations (Panel B). Posting growth is 100 times the log change in one plus postings from the July--September 2025 monthly average (shaded) to the month shown. Both panels use the agentic beta dated 26 September 2025 and include Fama-French 49 industry fixed effects and the nine firm characteristic controls. The bars show 95\% confidence intervals. Standard errors are clustered by firm in Panel B.

\medskip{}
\centering{}\includegraphics[width=\linewidth]{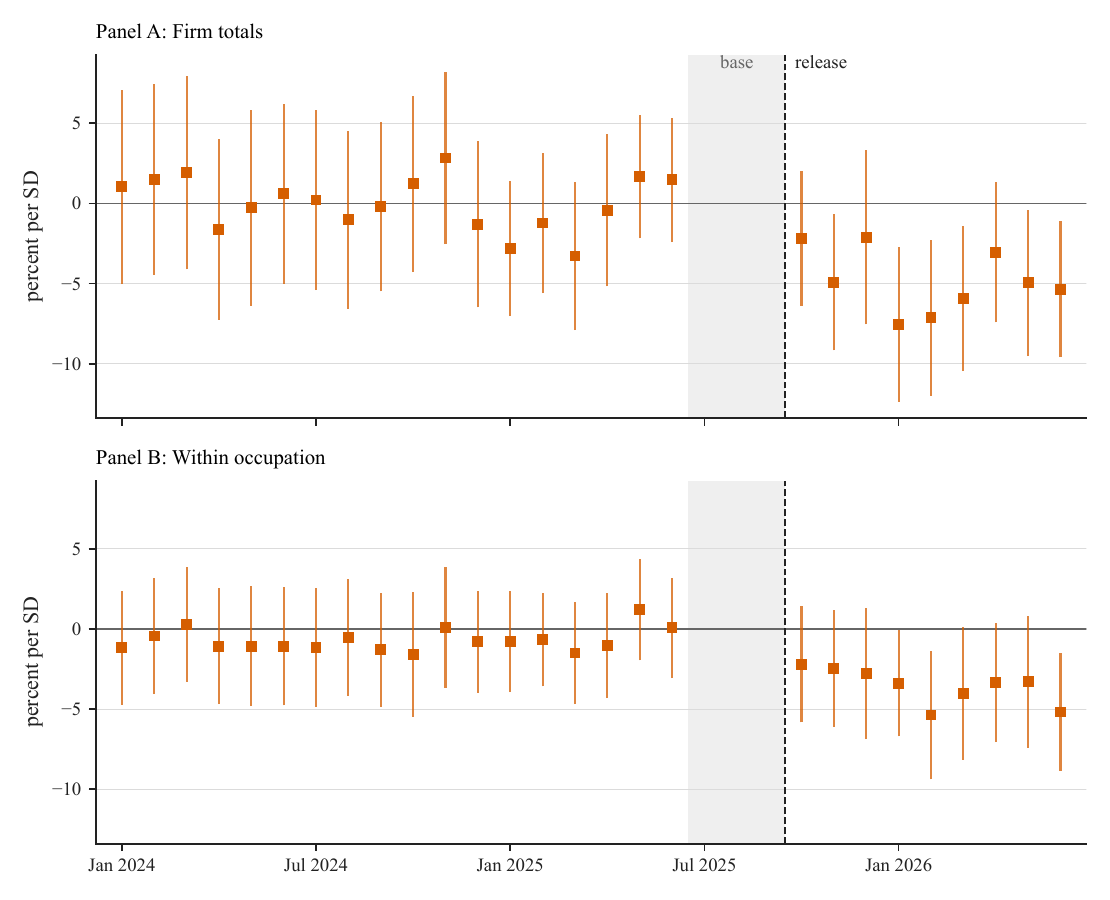}
\end{figure}

Finally, to assess the parallel-trends assumption and show when the labor demand response occurs, we examine the monthly coefficients in Figure~\ref{fig:S1}. Panel A shows the baseline specification, and Panel B uses the within-occupation specification. Specifically, we estimate specifications~(\ref{eq:posting_shock}) and~(\ref{eq:occupation_shock}) separately for each calendar month. In both panels, the 95 percent confidence intervals include zero in every pre-release month, consistent with the parallel-trends assumption. After the release, we find negative coefficients in every month from October 2025 through June 2026. In Panel A, the point estimates reach $-7.6$ in January 2026 and $-7.1$ in February, and remain at about $-5.4$ in June. Within occupations, the point estimates reach about $-5.4$ in February 2026 and $-5.2$ in June.

\subsection{Posting Growth by Job Type}\label{sec:posting_types}

In this subsection, we use detailed characteristics of job postings to examine how the relation between agentic AI exposure and firms' labor demand varies across job types. In Table~\ref{tab:cuts-shock}, we compare the relation between agentic AI exposure and firms' job posting growth across the nine posting types defined in Table~\ref{tab:cuts}. For each job type (``X''), we compare the relation between agentic AI exposure and growth in type-X postings with that between agentic AI exposure and growth in the firm's other postings (``non-X''). The Relative columns report the difference between the two coefficient estimates. We provide a detailed discussion of the results based on Panel B, columns (4) to (6), which use the specification including the AI beta, industry fixed effects, and firm characteristic controls. Results under the other specifications are qualitatively similar.

\begin{table}[!p]
\stepcounter{anch}
{\footnotesize\caption{{\small\textbf{Posting Growth after the Claude Code 2.0 Release by Job Type on Agentic Beta\label{tab:cuts-shock}}}}
\smallskip{}
}{\footnotesize\par}

{\footnotesize This table reports the regression results of Table~\ref{tab:S1} for six-month posting growth by job type. For each job type, X is the point estimate based on the growth in postings of that type and Not X is the point estimate based on the growth in the comparison group. Relative is the difference in the point estimates between the two specifications. Panel A includes the agentic beta alone in the regression specification. Panel B includes both the AI beta and the agentic beta in the regression specification. Both betas are dated 26 September 2025 and standardized across firms. Columns (1) to (3) include no controls. Columns (4) to (6) include Fama-French 49 industry fixed effects and the nine firm characteristic controls. The number of observations is reported. t-statistics are reported in parentheses. {*}, {*}{*}, and {*}{*}{*} denote statistical significance at the 10\%, 5\%, and 1\% levels, respectively. \par}

{\footnotesize\smallskip{}
}{\footnotesize\par}

\centering
{\footnotesize
\renewcommand{\arraystretch}{0.87}
\setlength{\tabcolsep}{5pt}
\begin{tabular}{lcccccc}
\hline
 & {\footnotesize (1)} & {\footnotesize (2)} & {\footnotesize (3)} & {\footnotesize (4)} & {\footnotesize (5)} & {\footnotesize (6)}\tabularnewline
 & \multicolumn{3}{c}{{\footnotesize No controls}} & \multicolumn{3}{c}{{\footnotesize With controls}}\tabularnewline
{\footnotesize Posting type} & {\footnotesize X} & {\footnotesize Not X} & {\footnotesize Relative} & {\footnotesize X} & {\footnotesize Not X} & {\footnotesize Relative}\tabularnewline
\hline
\multicolumn{7}{l}{{\footnotesize \textbf{Panel A: $\beta^{Agentic}$ alone}}}\tabularnewline
{\footnotesize All postings} & {\footnotesize $-4.785${*}{*}{*}} &  &  & {\footnotesize $-4.731${*}{*}{*}} &  & \tabularnewline
 & {\footnotesize $(-3.332)$} &  &  & {\footnotesize $(-2.979)$} &  & \tabularnewline
{\footnotesize Junior titles} & {\footnotesize $-4.125${*}{*}{*}} & {\footnotesize $-4.595${*}{*}{*}} & {\footnotesize $0.470$} & {\footnotesize $-3.526${*}{*}} & {\footnotesize $-4.566${*}{*}{*}} & {\footnotesize $1.040$}\tabularnewline
 & {\footnotesize $(-3.119)$} & {\footnotesize $(-3.208)$} & {\footnotesize $(0.335)$} & {\footnotesize $(-1.990)$} & {\footnotesize $(-2.881)$} & {\footnotesize $(0.627)$}\tabularnewline
{\footnotesize Senior titles} & {\footnotesize $-1.379$} & {\footnotesize $-5.061${*}{*}{*}} & {\footnotesize $3.682${*}{*}{*}} & {\footnotesize $-0.514$} & {\footnotesize $-5.094${*}{*}{*}} & {\footnotesize $4.581${*}{*}{*}}\tabularnewline
 & {\footnotesize $(-1.171)$} & {\footnotesize $(-3.515)$} & {\footnotesize $(3.164)$} & {\footnotesize $(-0.335)$} & {\footnotesize $(-3.208)$} & {\footnotesize $(3.205)$}\tabularnewline
{\footnotesize Asks 0--1 years} & {\footnotesize $-3.862${*}{*}{*}} & {\footnotesize $-3.919${*}{*}{*}} & {\footnotesize $0.057$} & {\footnotesize $-3.074${*}{*}} & {\footnotesize $-4.495${*}{*}} & {\footnotesize $1.422$}\tabularnewline
 & {\footnotesize $(-3.290)$} & {\footnotesize $(-2.656)$} & {\footnotesize $(0.037)$} & {\footnotesize $(-2.032)$} & {\footnotesize $(-2.527)$} & {\footnotesize $(0.705)$}\tabularnewline
{\footnotesize Asks 6+ years} & {\footnotesize $-1.076$} & {\footnotesize $-4.867${*}{*}{*}} & {\footnotesize $3.790${*}{*}} & {\footnotesize $-0.440$} & {\footnotesize $-5.207${*}{*}{*}} & {\footnotesize $4.768${*}{*}}\tabularnewline
 & {\footnotesize $(-0.735)$} & {\footnotesize $(-3.491)$} & {\footnotesize $(2.342)$} & {\footnotesize $(-0.230)$} & {\footnotesize $(-3.054)$} & {\footnotesize $(1.974)$}\tabularnewline
{\footnotesize Expects AI-tool use} & {\footnotesize $-0.225$} & {\footnotesize $-5.176${*}{*}{*}} & {\footnotesize $4.951${*}{*}} & {\footnotesize $-1.581$} & {\footnotesize $-5.118${*}{*}{*}} & {\footnotesize $3.536$}\tabularnewline
 & {\footnotesize $(-0.137)$} & {\footnotesize $(-3.527)$} & {\footnotesize $(2.370)$} & {\footnotesize $(-0.686)$} & {\footnotesize $(-3.068)$} & {\footnotesize $(1.333)$}\tabularnewline
{\footnotesize Remote or hybrid} & {\footnotesize $-2.004$} & {\footnotesize $-5.086${*}{*}{*}} & {\footnotesize $3.082${*}} & {\footnotesize $-2.143$} & {\footnotesize $-4.879${*}{*}{*}} & {\footnotesize $2.736$}\tabularnewline
 & {\footnotesize $(-1.321)$} & {\footnotesize $(-3.395)$} & {\footnotesize $(1.708)$} & {\footnotesize $(-1.019)$} & {\footnotesize $(-2.767)$} & {\footnotesize $(1.082)$}\tabularnewline
{\footnotesize Top half, Eloundou} & {\footnotesize $-3.504${*}{*}} & {\footnotesize $-5.127${*}{*}{*}} & {\footnotesize $1.624$} & {\footnotesize $-3.679${*}{*}} & {\footnotesize $-4.722${*}{*}} & {\footnotesize $1.043$}\tabularnewline
 & {\footnotesize $(-2.512)$} & {\footnotesize $(-3.486)$} & {\footnotesize $(1.159)$} & {\footnotesize $(-2.209)$} & {\footnotesize $(-2.431)$} & {\footnotesize $(0.551)$}\tabularnewline
{\footnotesize Top half, Webb AI} & {\footnotesize $-3.677${*}{*}} & {\footnotesize $-3.475${*}{*}} & {\footnotesize $-0.202$} & {\footnotesize $-2.645$} & {\footnotesize $-4.524${*}{*}} & {\footnotesize $1.879$}\tabularnewline
 & {\footnotesize $(-2.540)$} & {\footnotesize $(-2.456)$} & {\footnotesize $(-0.159)$} & {\footnotesize $(-1.509)$} & {\footnotesize $(-2.494)$} & {\footnotesize $(1.068)$}\tabularnewline
{\footnotesize Top half, Anthropic exposure} & {\footnotesize $-4.672${*}{*}{*}} & {\footnotesize $-3.511${*}{*}} & {\footnotesize $-1.161$} & {\footnotesize $-4.901${*}{*}{*}} & {\footnotesize $-1.803$} & {\footnotesize $-3.098$}\tabularnewline
 & {\footnotesize $(-3.306)$} & {\footnotesize $(-2.281)$} & {\footnotesize $(-0.735)$} & {\footnotesize $(-2.728)$} & {\footnotesize $(-0.900)$} & {\footnotesize $(-1.357)$}\tabularnewline
{\footnotesize Obs} & {\footnotesize 1,725} & {\footnotesize 1,725} & {\footnotesize 1,725} & {\footnotesize 1,394} & {\footnotesize 1,394} & {\footnotesize 1,394}\tabularnewline
\hline
\noalign{\vskip 6pt}
\multicolumn{7}{l}{{\footnotesize \textbf{Panel B: $\beta^{Agentic}$, controlling for $\beta^{AI}$}}}\tabularnewline
{\footnotesize All postings} & {\footnotesize $-5.020${*}{*}{*}} &  &  & {\footnotesize $-4.151${*}{*}} &  & \tabularnewline
 & {\footnotesize $(-2.852)$} &  &  & {\footnotesize $(-2.230)$} &  & \tabularnewline
{\footnotesize Junior titles} & {\footnotesize $-4.288${*}{*}{*}} & {\footnotesize $-4.862${*}{*}{*}} & {\footnotesize $0.574$} & {\footnotesize $-4.966${*}{*}} & {\footnotesize $-3.689${*}{*}} & {\footnotesize $-1.277$}\tabularnewline
 & {\footnotesize $(-2.652)$} & {\footnotesize $(-2.784)$} & {\footnotesize $(0.345)$} & {\footnotesize $(-2.340)$} & {\footnotesize $(-1.991)$} & {\footnotesize $(-0.642)$}\tabularnewline
{\footnotesize Senior titles} & {\footnotesize $-1.883$} & {\footnotesize $-5.213${*}{*}{*}} & {\footnotesize $3.330${*}{*}} & {\footnotesize $-0.391$} & {\footnotesize $-4.539${*}{*}} & {\footnotesize $4.148${*}{*}}\tabularnewline
 & {\footnotesize $(-1.348)$} & {\footnotesize $(-2.980)$} & {\footnotesize $(2.349)$} & {\footnotesize $(-0.224)$} & {\footnotesize $(-2.432)$} & {\footnotesize $(2.515)$}\tabularnewline
{\footnotesize Asks 0--1 years} & {\footnotesize $-3.873${*}{*}{*}} & {\footnotesize $-3.858${*}{*}} & {\footnotesize $-0.015$} & {\footnotesize $-3.343${*}} & {\footnotesize $-3.547${*}} & {\footnotesize $0.204$}\tabularnewline
 & {\footnotesize $(-2.635)$} & {\footnotesize $(-2.205)$} & {\footnotesize $(-0.008)$} & {\footnotesize $(-1.783)$} & {\footnotesize $(-1.758)$} & {\footnotesize $(0.086)$}\tabularnewline
{\footnotesize Asks 6+ years} & {\footnotesize $-0.546$} & {\footnotesize $-5.094${*}{*}{*}} & {\footnotesize $4.547${*}{*}} & {\footnotesize $1.488$} & {\footnotesize $-4.868${*}{*}} & {\footnotesize $6.356${*}{*}}\tabularnewline
 & {\footnotesize $(-0.331)$} & {\footnotesize $(-3.022)$} & {\footnotesize $(2.449)$} & {\footnotesize $(0.691)$} & {\footnotesize $(-2.486)$} & {\footnotesize $(2.308)$}\tabularnewline
{\footnotesize Expects AI-tool use} & {\footnotesize $-1.845$} & {\footnotesize $-5.136${*}{*}{*}} & {\footnotesize $3.290$} & {\footnotesize $-0.167$} & {\footnotesize $-5.233${*}{*}{*}} & {\footnotesize $5.066${*}}\tabularnewline
 & {\footnotesize $(-1.182)$} & {\footnotesize $(-2.868)$} & {\footnotesize $(1.410)$} & {\footnotesize $(-0.076)$} & {\footnotesize $(-2.607)$} & {\footnotesize $(1.742)$}\tabularnewline
{\footnotesize Remote or hybrid} & {\footnotesize $-1.268$} & {\footnotesize $-5.163${*}{*}{*}} & {\footnotesize $3.896${*}} & {\footnotesize $-1.748$} & {\footnotesize $-4.497${*}{*}} & {\footnotesize $2.748$}\tabularnewline
 & {\footnotesize $(-0.741)$} & {\footnotesize $(-2.825)$} & {\footnotesize $(1.890)$} & {\footnotesize $(-0.757)$} & {\footnotesize $(-2.158)$} & {\footnotesize $(0.962)$}\tabularnewline
{\footnotesize Top half, Eloundou} & {\footnotesize $-4.156${*}{*}} & {\footnotesize $-5.149${*}{*}{*}} & {\footnotesize $0.993$} & {\footnotesize $-3.378${*}} & {\footnotesize $-4.621${*}{*}} & {\footnotesize $1.243$}\tabularnewline
 & {\footnotesize $(-2.478)$} & {\footnotesize $(-2.795)$} & {\footnotesize $(0.594)$} & {\footnotesize $(-1.750)$} & {\footnotesize $(-2.048)$} & {\footnotesize $(0.572)$}\tabularnewline
{\footnotesize Top half, Webb AI} & {\footnotesize $-4.239${*}{*}} & {\footnotesize $-3.320${*}} & {\footnotesize $-0.919$} & {\footnotesize $-2.625$} & {\footnotesize $-3.605${*}} & {\footnotesize $0.980$}\tabularnewline
 & {\footnotesize $(-2.420)$} & {\footnotesize $(-1.934)$} & {\footnotesize $(-0.639)$} & {\footnotesize $(-1.301)$} & {\footnotesize $(-1.712)$} & {\footnotesize $(0.478)$}\tabularnewline
{\footnotesize Top half, Anthropic exposure} & {\footnotesize $-5.158${*}{*}{*}} & {\footnotesize $-4.205${*}{*}} & {\footnotesize $-0.954$} & {\footnotesize $-4.245${*}{*}} & {\footnotesize $-2.209$} & {\footnotesize $-2.036$}\tabularnewline
 & {\footnotesize $(-3.033)$} & {\footnotesize $(-2.238)$} & {\footnotesize $(-0.495)$} & {\footnotesize $(-2.086)$} & {\footnotesize $(-0.962)$} & {\footnotesize $(-0.783)$}\tabularnewline
{\footnotesize Obs} & {\footnotesize 1,725} & {\footnotesize 1,725} & {\footnotesize 1,725} & {\footnotesize 1,394} & {\footnotesize 1,394} & {\footnotesize 1,394}\tabularnewline
\hline
\end{tabular}
\par}
\end{table}

Our main finding is that the negative relation between agentic exposure and posting growth is weaker for senior positions and jobs requiring more experience. The coefficients for senior job postings and postings requiring six or more years of experience are $-0.39$ and $1.49$, respectively, and both are statistically insignificant. In contrast, a one-standard-deviation higher agentic beta is associated with about $4.54$ percent lower posting growth for non-senior positions and $4.87$ percent lower growth for jobs requiring zero to five years of experience. Both estimates are significant at the 5 percent level. The corresponding differences in the coefficient estimates are $4.15$ and $6.36$ percentage points, respectively, and both are significant at the 5 percent level.

For junior postings and jobs requiring zero to one year of experience, we find that greater agentic exposure is associated with lower posting growth. The estimates are about $-4.97$ and $-3.34$ percent, significant at the 5 and 10 percent levels, respectively. Using the sample mean of 32.7 junior postings per month as a reference, the junior-posting estimate implies about ten fewer postings out of 196 over six months, or 4.8 percent fewer postings, for a one-standard-deviation higher agentic beta.

The same pattern of smaller declines appears for jobs in which firms expect workers to use AI tools. The coefficient for postings expecting AI-tool use is only $-0.17$ and statistically insignificant. For other postings, the coefficient is $-5.23$, significant at the 1 percent level. The difference between the two is about $5.07$ percentage points, significant at the 10 percent level. The coefficient estimate difference between remote or hybrid jobs and on-site jobs is about $2.75$ percentage points and is statistically insignificant.

Moreover, we examine whether the occupational AI exposure measures used in the prior literature capture differences in the labor demand response to agentic AI. The estimated coefficient differences are about $1.24$ percentage points for job postings in the top half of \citet{Eloundou2024} exposure relative to the bottom half, $0.98$ for \citet{webb2020impact} exposure, and $-2.04$ for \citet{MassenkoffMcCrory2026} observed exposure. All three differences are statistically insignificant, and their signs vary across measures. In other words, the results show no consistent relation between these occupational exposure rankings and the posting-growth response to agentic exposure.

\begin{figure}[!htbp]
\caption{{\small\textbf{Postings by Month around the Release by Posting Type -- Agentic Beta\label{fig:S3}}}}
\medskip{}

This figure shows the monthly point estimates on the agentic beta for postings with senior titles, postings requiring six or more years of experience, and postings that expect the use of AI tools. Circles show the point estimate for the indicated job type and squares show the point estimate for its comparison group. For each group, posting growth is 100 times the log change in one plus postings from the July--September 2025 monthly average (shaded) to the month shown. The regressions use the agentic beta dated 26 September 2025, standardized across firms, and include Fama-French 49 industry fixed effects and the nine firm characteristic controls. The dashed line marks the 29 September 2025 release. The bars show 95\% confidence intervals.

\medskip{}
\centering{}\includegraphics[width=\linewidth]{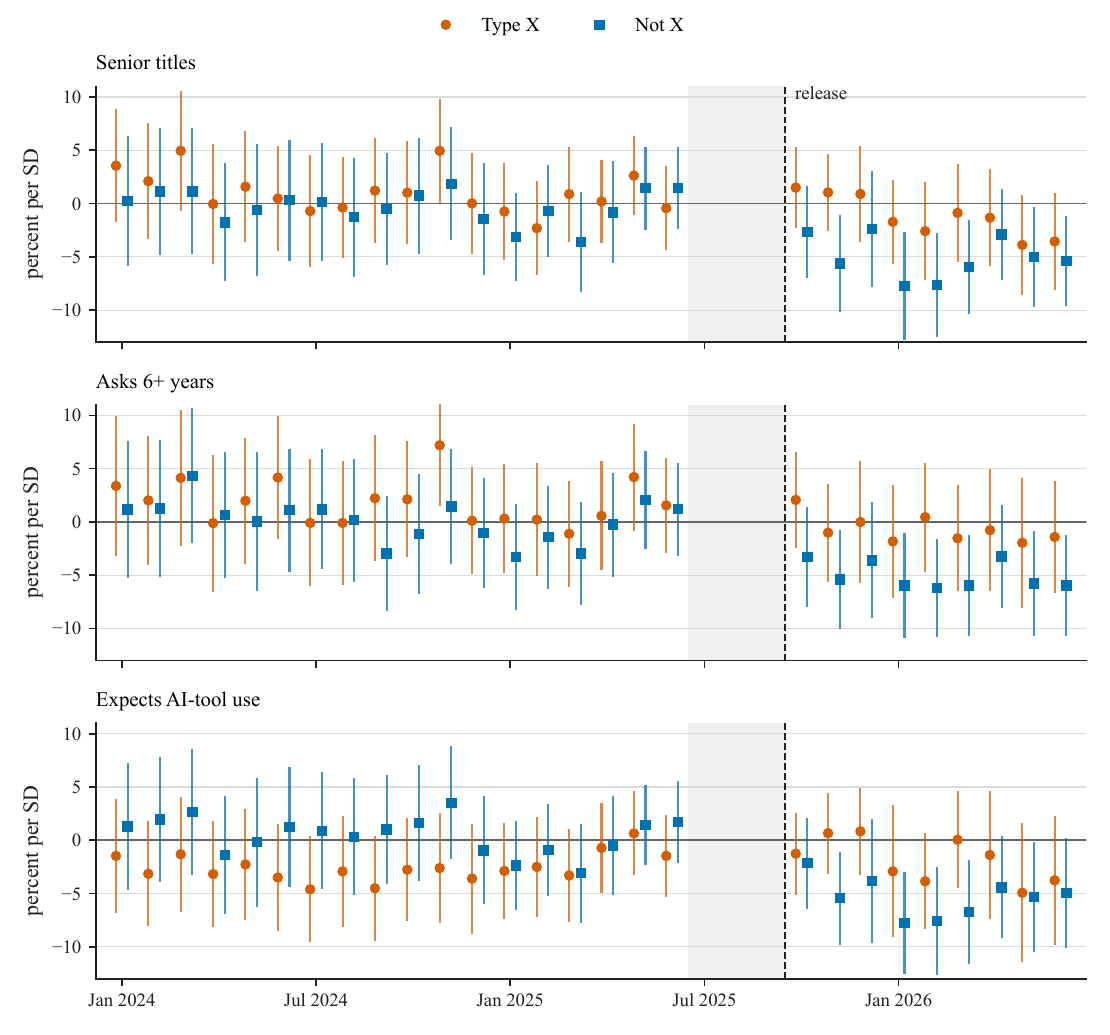}
\end{figure}

To illustrate how these differences evolve over time for the three job characteristics that matter in our context: seniority, experience, and AI-tool usage, Figure~\ref{fig:S3} shows monthly estimates. The monthly point estimates for senior and experienced postings remain closer to zero than those for their comparison groups in every month from October 2025 through June 2026. The 95 percent confidence intervals for both posting types include zero throughout this period. The declines are much larger and visible for their comparison groups subsequent to the release of Claude Code 2.0, whose coefficients reach about $-6$ to $-8$ percent in January--March 2026. For postings expecting AI-tool use, the coefficient is also generally closer to zero than that of the comparison group.

\begin{figure}[!htbp]
\caption{{\small\textbf{Posting Growth by Occupation Group -- Agentic Beta\label{fig:OCC}}}}
\medskip{}

This figure shows the coefficient estimates on the agentic beta using six-month job posting growth by occupation group, in percent per standard deviation of the beta. The first row uses all postings. The five groups are sorted by their share of the sample firms' postings. The numbers in parentheses are Standard Occupational Classification major group codes. Panel A uses the panel regression specification. Panel B uses the release-based event-study specification. The open circles show estimates with time fixed effects only in Panel A and no controls in Panel B. The filled circles show the corresponding estimates with Fama-French 49 industry fixed effects and the nine firm characteristic controls. The bars show 95\% confidence intervals. Standard errors are clustered by firm in Panel A.

\medskip{}
\centering{}\includegraphics[width=\linewidth]{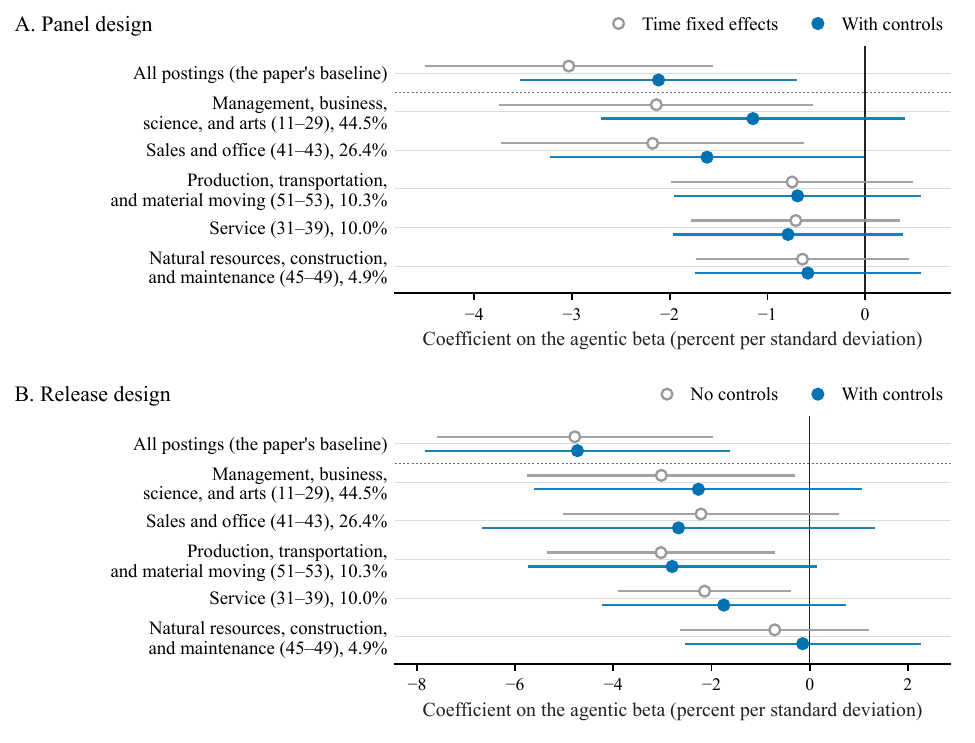}
\end{figure}

Finally, we examine whether the decline in labor demand associated with agentic exposure is concentrated in particular occupations. Figure~\ref{fig:OCC} reports six-month estimates for the panel regression approach and the event-study design, using growth in the firm's postings in each of the five high-level SOC occupation groups as the dependent variable. The two largest groups -- Management, business, science, and arts occupations and Sales and office occupations -- account for about 71 percent of postings. With industry fixed effects and firm characteristic controls, the point estimates are negative for all five groups in both specifications. In the panel specification, a one-standard-deviation increase in the agentic beta is associated with about $1.15$ percent lower posting growth in Management, business, science, and arts occupations, $1.62$ percent in Sales and office occupations, $0.69$ percent in Production, transportation, and material moving occupations, $0.79$ percent in Service occupations, and $0.59$ percent in Natural resources, construction, and maintenance occupations. Using the event-study specification, the corresponding estimates imply $2.27$, $2.67$, $2.80$, $1.75$, and $0.14$ percent lower posting growth for these five occupation groups, respectively. In summary, our results show that the negative effect on labor demand is widespread across different occupations.

\subsection{Posting Growth by Industry and Region}\label{sec:industry_region}

\begin{figure}[!htbp]
\caption{{\small\textbf{Posting Growth by Industry -- Agentic Beta\label{fig:SEC}}}}
\medskip{}

This figure shows the coefficient estimates on the agentic beta in separate regressions of six-month posting growth for firms in each of the five Fama-French industries. The first row includes all firms. Consumer goods and retail covers consumer durables, non-durables, wholesale, retail, and some services. High technology and telecom covers business equipment and telephone and television transmission. Health covers health care, medical equipment, and drugs. Other covers mines, construction, building materials, transportation, hotels, business services, entertainment, and finance. Industries are sorted by their share of the sample firms' postings. Panel A uses the panel regression specification. Panel B uses the release-based event-study specification. The open circles show estimates with time fixed effects only in Panel A and no controls in Panel B. The filled circles show the corresponding estimates with Fama-French 49 industry fixed effects and the nine firm characteristic controls. The bars show 95\% confidence intervals. Standard errors are clustered by firm in Panel A.

\medskip{}
\centering{}\includegraphics[width=\linewidth]{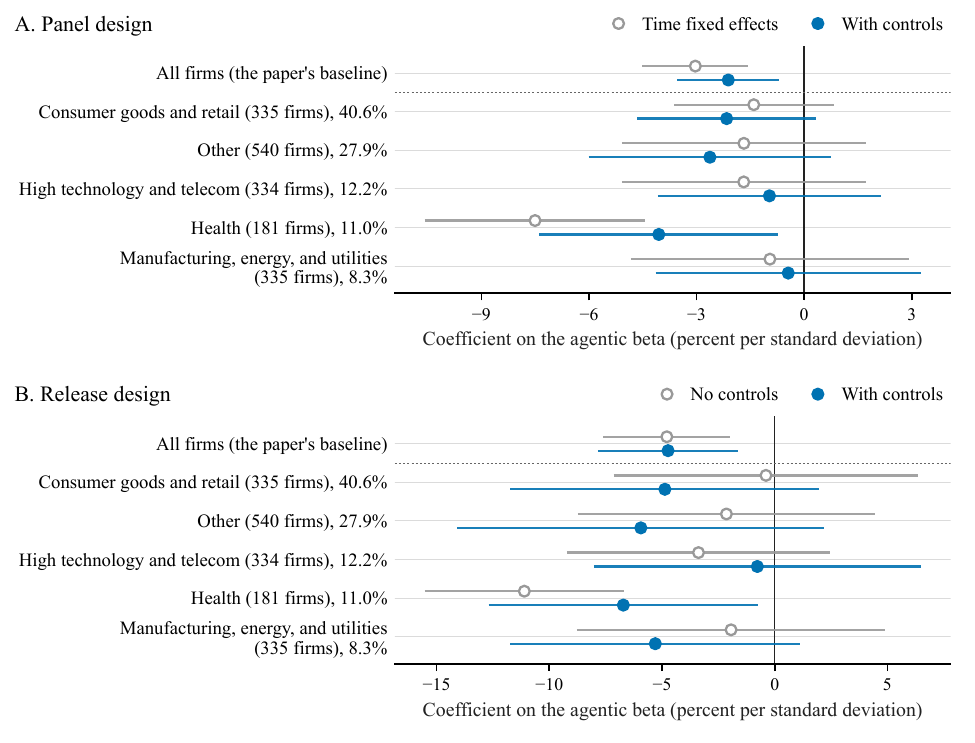}
\end{figure}

Beyond job types and occupations, we examine whether the decline in labor demand is concentrated in particular industries. We group firms into the five Fama-French industries and estimate the panel and release specifications separately for each industry. Figure~\ref{fig:SEC} reports the six-month estimates. Consumer goods and retail is the largest industry, corresponding to about 41 percent of the postings. With industry fixed effects and firm characteristic controls, the point estimates are negative for all five industries in both specifications. In the panel specification, a one-standard-deviation increase in the agentic beta is associated with about $2.16$ percent lower posting growth in Consumer goods and retail and $4.05$ percent lower growth in Health, significant at the 10 and 5 percent levels, respectively. In the release specification, the estimate for Health implies $6.71$ percent lower growth, significant at the 5 percent level. The point estimates for Other and for Manufacturing, energy, and utilities are negative in both specifications but not statistically significant. In contrast, the point estimates for High technology and telecom are small and statistically insignificant in both specifications. 

\begin{figure}[!htbp]
\caption{{\small\textbf{Posting Growth by Census Region -- Agentic Beta\label{fig:REG}}}}
\medskip{}

This figure shows the coefficient estimates on the agentic beta for six-month posting growth by Census region. Each posting is assigned to a region by the state where the job is located. The first row uses all postings. The four regions are sorted by their share of the sample firms' postings, shown next to each region. Panel A uses the panel regression specification. Panel B uses the release-based event-study specification. The open circles show estimates with time fixed effects only in Panel A and no controls in Panel B. The filled circles show the corresponding estimates with Fama-French 49 industry fixed effects and the nine firm characteristic controls. The bars show 95\% confidence intervals. Standard errors are clustered by firm in Panel A.

\medskip{}
\centering{}\includegraphics[width=\linewidth]{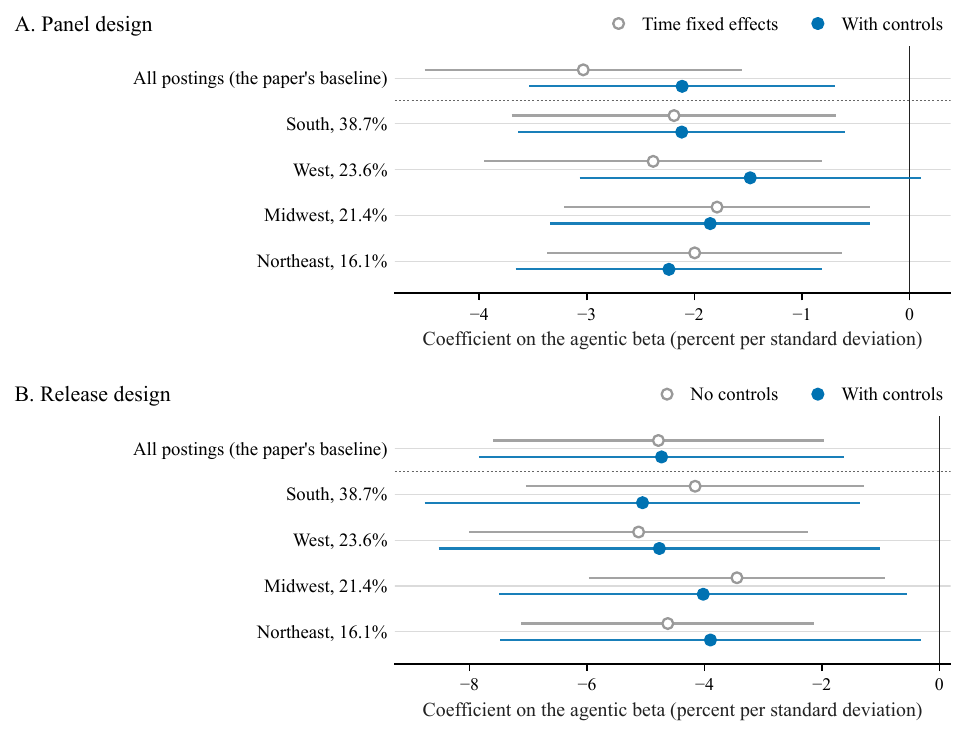}
\end{figure}

Furthermore, we assess geographic differences by examining the relation between agentic exposure and posting growth across regions. We assign each posting to one of the four Census regions based on the state in which the job is located and use growth in the firm's postings within each region as the dependent variable. Figure~\ref{fig:REG} reports the six-month estimates. The South accounts for about 39 percent of the postings, the West for 24 percent, the Midwest for 21 percent, and the Northeast for 16 percent. With industry fixed effects and firm characteristic controls, a one-standard-deviation increase in the agentic beta is associated with between $1.48$ and $2.24$ percent lower posting growth in the panel specification and between $3.90$ and $5.05$ percent lower growth in the event-study specification across the four regions. Seven of the eight estimates are significant at least at the 5 percent level, and the panel estimate for the West is significant at the 10 percent level. Overall, our results show that the negative relation between agentic exposure and labor demand is nationwide rather than concentrated in a single region.

\section{Additional Results and Robustness Checks}\label{sec:additional_results}

In this section, we complement the posting-growth evidence with results on occupational AI exposure, wages, and investment. Then, we test the robustness of the main findings.

\subsection{Occupational AI Exposure and Posting Growth}\label{sec:occupation_exposure}

The literature proposes several occupational measures of AI exposure---including \citet{Eloundou2024}, \citet{webb2020impact}, and the Anthropic Economic Index \citep{MassenkoffMcCrory2026}. We use these three measures, described in Section~\ref{sec:lightcast_data}, to assess whether they can predict changes in labor demand across firms after the release of Claude Code 2.0. First, we map these occupational measures to the firm level. Specifically, for each measure, we average the available occupational AI overlap across a firm's job postings in July--September 2025 and standardize the resulting firm-level AI overlap measure across firms. Then, we use the release-based event-study specification~(\ref{eq:posting_shock}) with one of the firm-level AI overlap measures as the independent variable.

Table~\ref{tab:O2} reports the results. Columns (1) and (2) use AI overlap alone, without and with industry fixed effects and firm characteristic controls, respectively. Columns (3) to (6) also include the agentic beta.\footnote{It is worth noting that the Anthropic measure contains post-release information. \citet{MassenkoffMcCrory2026} constructs it from Claude consumption in August and November 2025, with the November component postdating the 29 September 2025 release. This can be seen in footnote 5 of the report, published on 5 March 2026. Therefore, we treat it as a retrospective benchmark rather than an exposure measure available before the release. On the other hand, the original measures of \citet{Eloundou2024} and \citet{webb2020impact} predate the release.} The table shows that the relation between the AI overlap measures and the subsequent posting growth is generally insignificant when controls are included. When the AI overlap measures are used alone in Column (2), none of the coefficient estimates is significant at the 5 percent level. When the AI overlap measures are included alongside the agentic exposure measures in Columns (5) and (6), none of the coefficient estimates on the AI overlap measures is significant at the 5 percent level, while the coefficient estimates on the agentic exposure measures  remain significant at the 1 percent level at the six- and nine-month horizons.

\begin{table}[!htbp]
\stepcounter{anch}
{\scriptsize\caption{{\small\textbf{Posting Growth after the Claude Code 2.0 Release on Firm AI Overlap\label{tab:O2}}}}
\medskip{}
}{\scriptsize\par}

{\footnotesize This table reports regression results of posting growth after the 29 September 2025 release of Claude Code 2.0 on firm AI overlap measures. Panels A, B, and C use the \citet{Eloundou2024} GPT exposure beta, the \citet{webb2020impact} AI exposure percentile, and Anthropic Economic Index observed exposure, respectively. The dependent variable is growth in monthly postings from the three months before the release to the months after it, in percent, with one regression per horizon. Columns (1) and (2) include the AI overlap measure alone. Columns (3) and (4) include both the AI overlap measure and the agentic beta in the same regression. Columns (5) and (6) repeat the joint regression with controls. Columns (1), (3), and (4) include no controls. Columns (2), (5), and (6) add Fama-French 49 industry fixed effects and the nine firm characteristic controls. The number of observations and the adjusted $R^{2}$ of each regression are reported. t-statistics are reported in parentheses. {*}, {*}{*}, and {*}{*}{*} denote statistical significance at the 10\%, 5\%, and 1\% levels, respectively.\par}

{\scriptsize\medskip{}
}{\scriptsize\par}
\centering{}{\scriptsize{}%
\renewcommand{\arraystretch}{0.88}%
\setlength{\tabcolsep}{3pt}%
\begin{tabular}{lcccccccccc}
\hline
 & {\scriptsize (1)} & {\scriptsize (2)} & {\scriptsize (3)} & {\scriptsize (4)} & {\scriptsize (5)} & {\scriptsize (6)} & \multicolumn{4}{c}{{\scriptsize Adj $R^{2}$}}\tabularnewline
 & \multicolumn{2}{c}{{\scriptsize Overlap alone}} & \multicolumn{4}{c}{{\scriptsize Overlap and the agentic beta}} & {\scriptsize (1)} & {\scriptsize (2)} & {\scriptsize (3)--(4)} & {\scriptsize (5)--(6)}\tabularnewline
{\scriptsize Months after} & {\scriptsize No controls} & {\scriptsize With controls} & \multicolumn{2}{c}{{\scriptsize No controls}} & \multicolumn{2}{c}{{\scriptsize With controls}} &  &  &  & \tabularnewline
{\scriptsize the shock} &  &  & {\scriptsize Overlap} & {\scriptsize $\beta^{Agentic}$} & {\scriptsize Overlap} & {\scriptsize $\beta^{Agentic}$} &  &  &  & \tabularnewline
\hline
\multicolumn{11}{l}{{\scriptsize \textbf{Panel A: Eloundou GPT exposure}}}\tabularnewline
 &  &  &  &  &  &  &  &  &  & \tabularnewline
{\scriptsize 1} & {\scriptsize $1.520$} & {\scriptsize $2.305$} & {\scriptsize $1.489$} & {\scriptsize $-0.995$} & {\scriptsize $2.353$} & {\scriptsize $-2.274$} & {\scriptsize $0.000$} & {\scriptsize $0.002$} & {\scriptsize $0.000$} & {\scriptsize $0.002$}\tabularnewline
 & {\scriptsize $(1.002)$} & {\scriptsize $(0.959)$} & {\scriptsize $(0.982)$} & {\scriptsize $(-0.601)$} & {\scriptsize $(0.979)$} & {\scriptsize $(-1.057)$} &  &  &  & \tabularnewline
{\scriptsize 3} & {\scriptsize $1.976$} & {\scriptsize $2.768$} & {\scriptsize $1.889$} & {\scriptsize $-2.791${*}} & {\scriptsize $2.841$} & {\scriptsize $-3.486${*}} & {\scriptsize $0.001$} & {\scriptsize $0.008$} & {\scriptsize $0.004$} & {\scriptsize $0.011$}\tabularnewline
 & {\scriptsize $(1.605)$} & {\scriptsize $(1.454)$} & {\scriptsize $(1.540)$} & {\scriptsize $(-1.768)$} & {\scriptsize $(1.496)$} & {\scriptsize $(-1.818)$} &  &  &  & \tabularnewline
{\scriptsize 6} & {\scriptsize $2.047${*}} & {\scriptsize $2.786$} & {\scriptsize $1.897${*}} & {\scriptsize $-4.749${*}{*}{*}} & {\scriptsize $2.886$} & {\scriptsize $-4.811${*}{*}{*}} & {\scriptsize $0.001$} & {\scriptsize $0.038$} & {\scriptsize $0.010$} & {\scriptsize $0.045$}\tabularnewline
 & {\scriptsize $(1.771)$} & {\scriptsize $(1.570)$} & {\scriptsize $(1.649)$} & {\scriptsize $(-3.239)$} & {\scriptsize $(1.630)$} & {\scriptsize $(-3.019)$} &  &  &  & \tabularnewline
{\scriptsize 9} & {\scriptsize $1.824$} & {\scriptsize $1.930$} & {\scriptsize $1.670$} & {\scriptsize $-4.898${*}{*}{*}} & {\scriptsize $2.021$} & {\scriptsize $-4.380${*}{*}{*}} & {\scriptsize $0.001$} & {\scriptsize $0.064$} & {\scriptsize $0.010$} & {\scriptsize $0.070$}\tabularnewline
 & {\scriptsize $(1.557)$} & {\scriptsize $(1.116)$} & {\scriptsize $(1.434)$} & {\scriptsize $(-3.257)$} & {\scriptsize $(1.167)$} & {\scriptsize $(-2.753)$} &  &  &  & \tabularnewline
 &  &  &  &  &  &  &  &  &  & \tabularnewline
{\scriptsize Obs} & {\scriptsize 1,706} & {\scriptsize 1,384} & \multicolumn{2}{c}{{\scriptsize 1,706}} & \multicolumn{2}{c}{{\scriptsize 1,384}} &  &  &  & \tabularnewline
 &  &  &  &  &  &  &  &  &  & \tabularnewline
\hline
\multicolumn{11}{l}{{\scriptsize \textbf{Panel B: Webb AI exposure}}}\tabularnewline
 &  &  &  &  &  &  &  &  &  & \tabularnewline
{\scriptsize 1} & {\scriptsize $3.358${*}} & {\scriptsize $1.970$} & {\scriptsize $3.280${*}} & {\scriptsize $-0.406$} & {\scriptsize $1.812$} & {\scriptsize $-2.126$} & {\scriptsize $0.003$} & {\scriptsize $0.002$} & {\scriptsize $0.003$} & {\scriptsize $0.002$}\tabularnewline
 & {\scriptsize $(1.930)$} & {\scriptsize $(0.765)$} & {\scriptsize $(1.877)$} & {\scriptsize $(-0.244)$} & {\scriptsize $(0.703)$} & {\scriptsize $(-0.989)$} &  &  &  & \tabularnewline
{\scriptsize 3} & {\scriptsize $1.704$} & {\scriptsize $-0.100$} & {\scriptsize $1.201$} & {\scriptsize $-2.618${*}} & {\scriptsize $-0.358$} & {\scriptsize $-3.467${*}} & {\scriptsize $0.001$} & {\scriptsize $0.006$} & {\scriptsize $0.003$} & {\scriptsize $0.009$}\tabularnewline
 & {\scriptsize $(1.247)$} & {\scriptsize $(-0.055)$} & {\scriptsize $(0.878)$} & {\scriptsize $(-1.649)$} & {\scriptsize $(-0.195)$} & {\scriptsize $(-1.805)$} &  &  &  & \tabularnewline
{\scriptsize 6} & {\scriptsize $3.300${*}{*}{*}} & {\scriptsize $-0.919$} & {\scriptsize $2.468${*}} & {\scriptsize $-4.330${*}{*}{*}} & {\scriptsize $-1.280$} & {\scriptsize $-4.849${*}{*}{*}} & {\scriptsize $0.004$} & {\scriptsize $0.036$} & {\scriptsize $0.011$} & {\scriptsize $0.044$}\tabularnewline
 & {\scriptsize $(2.577)$} & {\scriptsize $(-0.533)$} & {\scriptsize $(1.895)$} & {\scriptsize $(-2.901)$} & {\scriptsize $(-0.737)$} & {\scriptsize $(-3.024)$} &  &  &  & \tabularnewline
{\scriptsize 9} & {\scriptsize $4.514${*}{*}{*}} & {\scriptsize $-1.015$} & {\scriptsize $3.701${*}{*}{*}} & {\scriptsize $-4.232${*}{*}{*}} & {\scriptsize $-1.344$} & {\scriptsize $-4.435${*}{*}{*}} & {\scriptsize $0.008$} & {\scriptsize $0.064$} & {\scriptsize $0.014$} & {\scriptsize $0.070$}\tabularnewline
 & {\scriptsize $(3.462)$} & {\scriptsize $(-0.572)$} & {\scriptsize $(2.778)$} & {\scriptsize $(-2.765)$} & {\scriptsize $(-0.747)$} & {\scriptsize $(-2.763)$} &  &  &  & \tabularnewline
 &  &  &  &  &  &  &  &  &  & \tabularnewline
{\scriptsize Obs} & {\scriptsize 1,706} & {\scriptsize 1,384} & \multicolumn{2}{c}{{\scriptsize 1,706}} & \multicolumn{2}{c}{{\scriptsize 1,384}} &  &  &  & \tabularnewline
 &  &  &  &  &  &  &  &  &  & \tabularnewline
\hline
\multicolumn{11}{l}{{\scriptsize \textbf{Panel C: Anthropic Economic Index observed exposure}}}\tabularnewline
 &  &  &  &  &  &  &  &  &  & \tabularnewline
{\scriptsize 1} & {\scriptsize $-2.839${*}} & {\scriptsize $-4.189${*}} & {\scriptsize $-2.799${*}} & {\scriptsize $-0.921$} & {\scriptsize $-4.052${*}} & {\scriptsize $-2.008$} & {\scriptsize $0.002$} & {\scriptsize $0.004$} & {\scriptsize $0.002$} & {\scriptsize $0.005$}\tabularnewline
 & {\scriptsize $(-1.762)$} & {\scriptsize $(-1.788)$} & {\scriptsize $(-1.742)$} & {\scriptsize $(-0.558)$} & {\scriptsize $(-1.728)$} & {\scriptsize $(-0.942)$} &  &  &  & \tabularnewline
{\scriptsize 3} & {\scriptsize $-1.101$} & {\scriptsize $-1.334$} & {\scriptsize $-0.980$} & {\scriptsize $-2.808${*}} & {\scriptsize $-1.103$} & {\scriptsize $-3.381${*}} & {\scriptsize $0.000$} & {\scriptsize $0.007$} & {\scriptsize $0.003$} & {\scriptsize $0.009$}\tabularnewline
 & {\scriptsize $(-0.834)$} & {\scriptsize $(-0.706)$} & {\scriptsize $(-0.745)$} & {\scriptsize $(-1.786)$} & {\scriptsize $(-0.586)$} & {\scriptsize $(-1.776)$} &  &  &  & \tabularnewline
{\scriptsize 6} & {\scriptsize $-1.214$} & {\scriptsize $-0.180$} & {\scriptsize $-1.009$} & {\scriptsize $-4.765${*}{*}{*}} & {\scriptsize $0.146$} & {\scriptsize $-4.777${*}{*}{*}} & {\scriptsize $0.000$} & {\scriptsize $0.036$} & {\scriptsize $0.009$} & {\scriptsize $0.043$}\tabularnewline
 & {\scriptsize $(-0.969)$} & {\scriptsize $(-0.104)$} & {\scriptsize $(-0.808)$} & {\scriptsize $(-3.263)$} & {\scriptsize $(0.084)$} & {\scriptsize $(-3.002)$} &  &  &  & \tabularnewline
{\scriptsize 9} & {\scriptsize $-2.130${*}} & {\scriptsize $-1.177$} & {\scriptsize $-1.920$} & {\scriptsize $-4.867${*}{*}{*}} & {\scriptsize $-0.883$} & {\scriptsize $-4.300${*}{*}{*}} & {\scriptsize $0.001$} & {\scriptsize $0.064$} & {\scriptsize $0.010$} & {\scriptsize $0.069$}\tabularnewline
 & {\scriptsize $(-1.684)$} & {\scriptsize $(-0.677)$} & {\scriptsize $(-1.521)$} & {\scriptsize $(-3.250)$} & {\scriptsize $(-0.503)$} & {\scriptsize $(-2.704)$} &  &  &  & \tabularnewline
 &  &  &  &  &  &  &  &  &  & \tabularnewline
{\scriptsize Obs} & {\scriptsize 1,706} & {\scriptsize 1,384} & \multicolumn{2}{c}{{\scriptsize 1,706}} & \multicolumn{2}{c}{{\scriptsize 1,384}} &  &  &  & \tabularnewline
 &  &  &  &  &  &  &  &  &  & \tabularnewline
\hline
\end{tabular}
}{\scriptsize\par}
\end{table}

\subsection{Wages}\label{sec:wages}

Alongside job posting volumes, we also examine whether firms with higher agentic betas change the wages offered in their job postings after the release of Claude Code 2.0. Specifically, we regress firms' posted-wage growth on their pre-release agentic betas using the release-based event-study specification~(\ref{eq:posting_shock}). For each firm, we calculate the mean log annualized salary across all postings that report a salary in each window. Posted-wage growth is calculated as 100 times the difference between this mean log annualized salary of postings over the months after the release and the average over July--September 2025.  

Table~\ref{tab:S6} presents the results. We find a positive relation between agentic betas and posted-wage growth at every horizon in Table~\ref{tab:S6}. The magnitude of the point estimates peaks over two to four months after the release. In the specification without controls, a one-standard-deviation increase in the agentic beta is associated with about $1.40$, $1.44$, and $1.34$ percent higher posted-wage growth over two, three, and four months, respectively. All three estimates are significant at the 1 percent level. With industry fixed effects and firm characteristic controls, the corresponding point estimates are $0.94$, $1.02$, and $0.96$. The only significant point estimate is over three months, significant at the 10 percent level. The estimates with controls become smaller at longer horizons, reaching about $0.60$ percent over six months and $0.42$ percent over nine months. Overall, we conclude that the negative relationship between agentic exposure and firms' labor demand is not accompanied by a decline in wages.

\begin{table}[!htbp]
\stepcounter{anch}
{\footnotesize\caption{{\small\textbf{Posted-Wage Growth after the Claude Code 2.0 Release on the Agentic Beta\label{tab:S6}}}}
\medskip{}
}{\footnotesize\par}

{\footnotesize This table reports the results for posted-wage growth. Posted-wage growth is 100 times the change in mean log annualized posted salary from the three months before the Claude Code 2.0 release to the months after it. We estimate a separate regression for each horizon from 1 to 9 months. The independent variable is the agentic beta dated 26 September 2025, standardized across firms. Column (1) includes no controls. Column (2) adds Fama-French 49 industry fixed effects and the nine firm characteristic controls. The number of observations and the adjusted $R^{2}$ of each regression are reported. t-statistics are reported in parentheses. {*}, {*}{*}, and {*}{*}{*} denote statistical significance at the 10\%, 5\%, and 1\% levels, respectively.\par}

{\footnotesize\medskip{}
}{\footnotesize\par}
\centering{}{\footnotesize{}%
\renewcommand{\arraystretch}{1.10}%
\begin{tabular}{lcccccc}
\hline 
 & {\footnotesize (1)} & {\footnotesize (2)} & \multicolumn{2}{c}{{\footnotesize Obs}} & \multicolumn{2}{c}{{\footnotesize Adj $R^{2}$}}\tabularnewline
{\footnotesize Months after the shock} & {\footnotesize No controls} & {\footnotesize With controls} & {\footnotesize (1)} & {\footnotesize (2)} & {\footnotesize (1)} & {\footnotesize (2)}\tabularnewline
\hline 
 &  &  &  &  &  & \tabularnewline
{\footnotesize 1} & {\footnotesize $0.802$} & {\footnotesize $0.341$} & {\footnotesize 1,539} & {\footnotesize 1,257} & {\footnotesize $0.001$} & {\footnotesize $0.009$}\tabularnewline
 & {\footnotesize $(1.341)$} & {\footnotesize $(0.448)$} &  &  &  & \tabularnewline
{\footnotesize 2} & {\footnotesize $1.399${*}{*}{*}} & {\footnotesize $0.937$} & {\footnotesize 1,571} & {\footnotesize 1,281} & {\footnotesize $0.004$} & {\footnotesize $0.015$}\tabularnewline
 & {\footnotesize $(2.734)$} & {\footnotesize $(1.499)$} &  &  &  & \tabularnewline
{\footnotesize 3} & {\footnotesize $1.438${*}{*}{*}} & {\footnotesize $1.017${*}} & {\footnotesize 1,585} & {\footnotesize 1,295} & {\footnotesize $0.005$} & {\footnotesize $0.012$}\tabularnewline
 & {\footnotesize $(2.954)$} & {\footnotesize $(1.674)$} &  &  &  & \tabularnewline
{\footnotesize 4} & {\footnotesize $1.345${*}{*}{*}} & {\footnotesize $0.957$} & {\footnotesize 1,598} & {\footnotesize 1,305} & {\footnotesize $0.004$} & {\footnotesize $0.006$}\tabularnewline
 & {\footnotesize $(2.705)$} & {\footnotesize $(1.581)$} &  &  &  & \tabularnewline
{\footnotesize 5} & {\footnotesize $0.987${*}} & {\footnotesize $0.555$} & {\footnotesize 1,603} & {\footnotesize 1,309} & {\footnotesize $0.002$} & {\footnotesize $-0.004$}\tabularnewline
 & {\footnotesize $(1.914)$} & {\footnotesize $(0.863)$} &  &  &  & \tabularnewline
{\footnotesize 6} & {\footnotesize $0.965${*}} & {\footnotesize $0.599$} & {\footnotesize 1,606} & {\footnotesize 1,311} & {\footnotesize $0.002$} & {\footnotesize $-0.002$}\tabularnewline
 & {\footnotesize $(1.838)$} & {\footnotesize $(0.928)$} &  &  &  & \tabularnewline
{\footnotesize 7} & {\footnotesize $0.892${*}} & {\footnotesize $0.592$} & {\footnotesize 1,610} & {\footnotesize 1,313} & {\footnotesize $0.002$} & {\footnotesize $-0.008$}\tabularnewline
 & {\footnotesize $(1.824)$} & {\footnotesize $(0.938)$} &  &  &  & \tabularnewline
{\footnotesize 8} & {\footnotesize $0.912${*}} & {\footnotesize $0.519$} & {\footnotesize 1,613} & {\footnotesize 1,315} & {\footnotesize $0.002$} & {\footnotesize $-0.007$}\tabularnewline
 & {\footnotesize $(1.917)$} & {\footnotesize $(0.852)$} &  &  &  & \tabularnewline
{\footnotesize 9} & {\footnotesize $0.731$} & {\footnotesize $0.425$} & {\footnotesize 1,614} & {\footnotesize 1,316} & {\footnotesize $0.001$} & {\footnotesize $-0.009$}\tabularnewline
 & {\footnotesize $(1.552)$} & {\footnotesize $(0.690)$} &  &  &  & \tabularnewline
 &  &  &  &  &  & \tabularnewline
\hline
\end{tabular}
}{\footnotesize\par}
\end{table}

\subsection{Firm Investment}\label{sec:firm_outcomes}

We next investigate other variables beyond labor demand. In particular, we study whether there is a relation between firms' AI exposure and investment. We again use the release-based event-study specification. We regress the subsequent growth in capital expenditures or R\&D spending on firms' AI and agentic betas dated 26 September 2025. For each investment measure, we sum spending over the firm's three fiscal quarters ending between October 2025 and June 2026 and compare it with spending over the same quarters a year earlier. Table~\ref{tab:S5} reports separate regressions with the agentic beta, the AI beta, and both betas together. Columns (1) to (3) have no controls, and columns (4) to (6) include industry fixed effects and firm characteristic controls. 

\begin{table}[!htbp]
\stepcounter{anch}
{\footnotesize\caption{{\small\textbf{Firm Investment Growth on the Agentic and AI Betas\label{tab:S5}}}}
\medskip{}
}{\footnotesize\par}

{\footnotesize This table reports the results for firm investment growth. Panel A measures capital-expenditure growth as 100 times the log change in total capital expenditures for firms with six quarters of data and positive totals in both windows. Quarterly capital expenditures are the change in year-to-date capital expenditures within the fiscal year. Panel B measures R\&D growth as 100 times the change in the log of one plus total R\&D expense in millions of dollars for firms with all six quarterly reports. We treat unreported quarterly R\&D expense as zero. Each dependent variable is winsorized at its 1st and 99th percentiles. Columns (1) and (4) include the agentic beta alone, columns (2) and (5) the AI beta alone, and columns (3) and (6) both betas. Columns (1) to (3) have no controls. Columns (4) to (6) add Fama-French 49 industry fixed effects and the nine firm characteristic controls. The number of observations and the adjusted $R^{2}$ of each regression are reported. t-statistics are reported in parentheses. {*}, {*}{*}, and {*}{*}{*} denote statistical significance at the 10\%, 5\%, and 1\% levels, respectively.\par}

{\footnotesize\medskip{}
}{\footnotesize\par}
\centering{}{\footnotesize{}%
\renewcommand{\arraystretch}{0.88}%
\begin{tabular}{lcccccc}
\hline 
 & {\footnotesize (1)} & {\footnotesize (2)} & {\footnotesize (3)} & {\footnotesize (4)} & {\footnotesize (5)} & {\footnotesize (6)}\tabularnewline
 & \multicolumn{3}{c}{{\footnotesize No controls}} & \multicolumn{3}{c}{{\footnotesize With controls}}\tabularnewline
\hline 
\multicolumn{7}{l}{{\footnotesize \textbf{Panel A: Capital expenditures}}}\tabularnewline
{\footnotesize $\beta^{Agentic}$} & {\footnotesize $4.238${*}{*}{*}} &  & {\footnotesize $-1.775$} & {\footnotesize $5.945${*}{*}{*}} &  & {\footnotesize $0.814$}\tabularnewline
 & {\footnotesize $(2.614)$} &  & {\footnotesize $(-0.922)$} & {\footnotesize $(3.016)$} &  & {\footnotesize $(0.333)$}\tabularnewline
{\footnotesize $\beta^{AI}$} &  & {\footnotesize $9.823${*}{*}{*}} & {\footnotesize $10.798${*}{*}{*}} &  & {\footnotesize $9.591${*}{*}{*}} & {\footnotesize $9.154${*}{*}{*}}\tabularnewline
 &  & {\footnotesize $(6.575)$} & {\footnotesize $(6.164)$} &  & {\footnotesize $(5.262)$} & {\footnotesize $(4.029)$}\tabularnewline
 &  &  &  &  &  & \tabularnewline
{\footnotesize Adj $R^{2}$} & {\footnotesize $0.006$} & {\footnotesize $0.034$} & {\footnotesize $0.034$} & {\footnotesize $0.058$} & {\footnotesize $0.074$} & {\footnotesize $0.073$}\tabularnewline
{\footnotesize Obs} & {\footnotesize 1,563} & {\footnotesize 1,563} & {\footnotesize 1,563} & {\footnotesize 1,323} & {\footnotesize 1,323} & {\footnotesize 1,323}\tabularnewline
\hline 
\multicolumn{7}{l}{{\footnotesize \textbf{Panel B: R\&D expense}}}\tabularnewline
{\footnotesize $\beta^{Agentic}$} & {\footnotesize $0.443$} &  & {\footnotesize $-0.709$} & {\footnotesize $0.398$} &  & {\footnotesize $0.059$}\tabularnewline
 & {\footnotesize $(1.030)$} &  & {\footnotesize $(-1.320)$} & {\footnotesize $(0.679)$} &  & {\footnotesize $(0.082)$}\tabularnewline
{\footnotesize $\beta^{AI}$} &  & {\footnotesize $1.683${*}{*}{*}} & {\footnotesize $2.072${*}{*}{*}} &  & {\footnotesize $0.641$} & {\footnotesize $0.609$}\tabularnewline
 &  & {\footnotesize $(4.213)$} & {\footnotesize $(4.150)$} &  & {\footnotesize $(1.189)$} & {\footnotesize $(0.922)$}\tabularnewline
 &  &  &  &  &  & \tabularnewline
{\footnotesize Adj $R^{2}$} & {\footnotesize $0.000$} & {\footnotesize $0.014$} & {\footnotesize $0.016$} & {\footnotesize $0.113$} & {\footnotesize $0.114$} & {\footnotesize $0.113$}\tabularnewline
{\footnotesize Obs} & {\footnotesize 1,651} & {\footnotesize 1,651} & {\footnotesize 1,651} & {\footnotesize 1,392} & {\footnotesize 1,392} & {\footnotesize 1,392}\tabularnewline
\hline 
\end{tabular}
}{\footnotesize\par}
\end{table}

We find that firms with higher AI betas have higher capital-expenditure growth. In Panel A, when both AI and agentic betas enter the regression with industry fixed effects and firm characteristic controls, the coefficient on the AI beta is $9.15$, significant at the 1 percent level. The coefficient on the agentic beta is $0.81$ and statistically insignificant. In terms of economic magnitude, a one-standard-deviation increase in the AI beta is associated with about 9.6 percent higher capital expenditures for a given prior-year level of spending. In contrast, we find no significant coefficient on either agentic exposure or AI exposure for R\&D expense when we include controls. 

\subsection{Alternative Beta Estimation Windows and Factor Models}\label{sec:beta_windows}

We next conduct several robustness checks for our baseline results. We begin the robustness checks by varying the window used to estimate agentic betas. Our baseline uses 26 weeks of stock returns. We re-estimate the agentic betas using specification~(\ref{eq:beta}) over windows of 20, 22, 30, and 34 weeks. We then regress firms' future job posting growth over three, six, and nine months on these alternative betas using the panel specification~(\ref{eq:posting_panel}). Each estimation window ends in the last week of month $t$, and each beta is standardized across firms within that month. Appendix Table~\ref{tab:BW} reports the results, including the 26-week baseline for comparison. 

The table shows that higher agentic betas predict lower posting growth across all estimation windows. Across all five estimation windows, the coefficients are negative at all three posting horizons and become larger in magnitude as the horizon increases from three to six to nine months. 

We also vary the factor model used to estimate agentic betas. Our baseline specification~(\ref{eq:beta}) controls for the excess market return only. We re-estimate the agentic betas after adding to specification~(\ref{eq:beta}) the remaining factors of the three-factor model of \citet{FamaFrench1993}, the four-factor model of \citet{Carhart1997}, and the five-factor model of \citet{FamaFrench2015}. We then regress firms' future job posting growth over three, six, and nine months on these alternative betas using the panel specification~(\ref{eq:posting_panel}). Appendix Table~\ref{tab:FM} reports the results, including the baseline for comparison.

The table shows that higher agentic betas predict lower posting growth under all factor models. Across all four factor models, the coefficients are negative at all three posting horizons and become larger in magnitude as the horizon increases from three to six to nine months. With industry fixed effects and firm characteristic controls, the six- and nine-month coefficient estimates are close to the baseline and significant at the 1 percent level under the three- and four-factor models. Under the five-factor model, they are smaller in magnitude and significant at the 5 percent level.

\subsection{Alternative Measures of Posting Growth}\label{sec:growth_measures}

We also test the robustness of the baseline results using two alternative measures of job posting growth. Our baseline measure in equation~(\ref{eq:growth}) is calculated as the difference between average monthly postings over the future window and the average over the three-month base window. The first alternative is the balanced growth rate of \citet{DavisHaltiwangerSchuh1996}, calculated as $200 \times (A - B)/(A + B)$, where $A$ is average monthly postings over the future window and $B$ is average monthly postings over the base window. This measure is bounded between $-200$ and $200$. Appendix Tables~\ref{tab:P1_balanced} and~\ref{tab:S1_balanced} report the panel and release-based results, respectively.

The second alternative measures growth from the three-month base window to a single future month by replacing average monthly postings over the future window with postings in month $t+h$. For the release-based regressions, this is the $h$-th month after the release. This measure retains more of the month-to-month variation in postings: its standard deviation in the panel sample is 1.3 to 1.6 times that of the baseline measure across the three horizons. Appendix Tables~\ref{tab:P1_single} and~\ref{tab:S1_single} report the panel and release-based results, respectively.

The estimates based on balanced growth are close to the baseline. In the panel regressions with time fixed effects, industry fixed effects, and firm characteristic controls, a one-standard-deviation increase in the agentic beta is associated with $0.88$, $2.00$, and $3.04$ percentage points lower balanced growth over three, six, and nine months, respectively. The corresponding baseline estimates in Table~\ref{tab:P1} imply about $0.87$, $2.11$, and $3.20$ percent lower posting growth. The six- and nine-month estimates based on balanced growth are significant at the 1 percent level.

Overall, we find that our baseline results are robust to using either of these two alternative measures of labor demand growth. This suggests that the baseline results are not simply driven by the way we define labor demand growth in the baseline specification.

\begin{table}[!htbp]
\stepcounter{anch}
{\footnotesize\caption{{\small\textbf{Posting Growth after the Claude Code 2.0 Release at Quantiles of Growth\label{tab:S4}}}}
\medskip{}
}{\footnotesize\par}

{\footnotesize This table reports regression results of posting growth after the release on the agentic beta dated 26 September 2025, standardized across firms. Column (1) reports the OLS estimate. Columns (2) to (6) report quantile regression estimates at the 10th, 25th, 50th, 75th, and 90th percentiles of posting growth. Panel A has no controls. Panel B adds Fama-French 49 industry fixed effects and the nine firm characteristic controls. Panel C also includes the AI beta of the same date, standardized across firms. t-statistics are reported in parentheses. {*}, {*}{*}, and {*}{*}{*} denote statistical significance at the 10\%, 5\%, and 1\% levels, respectively.\par}

{\footnotesize\medskip{}
}{\footnotesize\par}
\centering{}{\footnotesize{}%
\begin{tabular}{lcccccc}
\hline 
 & {\footnotesize (1)} & {\footnotesize (2)} & {\footnotesize (3)} & {\footnotesize (4)} & {\footnotesize (5)} & {\footnotesize (6)}\tabularnewline
 & {\footnotesize OLS} & \multicolumn{5}{c}{{\footnotesize Quantile of growth}}\tabularnewline
{\footnotesize Horizon (months)} &  & {\footnotesize 0.10} & {\footnotesize 0.25} & {\footnotesize 0.50} & {\footnotesize 0.75} & {\footnotesize 0.90}\tabularnewline
\hline 
 &  &  &  &  &  & \tabularnewline
\multicolumn{7}{l}{{\footnotesize \textbf{Panel A: $\beta^{Agentic}$, no controls}}}\tabularnewline
{\footnotesize 1} & {\footnotesize $-0.870$} & {\footnotesize $-2.661$} & {\footnotesize $-1.278$} & {\footnotesize $-1.238$} & {\footnotesize $0.992$} & {\footnotesize $-1.603$}\tabularnewline
 & {\footnotesize $(-0.534)$} & {\footnotesize $(-0.864)$} & {\footnotesize $(-0.845)$} & {\footnotesize $(-1.131)$} & {\footnotesize $(0.692)$} & {\footnotesize $(-0.724)$}\tabularnewline
{\footnotesize 3} & {\footnotesize $-2.689${*}} & {\footnotesize $-2.431$} & {\footnotesize $-2.344${*}} & {\footnotesize $-2.069${*}{*}} & {\footnotesize $-2.975${*}{*}} & {\footnotesize $-3.013$}\tabularnewline
 & {\footnotesize $(-1.736)$} & {\footnotesize $(-1.029)$} & {\footnotesize $(-1.878)$} & {\footnotesize $(-2.095)$} & {\footnotesize $(-2.166)$} & {\footnotesize $(-1.299)$}\tabularnewline
{\footnotesize 6} & {\footnotesize $-4.785${*}{*}{*}} & {\footnotesize $-3.140$} & {\footnotesize $-4.272${*}{*}{*}} & {\footnotesize $-3.985${*}{*}{*}} & {\footnotesize $-5.772${*}{*}{*}} & {\footnotesize $-7.940${*}{*}{*}}\tabularnewline
 & {\footnotesize $(-3.331)$} & {\footnotesize $(-1.292)$} & {\footnotesize $(-3.586)$} & {\footnotesize $(-4.228)$} & {\footnotesize $(-4.095)$} & {\footnotesize $(-2.915)$}\tabularnewline
{\footnotesize 9} & {\footnotesize $-4.772${*}{*}{*}} & {\footnotesize $-4.087${*}} & {\footnotesize $-4.150${*}{*}{*}} & {\footnotesize $-4.394${*}{*}{*}} & {\footnotesize $-5.630${*}{*}{*}} & {\footnotesize $-6.785${*}{*}{*}}\tabularnewline
 & {\footnotesize $(-3.198)$} & {\footnotesize $(-1.694)$} & {\footnotesize $(-4.067)$} & {\footnotesize $(-4.601)$} & {\footnotesize $(-3.565)$} & {\footnotesize $(-2.684)$}\tabularnewline
 &  &  &  &  &  & \tabularnewline
{\footnotesize Obs} & {\footnotesize 1,725} & {\footnotesize 1,725} & {\footnotesize 1,725} & {\footnotesize 1,725} & {\footnotesize 1,725} & {\footnotesize 1,725}\tabularnewline
 &  &  &  &  &  & \tabularnewline
\multicolumn{7}{l}{{\footnotesize \textbf{Panel B: $\beta^{Agentic}$, with controls}}}\tabularnewline
{\footnotesize 1} & {\footnotesize $-2.167$} & {\footnotesize $-3.017$} & {\footnotesize $-2.195$} & {\footnotesize $-1.575$} & {\footnotesize $0.638$} & {\footnotesize $0.131$}\tabularnewline
 & {\footnotesize $(-1.011)$} & {\footnotesize $(-1.412)$} & {\footnotesize $(-1.480)$} & {\footnotesize $(-1.190)$} & {\footnotesize $(0.438)$} & {\footnotesize $(0.080)$}\tabularnewline
{\footnotesize 3} & {\footnotesize $-3.383${*}} & {\footnotesize $-3.604${*}{*}} & {\footnotesize $-1.684$} & {\footnotesize $-2.074${*}} & {\footnotesize $-2.743${*}{*}} & {\footnotesize $-1.358$}\tabularnewline
 & {\footnotesize $(-1.772)$} & {\footnotesize $(-2.221)$} & {\footnotesize $(-1.333)$} & {\footnotesize $(-1.769)$} & {\footnotesize $(-2.043)$} & {\footnotesize $(-0.842)$}\tabularnewline
{\footnotesize 6} & {\footnotesize $-4.731${*}{*}{*}} & {\footnotesize $-4.419${*}{*}{*}} & {\footnotesize $-4.404${*}{*}{*}} & {\footnotesize $-3.633${*}{*}{*}} & {\footnotesize $-2.597${*}{*}} & {\footnotesize $-7.675${*}{*}{*}}\tabularnewline
 & {\footnotesize $(-2.979)$} & {\footnotesize $(-2.620)$} & {\footnotesize $(-3.468)$} & {\footnotesize $(-3.233)$} & {\footnotesize $(-1.996)$} & {\footnotesize $(-4.509)$}\tabularnewline
{\footnotesize 9} & {\footnotesize $-4.326${*}{*}{*}} & {\footnotesize $-5.988${*}{*}{*}} & {\footnotesize $-5.371${*}{*}{*}} & {\footnotesize $-4.325${*}{*}{*}} & {\footnotesize $-3.080${*}{*}} & {\footnotesize $-5.035${*}{*}{*}}\tabularnewline
 & {\footnotesize $(-2.732)$} & {\footnotesize $(-3.499)$} & {\footnotesize $(-4.259)$} & {\footnotesize $(-3.922)$} & {\footnotesize $(-2.266)$} & {\footnotesize $(-3.219)$}\tabularnewline
 &  &  &  &  &  & \tabularnewline
{\footnotesize Obs} & {\footnotesize 1,394} & {\footnotesize 1,394} & {\footnotesize 1,394} & {\footnotesize 1,394} & {\footnotesize 1,394} & {\footnotesize 1,394}\tabularnewline
 &  &  &  &  &  & \tabularnewline
\multicolumn{7}{l}{{\footnotesize \textbf{Panel C: $\beta^{Agentic}$, with controls and $\beta^{AI}$}}}\tabularnewline
{\footnotesize 1} & {\footnotesize $-0.683$} & {\footnotesize $-1.988$} & {\footnotesize $-1.262$} & {\footnotesize $1.057$} & {\footnotesize $2.448$} & {\footnotesize $-0.216$}\tabularnewline
 & {\footnotesize $(-0.281)$} & {\footnotesize $(-0.830)$} & {\footnotesize $(-0.690)$} & {\footnotesize $(0.666)$} & {\footnotesize $(1.477)$} & {\footnotesize $(-0.115)$}\tabularnewline
{\footnotesize 3} & {\footnotesize $-2.374$} & {\footnotesize $-0.628$} & {\footnotesize $-0.377$} & {\footnotesize $-1.434$} & {\footnotesize $-2.995${*}} & {\footnotesize $-0.081$}\tabularnewline
 & {\footnotesize $(-1.058)$} & {\footnotesize $(-0.281)$} & {\footnotesize $(-0.236)$} & {\footnotesize $(-1.015)$} & {\footnotesize $(-1.920)$} & {\footnotesize $(-0.045)$}\tabularnewline
{\footnotesize 6} & {\footnotesize $-4.151${*}{*}} & {\footnotesize $-4.814${*}{*}} & {\footnotesize $-3.959${*}{*}{*}} & {\footnotesize $-3.509${*}{*}{*}} & {\footnotesize $-2.586${*}} & {\footnotesize $-8.147${*}{*}{*}}\tabularnewline
 & {\footnotesize $(-2.230)$} & {\footnotesize $(-2.291)$} & {\footnotesize $(-2.675)$} & {\footnotesize $(-2.619)$} & {\footnotesize $(-1.728)$} & {\footnotesize $(-3.806)$}\tabularnewline
{\footnotesize 9} & {\footnotesize $-3.877${*}{*}} & {\footnotesize $-5.074${*}{*}} & {\footnotesize $-4.521${*}{*}{*}} & {\footnotesize $-3.400${*}{*}} & {\footnotesize $-2.927${*}} & {\footnotesize $-5.369${*}{*}{*}}\tabularnewline
 & {\footnotesize $(-2.092)$} & {\footnotesize $(-2.535)$} & {\footnotesize $(-3.013)$} & {\footnotesize $(-2.575)$} & {\footnotesize $(-1.833)$} & {\footnotesize $(-2.854)$}\tabularnewline
 &  &  &  &  &  & \tabularnewline
{\footnotesize Obs} & {\footnotesize 1,394} & {\footnotesize 1,394} & {\footnotesize 1,394} & {\footnotesize 1,394} & {\footnotesize 1,394} & {\footnotesize 1,394}\tabularnewline
 &  &  &  &  &  & \tabularnewline
\hline
\end{tabular}
}{\footnotesize\par}
\end{table}

\begin{table}[!htbp]
\stepcounter{anch}
{\footnotesize\caption{{\small\textbf{Posting Growth after the Claude Code 2.0 Release on the Open-Weight and Closed-Weight Agentic Betas\label{tab:S12}}}}
\medskip{}
}{\footnotesize\par}

{\footnotesize This table reports cross-sectional regression results of posting growth after the 29 September 2025 release of Claude Code 2.0 on the open-weight agentic beta in columns (1) and (2) and the closed-weight agentic beta in columns (3) and (4). Both betas are dated 26 September 2025 and standardized across firms. The dependent variable is growth in monthly postings from the three months before the release to the months after it, in percent. Columns (1) and (3) include no controls. Columns (2) and (4) add Fama-French 49 industry fixed effects and the nine firm characteristic controls. The number of observations and the adjusted $R^{2}$ of each regression are reported. t-statistics are reported in parentheses. {*}, {*}{*}, and {*}{*}{*} denote statistical significance at the 10\%, 5\%, and 1\% levels, respectively.\par}

{\footnotesize\medskip{}
}{\footnotesize\par}
\centering{}{\footnotesize{}%
\renewcommand{\arraystretch}{1.10}%
\setlength{\tabcolsep}{4pt}%
\begin{tabular}{lcccccccc}
\hline 
 & {\footnotesize (1)} & {\footnotesize (2)} & {\footnotesize (3)} & {\footnotesize (4)} & \multicolumn{4}{c}{{\footnotesize Adj $R^{2}$}}\tabularnewline
 & \multicolumn{2}{c}{{\footnotesize Open-weight $\beta^{Agentic}$}} & \multicolumn{2}{c}{{\footnotesize Closed-weight $\beta^{Agentic}$}} & \multicolumn{2}{c}{{\footnotesize Open-weight}} & \multicolumn{2}{c}{{\footnotesize Closed-weight}}\tabularnewline
{\footnotesize \shortstack[l]{Months after\\the shock}} & {\footnotesize \shortstack{No\\controls}} & {\footnotesize \shortstack{With\\controls}} & {\footnotesize \shortstack{No\\controls}} & {\footnotesize \shortstack{With\\controls}} & {\footnotesize (1)} & {\footnotesize (2)} & {\footnotesize (3)} & {\footnotesize (4)}\tabularnewline
\hline 
 &  &  &  &  &  &  &  & \tabularnewline
{\footnotesize 1} & {\footnotesize $0.610$} & {\footnotesize $0.484$} & {\footnotesize $-0.794$} & {\footnotesize $-2.081$} & {\footnotesize $0.000$} & {\footnotesize $0.001$} & {\footnotesize $0.000$} & {\footnotesize $0.002$}\tabularnewline
 & {\footnotesize $(0.409)$} & {\footnotesize $(0.291)$} & {\footnotesize $(-0.491)$} & {\footnotesize $(-0.979)$} &  &  &  & \tabularnewline
{\footnotesize 2} & {\footnotesize $-0.275$} & {\footnotesize $-0.222$} & {\footnotesize $-2.522${*}} & {\footnotesize $-3.658${*}{*}} & {\footnotesize $-0.001$} & {\footnotesize $0.003$} & {\footnotesize $0.002$} & {\footnotesize $0.007$}\tabularnewline
 & {\footnotesize $(-0.195)$} & {\footnotesize $(-0.151)$} & {\footnotesize $(-1.699)$} & {\footnotesize $(-2.006)$} &  &  &  & \tabularnewline
{\footnotesize 3} & {\footnotesize $-0.468$} & {\footnotesize $-0.173$} & {\footnotesize $-2.518$} & {\footnotesize $-3.313${*}} & {\footnotesize $0.000$} & {\footnotesize $0.006$} & {\footnotesize $0.002$} & {\footnotesize $0.009$}\tabularnewline
 & {\footnotesize $(-0.324)$} & {\footnotesize $(-0.110)$} & {\footnotesize $(-1.636)$} & {\footnotesize $(-1.773)$} &  &  &  & \tabularnewline
{\footnotesize 4} & {\footnotesize $-0.307$} & {\footnotesize $-0.102$} & {\footnotesize $-3.758${*}{*}} & {\footnotesize $-4.445${*}{*}} & {\footnotesize $-0.001$} & {\footnotesize $0.010$} & {\footnotesize $0.006$} & {\footnotesize $0.016$}\tabularnewline
 & {\footnotesize $(-0.217)$} & {\footnotesize $(-0.071)$} & {\footnotesize $(-2.495)$} & {\footnotesize $(-2.550)$} &  &  &  & \tabularnewline
{\footnotesize 5} & {\footnotesize $0.192$} & {\footnotesize $0.543$} & {\footnotesize $-4.406${*}{*}{*}} & {\footnotesize $-4.921${*}{*}{*}} & {\footnotesize $-0.001$} & {\footnotesize $0.025$} & {\footnotesize $0.008$} & {\footnotesize $0.033$}\tabularnewline
 & {\footnotesize $(0.133)$} & {\footnotesize $(0.384)$} & {\footnotesize $(-3.058)$} & {\footnotesize $(-3.117)$} &  &  &  & \tabularnewline
{\footnotesize 6} & {\footnotesize $-0.257$} & {\footnotesize $0.222$} & {\footnotesize $-4.617${*}{*}{*}} & {\footnotesize $-4.669${*}{*}{*}} & {\footnotesize $-0.001$} & {\footnotesize $0.036$} & {\footnotesize $0.009$} & {\footnotesize $0.043$}\tabularnewline
 & {\footnotesize $(-0.178)$} & {\footnotesize $(0.161)$} & {\footnotesize $(-3.220)$} & {\footnotesize $(-2.976)$} &  &  &  & \tabularnewline
{\footnotesize 7} & {\footnotesize $-0.369$} & {\footnotesize $0.117$} & {\footnotesize $-4.745${*}{*}{*}} & {\footnotesize $-4.300${*}{*}{*}} & {\footnotesize $-0.001$} & {\footnotesize $0.050$} & {\footnotesize $0.009$} & {\footnotesize $0.057$}\tabularnewline
 & {\footnotesize $(-0.259)$} & {\footnotesize $(0.086)$} & {\footnotesize $(-3.264)$} & {\footnotesize $(-2.728)$} &  &  &  & \tabularnewline
{\footnotesize 8} & {\footnotesize $-0.301$} & {\footnotesize $0.121$} & {\footnotesize $-4.430${*}{*}{*}} & {\footnotesize $-4.128${*}{*}{*}} & {\footnotesize $-0.001$} & {\footnotesize $0.057$} & {\footnotesize $0.008$} & {\footnotesize $0.063$}\tabularnewline
 & {\footnotesize $(-0.210)$} & {\footnotesize $(0.089)$} & {\footnotesize $(-2.978)$} & {\footnotesize $(-2.603)$} &  &  &  & \tabularnewline
{\footnotesize 9} & {\footnotesize $-0.192$} & {\footnotesize $0.213$} & {\footnotesize $-4.603${*}{*}{*}} & {\footnotesize $-4.260${*}{*}{*}} & {\footnotesize $-0.001$} & {\footnotesize $0.063$} & {\footnotesize $0.008$} & {\footnotesize $0.070$}\tabularnewline
 & {\footnotesize $(-0.132)$} & {\footnotesize $(0.158)$} & {\footnotesize $(-3.081)$} & {\footnotesize $(-2.717)$} &  &  &  & \tabularnewline
 &  &  &  &  &  &  &  & \tabularnewline
{\footnotesize Obs} & {\footnotesize 1,725} & {\footnotesize 1,394} & {\footnotesize 1,725} & {\footnotesize 1,394} &  &  &  & \tabularnewline
 &  &  &  &  &  &  &  & \tabularnewline
\hline
\end{tabular}
}{\footnotesize\par}
\end{table}

\subsection{Posting Growth across Firms and Models}\label{sec:response_patterns}

We also use quantile regressions to assess whether the negative relation between agentic exposure and labor demand extends across the distribution of firms' posting growth. These regressions relate subsequent posting growth to firms' pre-release agentic betas using the release-based event-study specification~(\ref{eq:posting_shock}). Table~\ref{tab:S4} reports the OLS estimate and the quantile regression estimates at the 10th, 25th, 50th, 75th, and 90th percentiles of the conditional posting-growth distribution. We focus on Panel C, which includes the AI beta, industry fixed effects, and firm characteristic controls. At each percentile, the coefficient gives the difference in posting growth associated with a one-standard-deviation increase in the agentic beta.

At the six-month horizon, we find a negative relation between agentic betas and posting growth at the bottom, middle, and top of this distribution. A one-standard-deviation increase in the agentic beta is associated with about $4.81$ percent lower growth at the 10th percentile, $3.51$ percent lower growth at the median, and $8.15$ percent lower growth at the 90th percentile. These estimates are significant at the 5, 1, and 1 percent levels, respectively. At the nine-month horizon, the median coefficient is $-3.40$, significant at the 5 percent level and close to the OLS estimate of $-3.88$. The coefficient at the 75th percentile is $-2.93$, significant at the 10 percent level.

Finally, we compare the labor demand response associated with exposure to open-weight and closed-weight agentic models. We use the release-based event-study specification~(\ref{eq:posting_shock}) by replacing the overall agentic beta with either the open-weight agentic beta or the closed-weight agentic beta. Each beta is dated 26 September 2025 and standardized across firms. We estimate a separate regression for each beta at each horizon from one to nine months. Table~\ref{tab:S12} reports the results.

We find that the negative relation with posting growth is concentrated in exposure to closed-weight agentic models. With industry fixed effects and firm characteristic controls, a one-standard-deviation increase in the closed-weight agentic beta is associated with about $4.67$ percent lower posting growth over six months and $4.26$ percent lower growth over nine months. The coefficients are significant at the 5 percent level at every horizon from four through nine months and at the 1 percent level from five through nine months. In contrast, we find that the corresponding coefficients on the open-weight agentic beta are $0.22$ and $0.21$, and both are statistically insignificant.

\clearpage
\begin{singlespace}
\bibliographystyle{jae}
\bibliography{references}
\end{singlespace}

\clearpage
\section*{Appendix}
\setcounter{table}{0}
\renewcommand{\thetable}{A\arabic{table}}
\begingroup
\begin{singlespace}
\footnotesize
\setlength{\LTpre}{0pt}
\setlength{\LTpost}{4pt}
\setlength{\tabcolsep}{3pt}
\renewcommand{\arraystretch}{1.02}
\stepcounter{anch}

\begin{longtable}{P{0.22\textwidth}P{0.54\textwidth}P{0.18\textwidth}}
\caption{\textbf{Variable Definitions}}\label{tab:variable_definitions}\\
\toprule
Name & Definition & Dataset used \\
\midrule
\endfirsthead
\toprule
Name & Definition & Dataset used \\
\midrule
\endhead
\midrule
\multicolumn{3}{r}{\textit{Continued on next page}} \\
\endfoot
\bottomrule
\endlastfoot

\multicolumn{3}{l}{\textit{AI exposure}} \\*
\addlinespace[2pt]
AI factor & First principal component of standardized weekly log growth in tokens, dollar expenditure, and the weekly average number of daily active users. & OpenRouter \\
Agentic factor & Weekly log growth of tool-call tokens, overall and separately for open-weight and closed-weight models. & OpenRouter \\
AI and agentic betas & Coefficients on the corresponding factors in 26-week stock-return regressions controlling for the market return, as in equation~(\ref{eq:beta}). & CRSP; OpenRouter; Fama-French \\

\addlinespace[5pt]
\multicolumn{3}{l}{\textit{Job postings}} \\*
\addlinespace[2pt]
Monthly postings & Number of new job postings by the firm's matched employers. & Lightcast \\
Posting growth & Growth in average monthly postings from the three-month base window to the next $h$ months, as in equation~(\ref{eq:growth}). & Lightcast \\
Junior titles & Sum of junior probabilities assigned by the seniority classifier. The comparison group is other scored postings. & Lightcast \\
Senior titles & Sum of senior probabilities assigned by the seniority classifier. The comparison group is other scored postings. & Lightcast \\
Asks 0--1 years & Postings requiring at most one year of experience, compared with postings requiring two or more years. & Lightcast \\
Asks 6+ years & Postings requiring six or more years of experience, compared with postings requiring zero to five years. & Lightcast \\
Expects AI-tool use & Postings containing phrases such as ``AI tools,'' ``leverage AI,'' or ``AI-assisted,'' compared with all other postings. & Lightcast \\
Remote or hybrid & Postings flagged as remote or hybrid, compared with on-site postings. & Lightcast \\
Top half, Eloundou & Postings in the most exposed occupations ranked by the Eloundou GPT exposure score. & Lightcast; \citet{Eloundou2024} \\
Top half, Webb AI & Postings in the most exposed occupations ranked by the Webb AI exposure percentile. & Lightcast; \citet{webb2020impact} \\
Top half, Anthropic & Postings in the most exposed occupations ranked by the Anthropic Economic Index measure of observed exposure. & Lightcast; \citet{MassenkoffMcCrory2026} \\
Relative posting growth & Posting growth for a given type minus posting growth for its comparison group. & Lightcast \\
Posted wage & Mean log annualized salary among postings that report pay. & Lightcast \\
Posted-wage growth & Change in mean log annualized posted salary, multiplied by 100. & Lightcast \\
Firm AI overlap & Posting-weighted average of the Eloundou, Webb, or Anthropic occupational exposure measure. & Lightcast; \citet{Eloundou2024}; \citet{webb2020impact}; \citet{MassenkoffMcCrory2026} \\

\noalign{\break}
\addlinespace[5pt]
\multicolumn{3}{l}{\textit{Firm outcomes}} \\*
\addlinespace[2pt]
Investment growth & Growth in three-quarter spending relative to the same quarters a year earlier, for capital expenditures and R\&D. & Compustat \\

\addlinespace[5pt]
\multicolumn{3}{l}{\textit{Firm controls}} \\*
\addlinespace[2pt]
Market beta & Coefficient on the market return in a 52-week stock-return regression. & CRSP; Fama-French \\
Return volatility & Standard deviation of weekly stock returns over 52 weeks. & CRSP \\
Momentum & Twelve-month stock return ending one month before month $t$. & CRSP \\
Log market equity & Natural logarithm of market equity. & CRSP \\
Sales growth & Growth in sales over two years. & Compustat \\
R\&D intensity & R\&D expense divided by sales. & Compustat \\
Book-to-market & Book equity divided by market equity. & Compustat; CRSP \\
Return on assets & Income divided by total assets. & Compustat \\
Net-loss indicator & Indicator for negative income. & Compustat \\
\end{longtable}
\end{singlespace}
\endgroup

\clearpage
\begin{table}[!htbp]
\stepcounter{anch}
{\footnotesize\caption{{\small\textbf{Task Types of OpenRouter Requests\label{tab:tasks}}}}
\medskip{}
}{\footnotesize\par}

{\footnotesize This table summarizes OpenRouter requests by task type. Each bold row is the total for the task types listed below it. Columns (1) to (3) show each row's share of all sampled requests, tokens, and expenditure, respectively, in percent. Column (4) shows the average number of tokens per sampled request. Within each row, columns (5) and (6) show the shares of tokens and expenditure for closed-weight models, and columns (7) and (8) show the corresponding shares for open-weight models.\par}

{\footnotesize\medskip{}
}{\footnotesize\par}
\centering{}{\footnotesize{}%
\renewcommand{\arraystretch}{1.15}%
\setlength{\tabcolsep}{3.5pt}%
\begin{tabular}{lcccccccc}
\hline 
 & {\footnotesize (1)} & {\footnotesize (2)} & {\footnotesize (3)} & {\footnotesize (4)} & {\footnotesize (5)} & {\footnotesize (6)} & {\footnotesize (7)} & {\footnotesize (8)}\tabularnewline
 & \multicolumn{3}{c}{{\footnotesize Share of the sample (\%)}} & {\footnotesize Tokens per} & \multicolumn{2}{c}{{\footnotesize \shortstack{Closed-weight\\models (\%)}}} & \multicolumn{2}{c}{{\footnotesize \shortstack{Open-weight\\models (\%)}}}\tabularnewline
{\footnotesize Task type} & {\footnotesize Requests} & {\footnotesize Tokens} & {\footnotesize Dollars} & {\footnotesize request} & {\footnotesize Tokens} & {\footnotesize Dollars} & {\footnotesize Tokens} & {\footnotesize Dollars}\tabularnewline
\hline 
 &  &  &  &  &  &  &  & \tabularnewline
{\footnotesize \textbf{Agent types}} & {\footnotesize $13.7$} & {\footnotesize $33.3$} & {\footnotesize $29.4$} & {\footnotesize 48,282} & {\footnotesize $24.0$} & {\footnotesize $74.1$} & {\footnotesize $72.2$} & {\footnotesize $25.1$}\tabularnewline
{\footnotesize \quad Workflow execution} & {\footnotesize $8.3$} & {\footnotesize $22.5$} & {\footnotesize $19.0$} & {\footnotesize 53,683} & {\footnotesize $24.1$} & {\footnotesize $74.1$} & {\footnotesize $72.3$} & {\footnotesize $25.1$}\tabularnewline
{\footnotesize \quad Multi-step planning} & {\footnotesize $2.7$} & {\footnotesize $6.7$} & {\footnotesize $6.9$} & {\footnotesize 49,228} & {\footnotesize $25.9$} & {\footnotesize $74.8$} & {\footnotesize $70.0$} & {\footnotesize $24.4$}\tabularnewline
{\footnotesize \quad Memory extraction} & {\footnotesize $1.0$} & {\footnotesize $1.8$} & {\footnotesize $0.8$} & {\footnotesize 35,853} & {\footnotesize $12.1$} & {\footnotesize $65.8$} & {\footnotesize $83.1$} & {\footnotesize $33.7$}\tabularnewline
{\footnotesize \quad Tool dispatch} & {\footnotesize $1.0$} & {\footnotesize $1.3$} & {\footnotesize $1.3$} & {\footnotesize 24,879} & {\footnotesize $29.7$} & {\footnotesize $75.1$} & {\footnotesize $66.0$} & {\footnotesize $23.8$}\tabularnewline
{\footnotesize \quad Web search} & {\footnotesize $0.7$} & {\footnotesize $1.0$} & {\footnotesize $1.3$} & {\footnotesize 30,274} & {\footnotesize $22.8$} & {\footnotesize $75.3$} & {\footnotesize $73.8$} & {\footnotesize $24.1$}\tabularnewline
 &  &  &  &  &  &  &  & \tabularnewline
{\footnotesize \textbf{Code types}} & {\footnotesize $10.8$} & {\footnotesize $33.2$} & {\footnotesize $27.1$} & {\footnotesize 61,007} & {\footnotesize $26.2$} & {\footnotesize $70.9$} & {\footnotesize $70.0$} & {\footnotesize $28.3$}\tabularnewline
{\footnotesize \quad General implementation} & {\footnotesize $3.3$} & {\footnotesize $10.6$} & {\footnotesize $8.8$} & {\footnotesize 63,945} & {\footnotesize $30.3$} & {\footnotesize $71.4$} & {\footnotesize $66.0$} & {\footnotesize $27.7$}\tabularnewline
{\footnotesize \quad Debugging} & {\footnotesize $2.0$} & {\footnotesize $6.6$} & {\footnotesize $5.1$} & {\footnotesize 64,491} & {\footnotesize $25.6$} & {\footnotesize $68.0$} & {\footnotesize $70.4$} & {\footnotesize $31.4$}\tabularnewline
{\footnotesize \quad File reading and writing} & {\footnotesize $1.7$} & {\footnotesize $5.1$} & {\footnotesize $4.0$} & {\footnotesize 60,786} & {\footnotesize $19.7$} & {\footnotesize $74.8$} & {\footnotesize $76.6$} & {\footnotesize $24.3$}\tabularnewline
{\footnotesize \quad Shell execution} & {\footnotesize $1.1$} & {\footnotesize $2.9$} & {\footnotesize $2.6$} & {\footnotesize 50,677} & {\footnotesize $28.2$} & {\footnotesize $62.6$} & {\footnotesize $68.3$} & {\footnotesize $36.1$}\tabularnewline
{\footnotesize \quad Front-end user interface} & {\footnotesize $0.7$} & {\footnotesize $2.8$} & {\footnotesize $2.3$} & {\footnotesize 77,974} & {\footnotesize $20.4$} & {\footnotesize $78.6$} & {\footnotesize $74.9$} & {\footnotesize $20.8$}\tabularnewline
{\footnotesize \quad Code review and security} & {\footnotesize $1.0$} & {\footnotesize $2.5$} & {\footnotesize $2.5$} & {\footnotesize 46,883} & {\footnotesize $30.0$} & {\footnotesize $68.8$} & {\footnotesize $67.1$} & {\footnotesize $30.7$}\tabularnewline
{\footnotesize \quad Repository scan} & {\footnotesize $0.5$} & {\footnotesize $1.4$} & {\footnotesize $1.0$} & {\footnotesize 55,448} & {\footnotesize $21.6$} & {\footnotesize $72.7$} & {\footnotesize $74.5$} & {\footnotesize $26.9$}\tabularnewline
{\footnotesize \quad DevOps configuration} & {\footnotesize $0.2$} & {\footnotesize $0.7$} & {\footnotesize $0.4$} & {\footnotesize 72,752} & {\footnotesize $26.3$} & {\footnotesize $68.8$} & {\footnotesize $69.9$} & {\footnotesize $30.6$}\tabularnewline
{\footnotesize \quad SQL and databases} & {\footnotesize $0.2$} & {\footnotesize $0.5$} & {\footnotesize $0.4$} & {\footnotesize 51,862} & {\footnotesize $26.8$} & {\footnotesize $76.3$} & {\footnotesize $69.3$} & {\footnotesize $23.0$}\tabularnewline
 &  &  &  &  &  &  &  & \tabularnewline
\hline
\end{tabular}
}{\footnotesize\par}
\end{table}
\clearpage
\begin{table}[!htbp]
\centering{\footnotesize
\renewcommand{\arraystretch}{1.15}%
\setlength{\tabcolsep}{3.5pt}%
\begin{tabular}{lcccccccc}
\hline 
 & {\footnotesize (1)} & {\footnotesize (2)} & {\footnotesize (3)} & {\footnotesize (4)} & {\footnotesize (5)} & {\footnotesize (6)} & {\footnotesize (7)} & {\footnotesize (8)}\tabularnewline
 & \multicolumn{3}{c}{{\footnotesize Share of the sample (\%)}} & {\footnotesize Tokens per} & \multicolumn{2}{c}{{\footnotesize \shortstack{Closed-weight\\models (\%)}}} & \multicolumn{2}{c}{{\footnotesize \shortstack{Open-weight\\models (\%)}}}\tabularnewline
{\footnotesize Task type} & {\footnotesize Requests} & {\footnotesize Tokens} & {\footnotesize Dollars} & {\footnotesize request} & {\footnotesize Tokens} & {\footnotesize Dollars} & {\footnotesize Tokens} & {\footnotesize Dollars}\tabularnewline
\hline 
 &  &  &  &  &  &  &  & \tabularnewline
{\footnotesize \textbf{Data types}} & {\footnotesize $18.8$} & {\footnotesize $6.4$} & {\footnotesize $9.2$} & {\footnotesize 6,783} & {\footnotesize $35.5$} & {\footnotesize $78.6$} & {\footnotesize $62.2$} & {\footnotesize $20.7$}\tabularnewline
{\footnotesize \quad Extraction} & {\footnotesize $14.2$} & {\footnotesize $4.8$} & {\footnotesize $6.4$} & {\footnotesize 6,677} & {\footnotesize $36.5$} & {\footnotesize $78.5$} & {\footnotesize $61.1$} & {\footnotesize $20.8$}\tabularnewline
{\footnotesize \quad Transformation} & {\footnotesize $4.6$} & {\footnotesize $1.6$} & {\footnotesize $2.8$} & {\footnotesize 7,112} & {\footnotesize $32.5$} & {\footnotesize $78.8$} & {\footnotesize $65.4$} & {\footnotesize $20.3$}\tabularnewline
 &  &  &  &  &  &  &  & \tabularnewline
{\footnotesize \textbf{Other types}} & {\footnotesize $56.7$} & {\footnotesize $27.1$} & {\footnotesize $34.4$} & {\footnotesize 9,477} & {\footnotesize $30.0$} & {\footnotesize $76.4$} & {\footnotesize $67.0$} & {\footnotesize $22.2$}\tabularnewline
{\footnotesize \quad Classification and tagging} & {\footnotesize $26.5$} & {\footnotesize $6.7$} & {\footnotesize $10.0$} & {\footnotesize 5,053} & {\footnotesize $39.1$} & {\footnotesize $81.1$} & {\footnotesize $55.7$} & {\footnotesize $17.6$}\tabularnewline
{\footnotesize \quad Role play and fiction} & {\footnotesize $9.0$} & {\footnotesize $4.0$} & {\footnotesize $3.4$} & {\footnotesize 8,777} & {\footnotesize $21.6$} & {\footnotesize $60.5$} & {\footnotesize $77.3$} & {\footnotesize $38.1$}\tabularnewline
{\footnotesize \quad Conversational reply} & {\footnotesize $2.6$} & {\footnotesize $3.1$} & {\footnotesize $2.5$} & {\footnotesize 23,743} & {\footnotesize $22.3$} & {\footnotesize $78.8$} & {\footnotesize $74.8$} & {\footnotesize $20.8$}\tabularnewline
{\footnotesize \quad Question answering} & {\footnotesize $2.9$} & {\footnotesize $2.7$} & {\footnotesize $3.8$} & {\footnotesize 18,460} & {\footnotesize $29.3$} & {\footnotesize $75.4$} & {\footnotesize $67.4$} & {\footnotesize $24.0$}\tabularnewline
{\footnotesize \quad Content writing} & {\footnotesize $5.2$} & {\footnotesize $2.6$} & {\footnotesize $4.9$} & {\footnotesize 9,881} & {\footnotesize $32.0$} & {\footnotesize $80.8$} & {\footnotesize $65.8$} & {\footnotesize $16.9$}\tabularnewline
{\footnotesize \quad Other} & {\footnotesize $1.3$} & {\footnotesize $2.0$} & {\footnotesize $1.8$} & {\footnotesize 30,234} & {\footnotesize $20.7$} & {\footnotesize $81.6$} & {\footnotesize $77.0$} & {\footnotesize $16.9$}\tabularnewline
{\footnotesize \quad Summarization} & {\footnotesize $2.4$} & {\footnotesize $1.4$} & {\footnotesize $1.4$} & {\footnotesize 11,825} & {\footnotesize $33.3$} & {\footnotesize $81.0$} & {\footnotesize $65.4$} & {\footnotesize $18.7$}\tabularnewline
{\footnotesize \quad Customer support} & {\footnotesize $2.0$} & {\footnotesize $1.3$} & {\footnotesize $1.4$} & {\footnotesize 12,853} & {\footnotesize $55.2$} & {\footnotesize $79.5$} & {\footnotesize $43.3$} & {\footnotesize $20.2$}\tabularnewline
{\footnotesize \quad Research report} & {\footnotesize $0.6$} & {\footnotesize $1.0$} & {\footnotesize $1.9$} & {\footnotesize 35,035} & {\footnotesize $28.2$} & {\footnotesize $81.7$} & {\footnotesize $67.7$} & {\footnotesize $17.6$}\tabularnewline
{\footnotesize \quad Security audit} & {\footnotesize $0.4$} & {\footnotesize $0.9$} & {\footnotesize $0.7$} & {\footnotesize 47,334} & {\footnotesize $10.0$} & {\footnotesize $42.2$} & {\footnotesize $87.3$} & {\footnotesize $56.8$}\tabularnewline
{\footnotesize \quad DevOps} & {\footnotesize $0.1$} & {\footnotesize $0.5$} & {\footnotesize $0.3$} & {\footnotesize 69,920} & {\footnotesize $17.1$} & {\footnotesize $70.3$} & {\footnotesize $77.5$} & {\footnotesize $29.3$}\tabularnewline
{\footnotesize \quad Finance and trading} & {\footnotesize $0.5$} & {\footnotesize $0.4$} & {\footnotesize $0.6$} & {\footnotesize 16,232} & {\footnotesize $20.3$} & {\footnotesize $68.7$} & {\footnotesize $75.8$} & {\footnotesize $30.4$}\tabularnewline
{\footnotesize \quad Translation} & {\footnotesize $2.8$} & {\footnotesize $0.3$} & {\footnotesize $0.4$} & {\footnotesize 2,029} & {\footnotesize $38.0$} & {\footnotesize $81.6$} & {\footnotesize $60.1$} & {\footnotesize $17.8$}\tabularnewline
{\footnotesize \quad Math} & {\footnotesize $0.4$} & {\footnotesize $0.2$} & {\footnotesize $0.7$} & {\footnotesize 7,687} & {\footnotesize $28.8$} & {\footnotesize $65.4$} & {\footnotesize $68.2$} & {\footnotesize $33.8$}\tabularnewline
{\footnotesize \quad Image prompting} & {\footnotesize $0.1$} & {\footnotesize $0.0$} & {\footnotesize $0.5$} & {\footnotesize 5,509} & {\footnotesize $60.5$} & {\footnotesize $54.1$} & {\footnotesize $36.5$} & {\footnotesize $21.4$}\tabularnewline
 &  &  &  &  &  &  &  & \tabularnewline
{\footnotesize All types} & {\footnotesize $100.0$} & {\footnotesize $100.0$} & {\footnotesize $100.0$} & {\footnotesize 19,838} & {\footnotesize $27.1$} & {\footnotesize $74.4$} & {\footnotesize $69.4$} & {\footnotesize $24.5$}\tabularnewline
 &  &  &  &  &  &  &  & \tabularnewline
\hline
\end{tabular}
}{\footnotesize\par}
\end{table}

\clearpage
\begin{table}[!htbp]
\stepcounter{anch}
{\footnotesize\caption{{\small\textbf{Future AI Mentions in SEC Filings on the AI and Agentic Betas with Controls\label{tab:mentions_controls}}}}
\medskip{}
}{\footnotesize\par}

{\footnotesize This table repeats the regressions of Table~\ref{tab:mentions} with industry fixed effects and firm characteristic controls. It reports panel regression results of future mentions of AI in firms' SEC filings on the trailing AI beta (Panel A), the trailing agentic beta (Panel B), and both betas in the same regression (Panel C). Mentions are sentences in the business section (Item 1 of Form 10-K) and in management's discussion and analysis (Item 7 of Form 10-K and Part I, Item 2 of Form 10-Q) of the annual and quarterly reports that a firm files in the six months after month $t$. Columns (1) to (3) use agentic mentions, and columns (4) to (6) use AI deployment mentions, both as defined in Table~\ref{tab:mentions}. The dependent variable is 100 times an indicator for at least one mention (Any mention), 100 times the log of one plus the number of mentions (Log count), or the change in Log count from the same six months one year earlier ($\Delta$ Log count). The beta is standardized within each month, so the coefficient estimates measure the change in the dependent variable per cross-sectional standard deviation of the beta. All regressions include Fama-French 49 industry fixed effects, time fixed effects, and the nine firm characteristic controls of Table~\ref{tab:P1}. The number of observations and the adjusted $R^{2}$ are reported. t-statistics are reported in parentheses. Standard errors are clustered by firm. {*}, {*}{*}, and {*}{*}{*} denote statistical significance at the 10\%, 5\%, and 1\% levels, respectively.\par}

{\footnotesize\medskip{}
}{\footnotesize\par}
\centering{}{\footnotesize{}%
\renewcommand{\arraystretch}{1.10}%
\begin{tabular}{lcccccc}
\hline 
 & {\footnotesize (1)} & {\footnotesize (2)} & {\footnotesize (3)} & {\footnotesize (4)} & {\footnotesize (5)} & {\footnotesize (6)}\tabularnewline
 & \multicolumn{3}{c}{{\footnotesize Agentic mentions}} & \multicolumn{3}{c}{{\footnotesize AI deployment mentions}}\tabularnewline
{\footnotesize Dependent variable} & {\footnotesize Any mention} & {\footnotesize Log count} & {\footnotesize $\Delta$ Log count} & {\footnotesize Any mention} & {\footnotesize Log count} & {\footnotesize $\Delta$ Log count}\tabularnewline
\hline 
\multicolumn{7}{l}{{\footnotesize \textbf{Panel A: $\beta^{AI}$}}}\tabularnewline
 &  &  &  &  &  & \tabularnewline
{\footnotesize $\beta^{AI}$} & {\footnotesize $0.037$} & {\footnotesize $-0.209$} & {\footnotesize $0.188$} & {\footnotesize $0.760$} & {\footnotesize $0.908$} & {\footnotesize $-0.061$}\tabularnewline
 & {\footnotesize $(0.136)$} & {\footnotesize $(-0.410)$} & {\footnotesize $(0.475)$} & {\footnotesize $(1.470)$} & {\footnotesize $(1.054)$} & {\footnotesize $(-0.101)$}\tabularnewline
 &  &  &  &  &  & \tabularnewline
{\footnotesize Adj $R^{2}$} & {\footnotesize $0.247$} & {\footnotesize $0.243$} & {\footnotesize $0.195$} & {\footnotesize $0.239$} & {\footnotesize $0.341$} & {\footnotesize $0.061$}\tabularnewline
{\footnotesize Obs} & {\footnotesize 16,249} & {\footnotesize 16,249} & {\footnotesize 16,217} & {\footnotesize 16,249} & {\footnotesize 16,249} & {\footnotesize 16,217}\tabularnewline
 &  &  &  &  &  & \tabularnewline
\hline 
\multicolumn{7}{l}{{\footnotesize \textbf{Panel B: $\beta^{Agentic}$}}}\tabularnewline
 &  &  &  &  &  & \tabularnewline
{\footnotesize $\beta^{Agentic}$} & {\footnotesize $0.832${*}{*}{*}} & {\footnotesize $1.848${*}{*}{*}} & {\footnotesize $1.423${*}{*}{*}} & {\footnotesize $1.126${*}{*}} & {\footnotesize $2.389${*}{*}{*}} & {\footnotesize $1.013${*}}\tabularnewline
 & {\footnotesize $(3.244)$} & {\footnotesize $(4.094)$} & {\footnotesize $(3.923)$} & {\footnotesize $(2.312)$} & {\footnotesize $(2.995)$} & {\footnotesize $(1.858)$}\tabularnewline
 &  &  &  &  &  & \tabularnewline
{\footnotesize Adj $R^{2}$} & {\footnotesize $0.249$} & {\footnotesize $0.245$} & {\footnotesize $0.197$} & {\footnotesize $0.239$} & {\footnotesize $0.342$} & {\footnotesize $0.061$}\tabularnewline
{\footnotesize Obs} & {\footnotesize 16,249} & {\footnotesize 16,249} & {\footnotesize 16,217} & {\footnotesize 16,249} & {\footnotesize 16,249} & {\footnotesize 16,217}\tabularnewline
 &  &  &  &  &  & \tabularnewline
\hline 
\multicolumn{7}{l}{{\footnotesize \textbf{Panel C: $\beta^{Agentic}$ and $\beta^{AI}$ in the same regression}}}\tabularnewline
 &  &  &  &  &  & \tabularnewline
{\footnotesize $\beta^{Agentic}$} & {\footnotesize $1.235${*}{*}{*}} & {\footnotesize $3.000${*}{*}{*}} & {\footnotesize $2.000${*}{*}{*}} & {\footnotesize $1.039${*}} & {\footnotesize $2.832${*}{*}{*}} & {\footnotesize $1.597${*}{*}}\tabularnewline
 & {\footnotesize $(3.746)$} & {\footnotesize $(4.509)$} & {\footnotesize $(4.226)$} & {\footnotesize $(1.673)$} & {\footnotesize $(2.584)$} & {\footnotesize $(2.251)$}\tabularnewline
{\footnotesize $\beta^{AI}$} & {\footnotesize $-0.691${*}{*}} & {\footnotesize $-1.976${*}{*}{*}} & {\footnotesize $-0.990${*}} & {\footnotesize $0.148$} & {\footnotesize $-0.759$} & {\footnotesize $-1.001$}\tabularnewline
 & {\footnotesize $(-2.009)$} & {\footnotesize $(-2.722)$} & {\footnotesize $(-1.938)$} & {\footnotesize $(0.225)$} & {\footnotesize $(-0.649)$} & {\footnotesize $(-1.287)$}\tabularnewline
 &  &  &  &  &  & \tabularnewline
{\footnotesize Adj $R^{2}$} & {\footnotesize $0.249$} & {\footnotesize $0.247$} & {\footnotesize $0.197$} & {\footnotesize $0.239$} & {\footnotesize $0.342$} & {\footnotesize $0.061$}\tabularnewline
{\footnotesize Obs} & {\footnotesize 16,249} & {\footnotesize 16,249} & {\footnotesize 16,217} & {\footnotesize 16,249} & {\footnotesize 16,249} & {\footnotesize 16,217}\tabularnewline
 &  &  &  &  &  & \tabularnewline
\hline
\end{tabular}
}{\footnotesize\par}
\end{table}

\clearpage
\begin{table}[!htbp]
\stepcounter{anch}
{\footnotesize\caption{{\small\textbf{Future AI Mentions by Filing Type on the AI and Agentic Betas\label{tab:mentions_filing}}}}
\medskip{}
}{\footnotesize\par}

{\footnotesize This table repeats the Log count regressions of Table~\ref{tab:mentions} separately by filing type. It reports panel regression results of future mentions of AI on the trailing AI beta (Panel A), the trailing agentic beta (Panel B), and both betas in the same regression (Panel C). Mentions are sentences in the filings of one type that a firm files in the six months after month $t$: annual reports (10-K) and quarterly reports (10-Q). Columns (1) and (2) use agentic mentions, and columns (3) and (4) use AI deployment mentions, both as defined in Table~\ref{tab:mentions}. The dependent variable is 100 times the log of one plus the number of mentions. A firm-month is included if the firm files at least one report of the type in the six months. The beta is standardized within each month, so the coefficient estimates measure the change in the dependent variable per cross-sectional standard deviation of the beta. All regressions include time fixed effects. The number of observations and the adjusted $R^{2}$ are reported. t-statistics are reported in parentheses. Standard errors are clustered by firm. {*}, {*}{*}, and {*}{*}{*} denote statistical significance at the 10\%, 5\%, and 1\% levels, respectively.\par}

{\footnotesize\medskip{}
}{\footnotesize\par}
\centering{}{\footnotesize{}%
\renewcommand{\arraystretch}{1.10}%
\begin{tabular}{lcccc}
\hline 
 & {\footnotesize (1)} & {\footnotesize (2)} & {\footnotesize (3)} & {\footnotesize (4)}\tabularnewline
 & \multicolumn{2}{c}{{\footnotesize Agentic mentions}} & \multicolumn{2}{c}{{\footnotesize AI deployment mentions}}\tabularnewline
{\footnotesize Filing type} & {\footnotesize 10-K} & {\footnotesize 10-Q} & {\footnotesize 10-K} & {\footnotesize 10-Q}\tabularnewline
\hline 
\multicolumn{5}{l}{{\footnotesize \textbf{Panel A: $\beta^{AI}$}}}\tabularnewline
 &  &  &  & \tabularnewline
{\footnotesize $\beta^{AI}$} & {\footnotesize $0.026$} & {\footnotesize $0.019$} & {\footnotesize $2.741${*}} & {\footnotesize $0.068$}\tabularnewline
 & {\footnotesize $(0.032)$} & {\footnotesize $(0.083)$} & {\footnotesize $(1.717)$} & {\footnotesize $(0.137)$}\tabularnewline
 &  &  &  & \tabularnewline
{\footnotesize Adj $R^{2}$} & {\footnotesize $0.004$} & {\footnotesize $0.001$} & {\footnotesize $0.002$} & {\footnotesize $0.005$}\tabularnewline
{\footnotesize Obs} & {\footnotesize 9,337} & {\footnotesize 16,109} & {\footnotesize 9,337} & {\footnotesize 16,109}\tabularnewline
 &  &  &  & \tabularnewline
\hline 
\multicolumn{5}{l}{{\footnotesize \textbf{Panel B: $\beta^{Agentic}$}}}\tabularnewline
 &  &  &  & \tabularnewline
{\footnotesize $\beta^{Agentic}$} & {\footnotesize $3.855${*}{*}{*}} & {\footnotesize $0.885${*}{*}{*}} & {\footnotesize $8.115${*}{*}{*}} & {\footnotesize $0.763$}\tabularnewline
 & {\footnotesize $(4.698)$} & {\footnotesize $(3.310)$} & {\footnotesize $(5.409)$} & {\footnotesize $(1.453)$}\tabularnewline
 &  &  &  & \tabularnewline
{\footnotesize Adj $R^{2}$} & {\footnotesize $0.011$} & {\footnotesize $0.003$} & {\footnotesize $0.010$} & {\footnotesize $0.005$}\tabularnewline
{\footnotesize Obs} & {\footnotesize 9,337} & {\footnotesize 16,109} & {\footnotesize 9,337} & {\footnotesize 16,109}\tabularnewline
 &  &  &  & \tabularnewline
\hline 
\multicolumn{5}{l}{{\footnotesize \textbf{Panel C: $\beta^{Agentic}$ and $\beta^{AI}$ in the same regression}}}\tabularnewline
 &  &  &  & \tabularnewline
{\footnotesize $\beta^{Agentic}$} & {\footnotesize $5.186${*}{*}{*}} & {\footnotesize $1.331${*}{*}{*}} & {\footnotesize $9.071${*}{*}{*}} & {\footnotesize $1.101${*}}\tabularnewline
 & {\footnotesize $(4.624)$} & {\footnotesize $(3.668)$} & {\footnotesize $(5.043)$} & {\footnotesize $(1.647)$}\tabularnewline
{\footnotesize $\beta^{AI}$} & {\footnotesize $-2.615${*}{*}} & {\footnotesize $-0.761${*}{*}} & {\footnotesize $-1.878$} & {\footnotesize $-0.577$}\tabularnewline
 & {\footnotesize $(-2.376)$} & {\footnotesize $(-2.413)$} & {\footnotesize $(-0.984)$} & {\footnotesize $(-0.903)$}\tabularnewline
 &  &  &  & \tabularnewline
{\footnotesize Adj $R^{2}$} & {\footnotesize $0.013$} & {\footnotesize $0.004$} & {\footnotesize $0.011$} & {\footnotesize $0.005$}\tabularnewline
{\footnotesize Obs} & {\footnotesize 9,337} & {\footnotesize 16,109} & {\footnotesize 9,337} & {\footnotesize 16,109}\tabularnewline
 &  &  &  & \tabularnewline
\hline
\end{tabular}
}{\footnotesize\par}
\end{table}

\clearpage
\begin{table}[!htbp]
\stepcounter{anch}
{\footnotesize\caption{{\small\textbf{Future AI Mentions in Earnings Calls on the AI and Agentic Betas\label{tab:mentions_calls}}}}
\medskip{}
}{\footnotesize\par}

{\footnotesize This table repeats the regressions of Table~\ref{tab:mentions} using firms' earnings conference calls in place of their SEC filings. It reports panel regression results of future mentions of AI by a firm's management in earnings calls on the trailing AI beta (Panel A), the trailing agentic beta (Panel B), and both betas in the same regression (Panel C). Mentions are sentences that the firm's management speaks, in the prepared remarks and in the answers to analysts' questions, in the earnings calls that the firm holds in the six months after month $t$. Sentences spoken by analysts and by the call operator are excluded. Columns (1) to (3) use agentic mentions, and columns (4) to (6) use AI deployment mentions, both as defined in Table~\ref{tab:mentions}. The dependent variable is 100 times an indicator for at least one mention (Any mention), 100 times the log of one plus the number of mentions (Log count), or the change in Log count from the same six months one year earlier ($\Delta$ Log count). A firm-month is included if the firm holds at least one earnings call in the six months. The beta is standardized within each month, so the coefficient estimates measure the change in the dependent variable per cross-sectional standard deviation of the beta. All regressions include time fixed effects. The number of observations and the adjusted $R^{2}$ are reported. t-statistics are reported in parentheses. Standard errors are clustered by firm. {*}, {*}{*}, and {*}{*}{*} denote statistical significance at the 10\%, 5\%, and 1\% levels, respectively.\par}

{\footnotesize\medskip{}
}{\footnotesize\par}
\centering{}{\footnotesize{}%
\renewcommand{\arraystretch}{1.10}%
\begin{tabular}{lcccccc}
\hline 
 & {\footnotesize (1)} & {\footnotesize (2)} & {\footnotesize (3)} & {\footnotesize (4)} & {\footnotesize (5)} & {\footnotesize (6)}\tabularnewline
 & \multicolumn{3}{c}{{\footnotesize Agentic mentions}} & \multicolumn{3}{c}{{\footnotesize AI deployment mentions}}\tabularnewline
{\footnotesize Dependent variable} & {\footnotesize Any mention} & {\footnotesize Log count} & {\footnotesize $\Delta$ Log count} & {\footnotesize Any mention} & {\footnotesize Log count} & {\footnotesize $\Delta$ Log count}\tabularnewline
\hline 
\multicolumn{7}{l}{{\footnotesize \textbf{Panel A: $\beta^{AI}$}}}\tabularnewline
 &  &  &  &  &  & \tabularnewline
{\footnotesize $\beta^{AI}$} & {\footnotesize $0.307$} & {\footnotesize $1.030$} & {\footnotesize $1.601${*}} & {\footnotesize $-0.247$} & {\footnotesize $-0.504$} & {\footnotesize $-1.368$}\tabularnewline
 & {\footnotesize $(0.568)$} & {\footnotesize $(0.950)$} & {\footnotesize $(1.932)$} & {\footnotesize $(-0.359)$} & {\footnotesize $(-0.335)$} & {\footnotesize $(-1.425)$}\tabularnewline
 &  &  &  &  &  & \tabularnewline
{\footnotesize Adj $R^{2}$} & {\footnotesize $0.004$} & {\footnotesize $0.003$} & {\footnotesize $0.001$} & {\footnotesize $0.003$} & {\footnotesize $0.004$} & {\footnotesize $0.007$}\tabularnewline
{\footnotesize Obs} & {\footnotesize 15,850} & {\footnotesize 15,850} & {\footnotesize 15,688} & {\footnotesize 15,850} & {\footnotesize 15,850} & {\footnotesize 15,688}\tabularnewline
 &  &  &  &  &  & \tabularnewline
\hline 
\multicolumn{7}{l}{{\footnotesize \textbf{Panel B: $\beta^{Agentic}$}}}\tabularnewline
 &  &  &  &  &  & \tabularnewline
{\footnotesize $\beta^{Agentic}$} & {\footnotesize $1.458${*}{*}{*}} & {\footnotesize $4.308${*}{*}{*}} & {\footnotesize $2.074${*}{*}} & {\footnotesize $1.696${*}{*}} & {\footnotesize $5.319${*}{*}{*}} & {\footnotesize $1.537$}\tabularnewline
 & {\footnotesize $(2.665)$} & {\footnotesize $(3.759)$} & {\footnotesize $(2.528)$} & {\footnotesize $(2.409)$} & {\footnotesize $(3.490)$} & {\footnotesize $(1.614)$}\tabularnewline
 &  &  &  &  &  & \tabularnewline
{\footnotesize Adj $R^{2}$} & {\footnotesize $0.005$} & {\footnotesize $0.006$} & {\footnotesize $0.002$} & {\footnotesize $0.004$} & {\footnotesize $0.006$} & {\footnotesize $0.007$}\tabularnewline
{\footnotesize Obs} & {\footnotesize 15,850} & {\footnotesize 15,850} & {\footnotesize 15,688} & {\footnotesize 15,850} & {\footnotesize 15,850} & {\footnotesize 15,688}\tabularnewline
 &  &  &  &  &  & \tabularnewline
\hline 
\multicolumn{7}{l}{{\footnotesize \textbf{Panel C: $\beta^{Agentic}$ and $\beta^{AI}$ in the same regression}}}\tabularnewline
 &  &  &  &  &  & \tabularnewline
{\footnotesize $\beta^{Agentic}$} & {\footnotesize $1.943${*}{*}{*}} & {\footnotesize $5.631${*}{*}{*}} & {\footnotesize $1.728$} & {\footnotesize $2.796${*}{*}{*}} & {\footnotesize $8.530${*}{*}{*}} & {\footnotesize $3.548${*}{*}{*}}\tabularnewline
 & {\footnotesize $(2.803)$} & {\footnotesize $(3.755)$} & {\footnotesize $(1.629)$} & {\footnotesize $(3.203)$} & {\footnotesize $(4.457)$} & {\footnotesize $(2.933)$}\tabularnewline
{\footnotesize $\beta^{AI}$} & {\footnotesize $-0.829$} & {\footnotesize $-2.263$} & {\footnotesize $0.591$} & {\footnotesize $-1.881${*}{*}} & {\footnotesize $-5.492${*}{*}{*}} & {\footnotesize $-3.442${*}{*}{*}}\tabularnewline
 & {\footnotesize $(-1.214)$} & {\footnotesize $(-1.591)$} & {\footnotesize $(0.553)$} & {\footnotesize $(-2.224)$} & {\footnotesize $(-2.922)$} & {\footnotesize $(-2.835)$}\tabularnewline
 &  &  &  &  &  & \tabularnewline
{\footnotesize Adj $R^{2}$} & {\footnotesize $0.005$} & {\footnotesize $0.007$} & {\footnotesize $0.002$} & {\footnotesize $0.005$} & {\footnotesize $0.008$} & {\footnotesize $0.008$}\tabularnewline
{\footnotesize Obs} & {\footnotesize 15,850} & {\footnotesize 15,850} & {\footnotesize 15,688} & {\footnotesize 15,850} & {\footnotesize 15,850} & {\footnotesize 15,688}\tabularnewline
 &  &  &  &  &  & \tabularnewline
\hline
\end{tabular}
}{\footnotesize\par}
\end{table}

\clearpage
\begin{table}[!htbp]
\stepcounter{anch}
{\footnotesize\caption{{\small\textbf{Future Posting Growth within Occupations on the Agentic Beta\label{tab:P2}}}}
\medskip{}
}{\footnotesize\par}

{\footnotesize This table reports panel regression results for future posting growth within occupations. Observations are at the occupation-firm level in each month. The dependent variable is growth in the firm's monthly postings in that occupation from the three months ending in month $t$ to the following 3, 6, or 9 months, in percent. The independent variable is the firm's trailing agentic beta, standardized within each month. Each occupation is weighted by its share of the firm's postings in the three months ending in month $t$. Columns (1) to (3) include occupation-by-time fixed effects. Columns (4) to (6) add Fama-French 49 industry fixed effects and the nine firm characteristic controls. The number of observations and the adjusted $R^{2}$ are reported. t-statistics are reported in parentheses. Standard errors are clustered by firm. {*}, {*}{*}, and {*}{*}{*} denote statistical significance at the 10\%, 5\%, and 1\% levels, respectively.\par}

{\footnotesize\medskip{}
}{\footnotesize\par}
\centering{}{\footnotesize{}%
\renewcommand{\arraystretch}{1.10}%
\begin{tabular}{lcccccc}
\hline 
 & {\footnotesize (1)} & {\footnotesize (2)} & {\footnotesize (3)} & {\footnotesize (4)} & {\footnotesize (5)} & {\footnotesize (6)}\tabularnewline
 & \multicolumn{3}{c}{{\footnotesize Occupation $\times$ time fixed effects}} & \multicolumn{3}{c}{{\footnotesize With controls}}\tabularnewline
{\footnotesize Horizon (months)} & {\footnotesize 3} & {\footnotesize 6} & {\footnotesize 9} & {\footnotesize 3} & {\footnotesize 6} & {\footnotesize 9}\tabularnewline
\hline 
 &  &  &  &  &  & \tabularnewline
{\footnotesize $\beta^{Agentic}$} & {\footnotesize $-0.769${*}} & {\footnotesize $-1.465${*}{*}} & {\footnotesize $-2.008${*}{*}} & {\footnotesize $-0.735$} & {\footnotesize $-1.669${*}{*}{*}} & {\footnotesize $-2.440${*}{*}{*}}\tabularnewline
 & {\footnotesize $(-1.674)$} & {\footnotesize $(-2.437)$} & {\footnotesize $(-2.555)$} & {\footnotesize $(-1.568)$} & {\footnotesize $(-2.682)$} & {\footnotesize $(-2.744)$}\tabularnewline
 &  &  &  &  &  & \tabularnewline
{\footnotesize Adj $R^{2}$} & {\footnotesize $0.056$} & {\footnotesize $0.061$} & {\footnotesize $0.048$} & {\footnotesize $0.064$} & {\footnotesize $0.072$} & {\footnotesize $0.063$}\tabularnewline
{\footnotesize Obs} & {\footnotesize 904,105} & {\footnotesize 628,738} & {\footnotesize 372,195} & {\footnotesize 893,100} & {\footnotesize 621,098} & {\footnotesize 367,809}\tabularnewline
 &  &  &  &  &  & \tabularnewline
\hline
\end{tabular}
}{\footnotesize\par}
\end{table}

\clearpage
\begin{table}[!htbp]
\stepcounter{anch}
{\footnotesize\caption{{\small\textbf{Posting Growth on the Non-Agentic Beta\label{tab:nonagentic}}}}
\medskip{}
}{\footnotesize\par}

{\footnotesize This table reports regression results of posting growth on the non-agentic beta. The non-agentic beta is estimated in the same way as the agentic beta, using weekly log growth in OpenRouter non-agentic tokens, defined as the tokens associated with responses that do not end in a tool call. Panels A and C include the non-agentic beta alone. Panels B and D include the agentic and non-agentic betas in the same regression. Panels A and B report panel regression results of future posting growth on the trailing betas. The dependent variable is the growth of monthly job postings from the three months ending in month $t$ to the following 3, 6, or 9 months, in percent. The betas are standardized within each month. Columns (1) to (3) include time fixed effects only. Columns (4) to (6) add Fama-French 49 industry fixed effects and the nine firm characteristic controls. Standard errors are clustered by firm. Panels C and D report regression results of posting growth after the 29 September 2025 release of Claude Code 2.0 on the betas dated 26 September 2025. The dependent variable is the growth of monthly postings from the three months before the release to the first 3, 6, or 9 months after it, in percent. The betas are standardized across firms. Columns (1) to (3) include no controls. Columns (4) to (6) add Fama-French 49 industry fixed effects and the nine firm characteristic controls. The number of observations and the adjusted $R^{2}$ are reported. t-statistics are reported in parentheses. {*}, {*}{*}, and {*}{*}{*} denote statistical significance at the 10\%, 5\%, and 1\% levels, respectively.\par}

{\footnotesize\medskip{}
}{\footnotesize\par}
\centering{}{\scriptsize{}%
\renewcommand{\arraystretch}{0.95}%
\begin{tabular}{lcccccc}
\hline 
\multicolumn{7}{l}{{\scriptsize \textbf{Panel A: Panel regressions, $\beta^{Non\text{-}agentic}$ alone}}}\tabularnewline
 & {\scriptsize (1)} & {\scriptsize (2)} & {\scriptsize (3)} & {\scriptsize (4)} & {\scriptsize (5)} & {\scriptsize (6)}\tabularnewline
 & \multicolumn{3}{c}{{\scriptsize Time fixed effects}} & \multicolumn{3}{c}{{\scriptsize With controls}}\tabularnewline
{\scriptsize Horizon (months)} & {\scriptsize 3} & {\scriptsize 6} & {\scriptsize 9} & {\scriptsize 3} & {\scriptsize 6} & {\scriptsize 9}\tabularnewline
\hline 
 &  &  &  &  &  & \tabularnewline
{\scriptsize $\beta^{Non\text{-}agentic}$} & {\scriptsize $-0.632$} & {\scriptsize $-0.790$} & {\scriptsize $-1.525$} & {\scriptsize $-0.621$} & {\scriptsize $-0.760$} & {\scriptsize $-1.112$}\tabularnewline
 & {\scriptsize $(-1.170)$} & {\scriptsize $(-1.017)$} & {\scriptsize $(-1.409)$} & {\scriptsize $(-1.158)$} & {\scriptsize $(-0.976)$} & {\scriptsize $(-1.063)$}\tabularnewline
 &  &  &  &  &  & \tabularnewline
{\scriptsize Adj $R^{2}$} & {\scriptsize $0.058$} & {\scriptsize $0.062$} & {\scriptsize $0.010$} & {\scriptsize $0.069$} & {\scriptsize $0.095$} & {\scriptsize $0.062$}\tabularnewline
{\scriptsize Obs} & {\scriptsize 17,090} & {\scriptsize 12,038} & {\scriptsize 6,920} & {\scriptsize 16,819} & {\scriptsize 11,847} & {\scriptsize 6,813}\tabularnewline
 &  &  &  &  &  & \tabularnewline
\hline 
\multicolumn{7}{l}{{\scriptsize \textbf{Panel B: Panel regressions, $\beta^{Agentic}$ and $\beta^{Non\text{-}agentic}$ in the same regression}}}\tabularnewline
 &  &  &  &  &  & \tabularnewline
{\scriptsize $\beta^{Agentic}$} & {\scriptsize $-1.541${*}{*}{*}} & {\scriptsize $-3.051${*}{*}{*}} & {\scriptsize $-4.285${*}{*}{*}} & {\scriptsize $-0.739$} & {\scriptsize $-2.072${*}{*}{*}} & {\scriptsize $-3.231${*}{*}{*}}\tabularnewline
 & {\scriptsize $(-2.602)$} & {\scriptsize $(-4.013)$} & {\scriptsize $(-3.816)$} & {\scriptsize $(-1.208)$} & {\scriptsize $(-2.858)$} & {\scriptsize $(-2.975)$}\tabularnewline
{\scriptsize $\beta^{Non\text{-}agentic}$} & {\scriptsize $-0.037$} & {\scriptsize $0.062$} & {\scriptsize $0.161$} & {\scriptsize $-0.319$} & {\scriptsize $-0.148$} & {\scriptsize $0.089$}\tabularnewline
 & {\scriptsize $(-0.066)$} & {\scriptsize $(0.081)$} & {\scriptsize $(0.145)$} & {\scriptsize $(-0.565)$} & {\scriptsize $(-0.191)$} & {\scriptsize $(0.084)$}\tabularnewline
 &  &  &  &  &  & \tabularnewline
{\scriptsize Adj $R^{2}$} & {\scriptsize $0.059$} & {\scriptsize $0.066$} & {\scriptsize $0.016$} & {\scriptsize $0.069$} & {\scriptsize $0.096$} & {\scriptsize $0.065$}\tabularnewline
{\scriptsize Obs} & {\scriptsize 17,090} & {\scriptsize 12,038} & {\scriptsize 6,920} & {\scriptsize 16,819} & {\scriptsize 11,847} & {\scriptsize 6,813}\tabularnewline
 &  &  &  &  &  & \tabularnewline
\hline 
\multicolumn{7}{l}{{\scriptsize \textbf{Panel C: Release-based regressions, $\beta^{Non\text{-}agentic}$ alone}}}\tabularnewline
 & {\scriptsize (1)} & {\scriptsize (2)} & {\scriptsize (3)} & {\scriptsize (4)} & {\scriptsize (5)} & {\scriptsize (6)}\tabularnewline
 & \multicolumn{3}{c}{{\scriptsize No controls}} & \multicolumn{3}{c}{{\scriptsize With controls}}\tabularnewline
{\scriptsize Months after the release} & {\scriptsize 3} & {\scriptsize 6} & {\scriptsize 9} & {\scriptsize 3} & {\scriptsize 6} & {\scriptsize 9}\tabularnewline
\hline 
 &  &  &  &  &  & \tabularnewline
{\scriptsize $\beta^{Non\text{-}agentic}$} & {\scriptsize $-1.437$} & {\scriptsize $-1.635$} & {\scriptsize $-1.527$} & {\scriptsize $-1.897$} & {\scriptsize $-2.517${*}} & {\scriptsize $-2.028$}\tabularnewline
 & {\scriptsize $(-1.057)$} & {\scriptsize $(-1.214)$} & {\scriptsize $(-1.087)$} & {\scriptsize $(-1.081)$} & {\scriptsize $(-1.658)$} & {\scriptsize $(-1.329)$}\tabularnewline
 &  &  &  &  &  & \tabularnewline
{\scriptsize Adj $R^{2}$} & {\scriptsize $0.000$} & {\scriptsize $0.001$} & {\scriptsize $0.000$} & {\scriptsize $0.007$} & {\scriptsize $0.038$} & {\scriptsize $0.065$}\tabularnewline
{\scriptsize Obs} & {\scriptsize 1,725} & {\scriptsize 1,725} & {\scriptsize 1,725} & {\scriptsize 1,394} & {\scriptsize 1,394} & {\scriptsize 1,394}\tabularnewline
 &  &  &  &  &  & \tabularnewline
\hline 
\multicolumn{7}{l}{{\scriptsize \textbf{Panel D: Release-based regressions, $\beta^{Agentic}$ and $\beta^{Non\text{-}agentic}$ in the same regression}}}\tabularnewline
 &  &  &  &  &  & \tabularnewline
{\scriptsize $\beta^{Agentic}$} & {\scriptsize $-2.558$} & {\scriptsize $-5.068${*}{*}{*}} & {\scriptsize $-5.114${*}{*}{*}} & {\scriptsize $-3.112$} & {\scriptsize $-4.433${*}{*}} & {\scriptsize $-4.219${*}{*}}\tabularnewline
 & {\scriptsize $(-1.530)$} & {\scriptsize $(-3.278)$} & {\scriptsize $(-3.154)$} & {\scriptsize $(-1.530)$} & {\scriptsize $(-2.571)$} & {\scriptsize $(-2.478)$}\tabularnewline
{\scriptsize $\beta^{Non\text{-}agentic}$} & {\scriptsize $-0.293$} & {\scriptsize $0.633$} & {\scriptsize $0.762$} & {\scriptsize $-0.567$} & {\scriptsize $-0.623$} & {\scriptsize $-0.225$}\tabularnewline
 & {\scriptsize $(-0.203)$} & {\scriptsize $(0.448)$} & {\scriptsize $(0.511)$} & {\scriptsize $(-0.308)$} & {\scriptsize $(-0.385)$} & {\scriptsize $(-0.140)$}\tabularnewline
 &  &  &  &  &  & \tabularnewline
{\scriptsize Adj $R^{2}$} & {\scriptsize $0.002$} & {\scriptsize $0.009$} & {\scriptsize $0.009$} & {\scriptsize $0.008$} & {\scriptsize $0.043$} & {\scriptsize $0.069$}\tabularnewline
{\scriptsize Obs} & {\scriptsize 1,725} & {\scriptsize 1,725} & {\scriptsize 1,725} & {\scriptsize 1,394} & {\scriptsize 1,394} & {\scriptsize 1,394}\tabularnewline
 &  &  &  &  &  & \tabularnewline
\hline
\end{tabular}
}{\scriptsize\par}
\end{table}

\clearpage
\begin{table}[!htbp]
\stepcounter{anch}
{\scriptsize\caption{{\small\textbf{Future Posting Growth on the Agentic Beta from 20- to 34-Week Rolling Windows\label{tab:BW}}}}
\medskip{}
}{\scriptsize\par}

{\footnotesize This table reports panel regression results for future posting growth on the trailing agentic beta using different beta estimation windows. Panels A to E use 20, 22, 26, 30, and 34 weeks, respectively, ending in the last week of month $t$. The dependent variable is growth in monthly job postings from the three months ending in month $t$ to the following 3, 6, or 9 months, in percent. Each beta is standardized within each month. Columns (1) to (3) include time fixed effects only. Columns (4) to (6) add Fama-French 49 industry fixed effects and the nine firm characteristic controls. The number of observations and the adjusted $R^{2}$ are reported. t-statistics are reported in parentheses. Standard errors are clustered by firm. {*}, {*}{*}, and {*}{*}{*} denote statistical significance at the 10\%, 5\%, and 1\% levels, respectively.\par}

{\scriptsize\medskip{}
}{\scriptsize\par}
\centering{}{\scriptsize{}%
\begin{tabular}{lcccccc}
\hline 
 & {\scriptsize (1)} & {\scriptsize (2)} & {\scriptsize (3)} & {\scriptsize (4)} & {\scriptsize (5)} & {\scriptsize (6)}\tabularnewline
 & \multicolumn{3}{c}{{\scriptsize Time fixed effects}} & \multicolumn{3}{c}{{\scriptsize With controls}}\tabularnewline
{\scriptsize Horizon (months)} & {\scriptsize 3} & {\scriptsize 6} & {\scriptsize 9} & {\scriptsize 3} & {\scriptsize 6} & {\scriptsize 9}\tabularnewline
\hline 
{\scriptsize \textbf{Panel A: 20-week window}} &  &  &  &  &  & \tabularnewline
 &  &  &  &  &  & \tabularnewline
{\scriptsize $\beta^{Agentic}$} & {\scriptsize $-0.986${*}} & {\scriptsize $-2.398${*}{*}{*}} & {\scriptsize $-2.535${*}{*}{*}} & {\scriptsize $-0.249$} & {\scriptsize $-1.534${*}{*}} & {\scriptsize $-1.932${*}{*}}\tabularnewline
 & {\scriptsize $(-1.768)$} & {\scriptsize $(-3.298)$} & {\scriptsize $(-2.625)$} & {\scriptsize $(-0.443)$} & {\scriptsize $(-2.267)$} & {\scriptsize $(-2.142)$}\tabularnewline
 &  &  &  &  &  & \tabularnewline
{\scriptsize Adj $R^{2}$} & {\scriptsize $0.058$} & {\scriptsize $0.064$} & {\scriptsize $0.012$} & {\scriptsize $0.069$} & {\scriptsize $0.095$} & {\scriptsize $0.063$}\tabularnewline
{\scriptsize Obs} & {\scriptsize 17,090} & {\scriptsize 12,038} & {\scriptsize 6,920} & {\scriptsize 16,819} & {\scriptsize 11,847} & {\scriptsize 6,813}\tabularnewline
 &  &  &  &  &  & \tabularnewline
\hline 
{\scriptsize \textbf{Panel B: 22-week window}} &  &  &  &  &  & \tabularnewline
 &  &  &  &  &  & \tabularnewline
{\scriptsize $\beta^{Agentic}$} & {\scriptsize $-1.408${*}{*}} & {\scriptsize $-2.998${*}{*}{*}} & {\scriptsize $-3.985${*}{*}{*}} & {\scriptsize $-0.718$} & {\scriptsize $-2.042${*}{*}{*}} & {\scriptsize $-2.857${*}{*}{*}}\tabularnewline
 & {\scriptsize $(-2.507)$} & {\scriptsize $(-4.064)$} & {\scriptsize $(-3.769)$} & {\scriptsize $(-1.259)$} & {\scriptsize $(-2.917)$} & {\scriptsize $(-2.764)$}\tabularnewline
 &  &  &  &  &  & \tabularnewline
{\scriptsize Adj $R^{2}$} & {\scriptsize $0.058$} & {\scriptsize $0.066$} & {\scriptsize $0.016$} & {\scriptsize $0.069$} & {\scriptsize $0.096$} & {\scriptsize $0.065$}\tabularnewline
{\scriptsize Obs} & {\scriptsize 17,090} & {\scriptsize 12,038} & {\scriptsize 6,920} & {\scriptsize 16,819} & {\scriptsize 11,847} & {\scriptsize 6,813}\tabularnewline
 &  &  &  &  &  & \tabularnewline
\hline
{\scriptsize \textbf{Panel C: 26-week window}} &  &  &  &  &  & \tabularnewline
 &  &  &  &  &  & \tabularnewline
{\scriptsize $\beta^{Agentic}$} & {\scriptsize $-1.556${*}{*}{*}} & {\scriptsize $-3.033${*}{*}{*}} & {\scriptsize $-4.222${*}{*}{*}} & {\scriptsize $-0.869$} & {\scriptsize $-2.114${*}{*}{*}} & {\scriptsize $-3.198${*}{*}{*}}\tabularnewline
 & {\scriptsize $(-2.779)$} & {\scriptsize $(-4.031)$} & {\scriptsize $(-3.960)$} & {\scriptsize $(-1.506)$} & {\scriptsize $(-2.916)$} & {\scriptsize $(-3.023)$}\tabularnewline
 &  &  &  &  &  & \tabularnewline
{\scriptsize Adj $R^{2}$} & {\scriptsize $0.059$} & {\scriptsize $0.066$} & {\scriptsize $0.017$} & {\scriptsize $0.069$} & {\scriptsize $0.096$} & {\scriptsize $0.065$}\tabularnewline
{\scriptsize Obs} & {\scriptsize 17,090} & {\scriptsize 12,038} & {\scriptsize 6,920} & {\scriptsize 16,819} & {\scriptsize 11,847} & {\scriptsize 6,813}\tabularnewline
 &  &  &  &  &  & \tabularnewline
\hline 
{\scriptsize \textbf{Panel D: 30-week window}} &  &  &  &  &  & \tabularnewline
 &  &  &  &  &  & \tabularnewline
{\scriptsize $\beta^{Agentic}$} & {\scriptsize $-1.671${*}{*}{*}} & {\scriptsize $-3.058${*}{*}{*}} & {\scriptsize $-3.963${*}{*}{*}} & {\scriptsize $-1.038${*}} & {\scriptsize $-2.315${*}{*}{*}} & {\scriptsize $-3.490${*}{*}{*}}\tabularnewline
 & {\scriptsize $(-3.192)$} & {\scriptsize $(-4.053)$} & {\scriptsize $(-3.727)$} & {\scriptsize $(-1.932)$} & {\scriptsize $(-3.175)$} & {\scriptsize $(-3.273)$}\tabularnewline
 &  &  &  &  &  & \tabularnewline
{\scriptsize Adj $R^{2}$} & {\scriptsize $0.059$} & {\scriptsize $0.066$} & {\scriptsize $0.016$} & {\scriptsize $0.070$} & {\scriptsize $0.096$} & {\scriptsize $0.066$}\tabularnewline
{\scriptsize Obs} & {\scriptsize 17,089} & {\scriptsize 12,037} & {\scriptsize 6,919} & {\scriptsize 16,819} & {\scriptsize 11,847} & {\scriptsize 6,813}\tabularnewline
 &  &  &  &  &  & \tabularnewline
\hline 
{\scriptsize \textbf{Panel E: 34-week window}} &  &  &  &  &  & \tabularnewline
 &  &  &  &  &  & \tabularnewline
{\scriptsize $\beta^{Agentic}$} & {\scriptsize $-1.420${*}{*}{*}} & {\scriptsize $-2.759${*}{*}{*}} & {\scriptsize $-4.188${*}{*}{*}} & {\scriptsize $-0.754$} & {\scriptsize $-2.065${*}{*}{*}} & {\scriptsize $-3.933${*}{*}{*}}\tabularnewline
 & {\scriptsize $(-2.775)$} & {\scriptsize $(-3.532)$} & {\scriptsize $(-3.731)$} & {\scriptsize $(-1.443)$} & {\scriptsize $(-2.667)$} & {\scriptsize $(-3.382)$}\tabularnewline
 &  &  &  &  &  & \tabularnewline
{\scriptsize Adj $R^{2}$} & {\scriptsize $0.058$} & {\scriptsize $0.065$} & {\scriptsize $0.016$} & {\scriptsize $0.069$} & {\scriptsize $0.096$} & {\scriptsize $0.067$}\tabularnewline
{\scriptsize Obs} & {\scriptsize 17,088} & {\scriptsize 12,036} & {\scriptsize 6,919} & {\scriptsize 16,819} & {\scriptsize 11,847} & {\scriptsize 6,813}\tabularnewline
 &  &  &  &  &  & \tabularnewline
\hline
\end{tabular}
}{\scriptsize\par}
\end{table}

\clearpage
\begin{table}[!htbp]
\stepcounter{anch}
{\scriptsize\caption{{\small\textbf{Future Posting Growth on the Agentic Beta from Alternative Factor Models\label{tab:FM}}}}
\medskip{}
}{\scriptsize\par}

{\footnotesize This table reports panel regression results for future posting growth on the trailing agentic beta using different factor models to estimate the beta. Panel A uses the baseline specification, which controls for the excess market return. Panels B to D also control for the other factors of the \citet{FamaFrench1993} three-factor model, the \citet{Carhart1997} four-factor model, and the \citet{FamaFrench2015} five-factor model, respectively. The dependent variable is growth in monthly job postings from the three months ending in month $t$ to the following 3, 6, or 9 months, in percent. Each beta is standardized within each month. Columns (1) to (3) include time fixed effects only. Columns (4) to (6) add Fama-French 49 industry fixed effects and the nine firm characteristic controls. The number of observations and the adjusted $R^{2}$ are reported. t-statistics are reported in parentheses. Standard errors are clustered by firm. {*}, {*}{*}, and {*}{*}{*} denote statistical significance at the 10\%, 5\%, and 1\% levels, respectively.\par}

{\scriptsize\medskip{}
}{\scriptsize\par}
\centering{}{\scriptsize{}%
\begin{tabular}{lcccccc}
\hline
 & {\scriptsize (1)} & {\scriptsize (2)} & {\scriptsize (3)} & {\scriptsize (4)} & {\scriptsize (5)} & {\scriptsize (6)}\tabularnewline
 & \multicolumn{3}{c}{{\scriptsize Time fixed effects}} & \multicolumn{3}{c}{{\scriptsize With controls}}\tabularnewline
{\scriptsize Horizon (months)} & {\scriptsize 3} & {\scriptsize 6} & {\scriptsize 9} & {\scriptsize 3} & {\scriptsize 6} & {\scriptsize 9}\tabularnewline
\hline
\multicolumn{7}{l}{{\scriptsize \textbf{Panel A: Market model}}}\tabularnewline
 &  &  &  &  &  & \tabularnewline
{\scriptsize $\beta^{Agentic}$} & {\scriptsize $-1.556${*}{*}{*}} & {\scriptsize $-3.033${*}{*}{*}} & {\scriptsize $-4.222${*}{*}{*}} & {\scriptsize $-0.869$} & {\scriptsize $-2.114${*}{*}{*}} & {\scriptsize $-3.198${*}{*}{*}}\tabularnewline
 & {\scriptsize $(-2.779)$} & {\scriptsize $(-4.031)$} & {\scriptsize $(-3.960)$} & {\scriptsize $(-1.506)$} & {\scriptsize $(-2.916)$} & {\scriptsize $(-3.023)$}\tabularnewline
 &  &  &  &  &  & \tabularnewline
{\scriptsize Adj $R^{2}$} & {\scriptsize $0.059$} & {\scriptsize $0.066$} & {\scriptsize $0.017$} & {\scriptsize $0.069$} & {\scriptsize $0.096$} & {\scriptsize $0.065$}\tabularnewline
{\scriptsize Obs} & {\scriptsize 17,090} & {\scriptsize 12,038} & {\scriptsize 6,920} & {\scriptsize 16,819} & {\scriptsize 11,847} & {\scriptsize 6,813}\tabularnewline
 &  &  &  &  &  & \tabularnewline
\hline
\multicolumn{7}{l}{{\scriptsize \textbf{Panel B: Fama-French three-factor model}}}\tabularnewline
 &  &  &  &  &  & \tabularnewline
{\scriptsize $\beta^{Agentic}$} & {\scriptsize $-1.567${*}{*}{*}} & {\scriptsize $-2.948${*}{*}{*}} & {\scriptsize $-4.402${*}{*}{*}} & {\scriptsize $-1.002${*}} & {\scriptsize $-2.019${*}{*}{*}} & {\scriptsize $-3.061${*}{*}{*}}\tabularnewline
 & {\scriptsize $(-2.862)$} & {\scriptsize $(-3.991)$} & {\scriptsize $(-4.265)$} & {\scriptsize $(-1.792)$} & {\scriptsize $(-2.866)$} & {\scriptsize $(-3.022)$}\tabularnewline
 &  &  &  &  &  & \tabularnewline
{\scriptsize Adj $R^{2}$} & {\scriptsize $0.059$} & {\scriptsize $0.065$} & {\scriptsize $0.017$} & {\scriptsize $0.069$} & {\scriptsize $0.096$} & {\scriptsize $0.065$}\tabularnewline
{\scriptsize Obs} & {\scriptsize 17,090} & {\scriptsize 12,038} & {\scriptsize 6,920} & {\scriptsize 16,819} & {\scriptsize 11,847} & {\scriptsize 6,813}\tabularnewline
 &  &  &  &  &  & \tabularnewline
\hline
\multicolumn{7}{l}{{\scriptsize \textbf{Panel C: Carhart four-factor model}}}\tabularnewline
 &  &  &  &  &  & \tabularnewline
{\scriptsize $\beta^{Agentic}$} & {\scriptsize $-1.409${*}{*}{*}} & {\scriptsize $-2.938${*}{*}{*}} & {\scriptsize $-4.354${*}{*}{*}} & {\scriptsize $-0.894$} & {\scriptsize $-2.108${*}{*}{*}} & {\scriptsize $-3.049${*}{*}{*}}\tabularnewline
 & {\scriptsize $(-2.578)$} & {\scriptsize $(-4.020)$} & {\scriptsize $(-4.242)$} & {\scriptsize $(-1.629)$} & {\scriptsize $(-3.078)$} & {\scriptsize $(-3.055)$}\tabularnewline
 &  &  &  &  &  & \tabularnewline
{\scriptsize Adj $R^{2}$} & {\scriptsize $0.058$} & {\scriptsize $0.065$} & {\scriptsize $0.017$} & {\scriptsize $0.069$} & {\scriptsize $0.096$} & {\scriptsize $0.065$}\tabularnewline
{\scriptsize Obs} & {\scriptsize 17,090} & {\scriptsize 12,038} & {\scriptsize 6,920} & {\scriptsize 16,819} & {\scriptsize 11,847} & {\scriptsize 6,813}\tabularnewline
 &  &  &  &  &  & \tabularnewline
\hline
\multicolumn{7}{l}{{\scriptsize \textbf{Panel D: Fama-French five-factor model}}}\tabularnewline
 &  &  &  &  &  & \tabularnewline
{\scriptsize $\beta^{Agentic}$} & {\scriptsize $-1.277${*}{*}} & {\scriptsize $-2.526${*}{*}{*}} & {\scriptsize $-3.700${*}{*}{*}} & {\scriptsize $-0.643$} & {\scriptsize $-1.463${*}{*}} & {\scriptsize $-2.166${*}{*}}\tabularnewline
 & {\scriptsize $(-2.333)$} & {\scriptsize $(-3.347)$} & {\scriptsize $(-3.401)$} & {\scriptsize $(-1.154)$} & {\scriptsize $(-1.981)$} & {\scriptsize $(-1.962)$}\tabularnewline
 &  &  &  &  &  & \tabularnewline
{\scriptsize Adj $R^{2}$} & {\scriptsize $0.058$} & {\scriptsize $0.065$} & {\scriptsize $0.015$} & {\scriptsize $0.069$} & {\scriptsize $0.095$} & {\scriptsize $0.064$}\tabularnewline
{\scriptsize Obs} & {\scriptsize 17,090} & {\scriptsize 12,038} & {\scriptsize 6,920} & {\scriptsize 16,819} & {\scriptsize 11,847} & {\scriptsize 6,813}\tabularnewline
 &  &  &  &  &  & \tabularnewline
\hline
\end{tabular}
}{\scriptsize\par}
\end{table}

\clearpage
\begin{table}[!htbp]
\stepcounter{anch}
{\footnotesize\caption{{\small\textbf{Future Posting Growth on the AI and Agentic Betas -- Balanced Growth\label{tab:P1_balanced}}}}
\medskip{}
}{\footnotesize\par}

{\footnotesize This table reports panel regression results for balanced growth in job postings. Panel A includes the trailing AI beta, Panel B includes the trailing agentic beta, and Panel C includes both betas in the same regression. The dependent variable is $200 \times (A - B)/(A + B)$, where $A$ is average monthly postings over the following 3, 6, or 9 months and $B$ is average monthly postings over the three months ending in month $t$. Balanced growth ranges from $-200$ to $200$ and is defined when at least one window has postings. The betas are standardized within each month, so the coefficients measure percentage-point differences in balanced growth per standard deviation of the beta. Columns (1) to (3) include time fixed effects only. Columns (4) to (6) add Fama-French 49 industry fixed effects and the nine firm characteristic controls measured at the end of month $t$. The number of observations and the adjusted $R^{2}$ are reported. t-statistics are reported in parentheses. Standard errors are clustered by firm. {*}, {*}{*}, and {*}{*}{*} denote statistical significance at the 10\%, 5\%, and 1\% levels, respectively.\par}

{\footnotesize\medskip{}
}{\footnotesize\par}
\centering{}{\footnotesize{}%
\renewcommand{\arraystretch}{1.10}%
\begin{tabular}{lcccccc}
\hline 
 & {\footnotesize (1)} & {\footnotesize (2)} & {\footnotesize (3)} & {\footnotesize (4)} & {\footnotesize (5)} & {\footnotesize (6)}\tabularnewline
 & \multicolumn{3}{c}{{\footnotesize Time fixed effects}} & \multicolumn{3}{c}{{\footnotesize With controls}}\tabularnewline
{\footnotesize Horizon (months)} & {\footnotesize 3} & {\footnotesize 6} & {\footnotesize 9} & {\footnotesize 3} & {\footnotesize 6} & {\footnotesize 9}\tabularnewline
\hline 
\multicolumn{7}{l}{{\footnotesize \textbf{Panel A: $\beta^{AI}$}}}\tabularnewline
 &  &  &  &  &  & \tabularnewline
{\footnotesize $\beta^{AI}$} & {\footnotesize $-0.550$} & {\footnotesize $-0.988$} & {\footnotesize $-2.265${*}{*}} & {\footnotesize $-0.253$} & {\footnotesize $-0.314$} & {\footnotesize $-1.656$}\tabularnewline
 & {\footnotesize $(-1.074)$} & {\footnotesize $(-1.365)$} & {\footnotesize $(-2.352)$} & {\footnotesize $(-0.477)$} & {\footnotesize $(-0.431)$} & {\footnotesize $(-1.606)$}\tabularnewline
 &  &  &  &  &  & \tabularnewline
{\footnotesize Adj $R^{2}$} & {\footnotesize $0.061$} & {\footnotesize $0.065$} & {\footnotesize $0.012$} & {\footnotesize $0.072$} & {\footnotesize $0.098$} & {\footnotesize $0.067$}\tabularnewline
{\footnotesize Obs} & {\footnotesize 16,925} & {\footnotesize 11,947} & {\footnotesize 6,875} & {\footnotesize 16,656} & {\footnotesize 11,756} & {\footnotesize 6,768}\tabularnewline
 &  &  &  &  &  & \tabularnewline
\hline 
\multicolumn{7}{l}{{\footnotesize \textbf{Panel B: $\beta^{Agentic}$}}}\tabularnewline
 &  &  &  &  &  & \tabularnewline
{\footnotesize $\beta^{Agentic}$} & {\footnotesize $-1.597${*}{*}{*}} & {\footnotesize $-3.144${*}{*}{*}} & {\footnotesize $-4.370${*}{*}{*}} & {\footnotesize $-0.884$} & {\footnotesize $-2.002${*}{*}{*}} & {\footnotesize $-3.038${*}{*}{*}}\tabularnewline
 & {\footnotesize $(-2.912)$} & {\footnotesize $(-4.041)$} & {\footnotesize $(-3.983)$} & {\footnotesize $(-1.560)$} & {\footnotesize $(-2.674)$} & {\footnotesize $(-2.789)$}\tabularnewline
 &  &  &  &  &  & \tabularnewline
{\footnotesize Adj $R^{2}$} & {\footnotesize $0.062$} & {\footnotesize $0.068$} & {\footnotesize $0.018$} & {\footnotesize $0.072$} & {\footnotesize $0.099$} & {\footnotesize $0.069$}\tabularnewline
{\footnotesize Obs} & {\footnotesize 16,925} & {\footnotesize 11,947} & {\footnotesize 6,875} & {\footnotesize 16,656} & {\footnotesize 11,756} & {\footnotesize 6,768}\tabularnewline
 &  &  &  &  &  & \tabularnewline
\hline 
\multicolumn{7}{l}{{\footnotesize \textbf{Panel C: $\beta^{Agentic}$ and $\beta^{AI}$ in the same regression}}}\tabularnewline
 &  &  &  &  &  & \tabularnewline
{\footnotesize $\beta^{Agentic}$} & {\footnotesize $-1.940${*}{*}{*}} & {\footnotesize $-3.456${*}{*}{*}} & {\footnotesize $-4.458${*}{*}{*}} & {\footnotesize $-1.123$} & {\footnotesize $-2.394${*}{*}{*}} & {\footnotesize $-3.086${*}{*}}\tabularnewline
 & {\footnotesize $(-2.806)$} & {\footnotesize $(-4.006)$} & {\footnotesize $(-3.391)$} & {\footnotesize $(-1.560)$} & {\footnotesize $(-2.852)$} & {\footnotesize $(-2.379)$}\tabularnewline
{\footnotesize $\beta^{AI}$} & {\footnotesize $0.586$} & {\footnotesize $0.655$} & {\footnotesize $0.163$} & {\footnotesize $0.409$} & {\footnotesize $0.844$} & {\footnotesize $0.094$}\tabularnewline
 & {\footnotesize $(0.914)$} & {\footnotesize $(0.837)$} & {\footnotesize $(0.144)$} & {\footnotesize $(0.607)$} & {\footnotesize $(1.043)$} & {\footnotesize $(0.078)$}\tabularnewline
 &  &  &  &  &  & \tabularnewline
{\footnotesize Adj $R^{2}$} & {\footnotesize $0.062$} & {\footnotesize $0.068$} & {\footnotesize $0.018$} & {\footnotesize $0.072$} & {\footnotesize $0.100$} & {\footnotesize $0.069$}\tabularnewline
{\footnotesize Obs} & {\footnotesize 16,925} & {\footnotesize 11,947} & {\footnotesize 6,875} & {\footnotesize 16,656} & {\footnotesize 11,756} & {\footnotesize 6,768}\tabularnewline
 &  &  &  &  &  & \tabularnewline
\hline
\end{tabular}
}{\footnotesize\par}
\end{table}

\clearpage
\begin{table}[!htbp]
\stepcounter{anch}
{\footnotesize\caption{{\small\textbf{Posting Growth after the Release on the Agentic and AI Betas -- Balanced Growth\label{tab:S1_balanced}}}}
\medskip{}
}{\footnotesize\par}

{\footnotesize This table reports regression results for balanced growth in job postings after the 29 September 2025 release of Claude Code 2.0. At each horizon $h$ from 1 to 9 months, the dependent variable is $200 \times (A - B)/(A + B)$, where $A$ is average monthly postings over the first $h$ months after the release and $B$ is average monthly postings over the three months before the release. Balanced growth ranges from $-200$ to $200$ and is defined when at least one window has postings. The agentic and AI betas are dated 26 September 2025 and standardized across firms. Panel A includes the agentic beta alone. Column (1) includes no controls and column (2) adds Fama-French 49 industry fixed effects and the nine firm characteristic controls. Panel B includes both betas in the same regression. Columns (1) and (2) report the two coefficients without controls, and columns (3) and (4) report them with the same industry fixed effects and firm characteristic controls. The last four columns report the adjusted $R^{2}$ and the number of observations for each regression. t-statistics are reported in parentheses. {*}, {*}{*}, and {*}{*}{*} denote statistical significance at the 10\%, 5\%, and 1\% levels, respectively.\par}

{\footnotesize\medskip{}
}{\footnotesize\par}
\centering{}{\scriptsize{}%
\renewcommand{\arraystretch}{0.85}%
\begin{tabular}{lcccccccc}
\hline 
\multicolumn{9}{l}{{\scriptsize \textbf{Panel A: $\beta^{Agentic}$ alone}}}\tabularnewline
 & \multicolumn{2}{c}{{\scriptsize (1)}} & \multicolumn{2}{c}{{\scriptsize (2)}} & \multicolumn{2}{c}{{\scriptsize Adj $R^{2}$}} & \multicolumn{2}{c}{{\scriptsize Obs}}\tabularnewline
{\scriptsize Months after the release} & \multicolumn{2}{c}{{\scriptsize No controls}} & \multicolumn{2}{c}{{\scriptsize With controls}} & {\scriptsize (1)} & {\scriptsize (2)} & {\scriptsize (1)} & {\scriptsize (2)}\tabularnewline
\hline 
 &  &  &  &  &  &  &  & \tabularnewline
{\scriptsize 1} & \multicolumn{2}{c}{{\scriptsize $0.087$}} & \multicolumn{2}{c}{{\scriptsize $-1.159$}} & {\scriptsize $-0.001$} & {\scriptsize $0.001$} & {\scriptsize 1,708} & {\scriptsize 1,385}\tabularnewline
 & \multicolumn{2}{c}{{\scriptsize $(0.056)$}} & \multicolumn{2}{c}{{\scriptsize $(-0.608)$}} &  &  &  & \tabularnewline
{\scriptsize 2} & \multicolumn{2}{c}{{\scriptsize $-2.094$}} & \multicolumn{2}{c}{{\scriptsize $-3.260${*}{*}}} & {\scriptsize $0.001$} & {\scriptsize $0.006$} & {\scriptsize 1,709} & {\scriptsize 1,386}\tabularnewline
 & \multicolumn{2}{c}{{\scriptsize $(-1.476)$}} & \multicolumn{2}{c}{{\scriptsize $(-1.977)$}} &  &  &  & \tabularnewline
{\scriptsize 3} & \multicolumn{2}{c}{{\scriptsize $-2.098$}} & \multicolumn{2}{c}{{\scriptsize $-2.843${*}}} & {\scriptsize $0.001$} & {\scriptsize $0.006$} & {\scriptsize 1,710} & {\scriptsize 1,386}\tabularnewline
 & \multicolumn{2}{c}{{\scriptsize $(-1.454)$}} & \multicolumn{2}{c}{{\scriptsize $(-1.700)$}} &  &  &  & \tabularnewline
{\scriptsize 4} & \multicolumn{2}{c}{{\scriptsize $-3.951${*}{*}{*}}} & \multicolumn{2}{c}{{\scriptsize $-4.134${*}{*}{*}}} & {\scriptsize $0.006$} & {\scriptsize $0.012$} & {\scriptsize 1,712} & {\scriptsize 1,386}\tabularnewline
 & \multicolumn{2}{c}{{\scriptsize $(-2.696)$}} & \multicolumn{2}{c}{{\scriptsize $(-2.588)$}} &  &  &  & \tabularnewline
{\scriptsize 5} & \multicolumn{2}{c}{{\scriptsize $-5.032${*}{*}{*}}} & \multicolumn{2}{c}{{\scriptsize $-4.722${*}{*}{*}}} & {\scriptsize $0.010$} & {\scriptsize $0.030$} & {\scriptsize 1,713} & {\scriptsize 1,386}\tabularnewline
 & \multicolumn{2}{c}{{\scriptsize $(-3.409)$}} & \multicolumn{2}{c}{{\scriptsize $(-3.033)$}} &  &  &  & \tabularnewline
{\scriptsize 6} & \multicolumn{2}{c}{{\scriptsize $-5.316${*}{*}{*}}} & \multicolumn{2}{c}{{\scriptsize $-4.608${*}{*}{*}}} & {\scriptsize $0.012$} & {\scriptsize $0.040$} & {\scriptsize 1,713} & {\scriptsize 1,386}\tabularnewline
 & \multicolumn{2}{c}{{\scriptsize $(-3.624)$}} & \multicolumn{2}{c}{{\scriptsize $(-2.971)$}} &  &  &  & \tabularnewline
{\scriptsize 7} & \multicolumn{2}{c}{{\scriptsize $-5.351${*}{*}{*}}} & \multicolumn{2}{c}{{\scriptsize $-4.280${*}{*}{*}}} & {\scriptsize $0.012$} & {\scriptsize $0.048$} & {\scriptsize 1,713} & {\scriptsize 1,386}\tabularnewline
 & \multicolumn{2}{c}{{\scriptsize $(-3.608)$}} & \multicolumn{2}{c}{{\scriptsize $(-2.741)$}} &  &  &  & \tabularnewline
{\scriptsize 8} & \multicolumn{2}{c}{{\scriptsize $-4.991${*}{*}{*}}} & \multicolumn{2}{c}{{\scriptsize $-4.207${*}{*}{*}}} & {\scriptsize $0.010$} & {\scriptsize $0.054$} & {\scriptsize 1,714} & {\scriptsize 1,386}\tabularnewline
 & \multicolumn{2}{c}{{\scriptsize $(-3.280)$}} & \multicolumn{2}{c}{{\scriptsize $(-2.674)$}} &  &  &  & \tabularnewline
{\scriptsize 9} & \multicolumn{2}{c}{{\scriptsize $-5.151${*}{*}{*}}} & \multicolumn{2}{c}{{\scriptsize $-4.345${*}{*}{*}}} & {\scriptsize $0.011$} & {\scriptsize $0.060$} & {\scriptsize 1,714} & {\scriptsize 1,386}\tabularnewline
 & \multicolumn{2}{c}{{\scriptsize $(-3.390)$}} & \multicolumn{2}{c}{{\scriptsize $(-2.797)$}} &  &  &  & \tabularnewline
 &  &  &  &  &  &  &  & \tabularnewline
\hline 
\multicolumn{9}{l}{{\scriptsize \textbf{Panel B: $\beta^{Agentic}$ and $\beta^{AI}$ in the same regression}}}\tabularnewline
 & {\scriptsize (1)} & {\scriptsize (2)} & {\scriptsize (3)} & {\scriptsize (4)} & \multicolumn{2}{c}{{\scriptsize Adj $R^{2}$}} & \multicolumn{2}{c}{{\scriptsize Obs}}\tabularnewline
 & \multicolumn{2}{c}{{\scriptsize No controls}} & \multicolumn{2}{c}{{\scriptsize With controls}} &  &  &  & \tabularnewline
{\scriptsize Months after the release} & {\scriptsize $\beta^{Agentic}$} & {\scriptsize $\beta^{AI}$} & {\scriptsize $\beta^{Agentic}$} & {\scriptsize $\beta^{AI}$} & {\scriptsize (1)--(2)} & {\scriptsize (3)--(4)} & {\scriptsize (1)--(2)} & {\scriptsize (3)--(4)}\tabularnewline
\hline 
 &  &  &  &  &  &  &  & \tabularnewline
{\scriptsize 1} & {\scriptsize $0.221$} & {\scriptsize $-0.250$} & {\scriptsize $0.297$} & {\scriptsize $-2.622$} & {\scriptsize $-0.001$} & {\scriptsize $0.001$} & {\scriptsize 1,708} & {\scriptsize 1,385}\tabularnewline
 & {\scriptsize $(0.119)$} & {\scriptsize $(-0.153)$} & {\scriptsize $(0.134)$} & {\scriptsize $(-1.309)$} &  &  &  & \tabularnewline
{\scriptsize 2} & {\scriptsize $-2.320$} & {\scriptsize $0.421$} & {\scriptsize $-2.415$} & {\scriptsize $-1.522$} & {\scriptsize $0.001$} & {\scriptsize $0.006$} & {\scriptsize 1,709} & {\scriptsize 1,386}\tabularnewline
 & {\scriptsize $(-1.320)$} & {\scriptsize $(0.294)$} & {\scriptsize $(-1.208)$} & {\scriptsize $(-0.862)$} &  &  &  & \tabularnewline
{\scriptsize 3} & {\scriptsize $-1.826$} & {\scriptsize $-0.508$} & {\scriptsize $-1.494$} & {\scriptsize $-2.430$} & {\scriptsize $0.001$} & {\scriptsize $0.007$} & {\scriptsize 1,710} & {\scriptsize 1,386}\tabularnewline
 & {\scriptsize $(-1.022)$} & {\scriptsize $(-0.347)$} & {\scriptsize $(-0.747)$} & {\scriptsize $(-1.375)$} &  &  &  & \tabularnewline
{\scriptsize 4} & {\scriptsize $-3.927${*}{*}} & {\scriptsize $-0.045$} & {\scriptsize $-2.602$} & {\scriptsize $-2.759$} & {\scriptsize $0.006$} & {\scriptsize $0.014$} & {\scriptsize 1,712} & {\scriptsize 1,386}\tabularnewline
 & {\scriptsize $(-2.189)$} & {\scriptsize $(-0.031)$} & {\scriptsize $(-1.407)$} & {\scriptsize $(-1.626)$} &  &  &  & \tabularnewline
{\scriptsize 5} & {\scriptsize $-5.413${*}{*}{*}} & {\scriptsize $0.715$} & {\scriptsize $-3.868${*}{*}} & {\scriptsize $-1.538$} & {\scriptsize $0.010$} & {\scriptsize $0.030$} & {\scriptsize 1,713} & {\scriptsize 1,386}\tabularnewline
 & {\scriptsize $(-2.985)$} & {\scriptsize $(0.493)$} & {\scriptsize $(-2.117)$} & {\scriptsize $(-0.907)$} &  &  &  & \tabularnewline
{\scriptsize 6} & {\scriptsize $-5.626${*}{*}{*}} & {\scriptsize $0.582$} & {\scriptsize $-3.950${*}{*}} & {\scriptsize $-1.185$} & {\scriptsize $0.011$} & {\scriptsize $0.040$} & {\scriptsize 1,713} & {\scriptsize 1,386}\tabularnewline
 & {\scriptsize $(-3.105)$} & {\scriptsize $(0.397)$} & {\scriptsize $(-2.169)$} & {\scriptsize $(-0.702)$} &  &  &  & \tabularnewline
{\scriptsize 7} & {\scriptsize $-5.590${*}{*}{*}} & {\scriptsize $0.447$} & {\scriptsize $-3.878${*}{*}} & {\scriptsize $-0.724$} & {\scriptsize $0.011$} & {\scriptsize $0.048$} & {\scriptsize 1,713} & {\scriptsize 1,386}\tabularnewline
 & {\scriptsize $(-3.073)$} & {\scriptsize $(0.303)$} & {\scriptsize $(-2.131)$} & {\scriptsize $(-0.428)$} &  &  &  & \tabularnewline
{\scriptsize 8} & {\scriptsize $-5.166${*}{*}{*}} & {\scriptsize $0.328$} & {\scriptsize $-3.763${*}{*}} & {\scriptsize $-0.799$} & {\scriptsize $0.010$} & {\scriptsize $0.054$} & {\scriptsize 1,714} & {\scriptsize 1,386}\tabularnewline
 & {\scriptsize $(-2.785)$} & {\scriptsize $(0.221)$} & {\scriptsize $(-2.048)$} & {\scriptsize $(-0.474)$} &  &  &  & \tabularnewline
{\scriptsize 9} & {\scriptsize $-5.234${*}{*}{*}} & {\scriptsize $0.154$} & {\scriptsize $-3.834${*}{*}} & {\scriptsize $-0.920$} & {\scriptsize $0.010$} & {\scriptsize $0.060$} & {\scriptsize 1,714} & {\scriptsize 1,386}\tabularnewline
 & {\scriptsize $(-2.825)$} & {\scriptsize $(0.104)$} & {\scriptsize $(-2.099)$} & {\scriptsize $(-0.548)$} &  &  &  & \tabularnewline
 &  &  &  &  &  &  &  & \tabularnewline
\hline
\end{tabular}
}{\scriptsize\par}
\end{table}

\clearpage
\begin{table}[!htbp]
\stepcounter{anch}
{\footnotesize\caption{{\small\textbf{Future Posting Growth on the AI and Agentic Betas -- Single-Month Growth\label{tab:P1_single}}}}
\medskip{}
}{\footnotesize\par}

{\footnotesize This table reports panel regression results for posting growth from the three-month base period to a single future month. Panel A includes the trailing AI beta, Panel B includes the trailing agentic beta, and Panel C includes both betas in the same regression. The dependent variable is $100 \times [\ln(1 + x) - \ln(1 + y)]$, where $x$ is the number of postings in the month 3, 6, or 9 months after month $t$, and $y$ is average monthly postings over the three months ending in month $t$. The betas are standardized within each month, so the coefficients measure approximate percent growth per cross-sectional standard deviation of the beta. Columns (1) to (3) include time fixed effects only. Columns (4) to (6) add Fama-French 49 industry fixed effects and the nine firm characteristic controls measured at the end of month $t$. The number of observations and the adjusted $R^{2}$ are reported. t-statistics are reported in parentheses. Standard errors are clustered by firm. {*}, {*}{*}, and {*}{*}{*} denote statistical significance at the 10\%, 5\%, and 1\% levels, respectively.\par}

{\footnotesize\medskip{}
}{\footnotesize\par}
\centering{}{\footnotesize{}%
\renewcommand{\arraystretch}{1.10}%
\begin{tabular}{lcccccc}
\hline 
 & {\footnotesize (1)} & {\footnotesize (2)} & {\footnotesize (3)} & {\footnotesize (4)} & {\footnotesize (5)} & {\footnotesize (6)}\tabularnewline
 & \multicolumn{3}{c}{{\footnotesize Time fixed effects}} & \multicolumn{3}{c}{{\footnotesize With controls}}\tabularnewline
{\footnotesize Horizon (months)} & {\footnotesize 3} & {\footnotesize 6} & {\footnotesize 9} & {\footnotesize 3} & {\footnotesize 6} & {\footnotesize 9}\tabularnewline
\hline 
\multicolumn{7}{l}{{\footnotesize \textbf{Panel A: $\beta^{AI}$}}}\tabularnewline
 &  &  &  &  &  & \tabularnewline
{\footnotesize $\beta^{AI}$} & {\footnotesize $-0.638$} & {\footnotesize $-1.742${*}} & {\footnotesize $-2.967${*}{*}} & {\footnotesize $-0.518$} & {\footnotesize $-1.348$} & {\footnotesize $-3.102${*}{*}}\tabularnewline
 & {\footnotesize $(-0.986)$} & {\footnotesize $(-1.776)$} & {\footnotesize $(-2.365)$} & {\footnotesize $(-0.773)$} & {\footnotesize $(-1.330)$} & {\footnotesize $(-2.157)$}\tabularnewline
 &  &  &  &  &  & \tabularnewline
{\footnotesize Adj $R^{2}$} & {\footnotesize $0.063$} & {\footnotesize $0.022$} & {\footnotesize $0.004$} & {\footnotesize $0.077$} & {\footnotesize $0.062$} & {\footnotesize $0.062$}\tabularnewline
{\footnotesize Obs} & {\footnotesize 17,090} & {\footnotesize 12,038} & {\footnotesize 6,920} & {\footnotesize 16,819} & {\footnotesize 11,847} & {\footnotesize 6,813}\tabularnewline
 &  &  &  &  &  & \tabularnewline
\hline 
\multicolumn{7}{l}{{\footnotesize \textbf{Panel B: $\beta^{Agentic}$}}}\tabularnewline
 &  &  &  &  &  & \tabularnewline
{\footnotesize $\beta^{Agentic}$} & {\footnotesize $-2.221${*}{*}{*}} & {\footnotesize $-4.658${*}{*}{*}} & {\footnotesize $-4.511${*}{*}{*}} & {\footnotesize $-1.450${*}{*}} & {\footnotesize $-3.490${*}{*}{*}} & {\footnotesize $-3.259${*}{*}}\tabularnewline
 & {\footnotesize $(-3.108)$} & {\footnotesize $(-4.421)$} & {\footnotesize $(-3.068)$} & {\footnotesize $(-1.972)$} & {\footnotesize $(-3.414)$} & {\footnotesize $(-2.232)$}\tabularnewline
 &  &  &  &  &  & \tabularnewline
{\footnotesize Adj $R^{2}$} & {\footnotesize $0.064$} & {\footnotesize $0.026$} & {\footnotesize $0.006$} & {\footnotesize $0.077$} & {\footnotesize $0.064$} & {\footnotesize $0.063$}\tabularnewline
{\footnotesize Obs} & {\footnotesize 17,090} & {\footnotesize 12,038} & {\footnotesize 6,920} & {\footnotesize 16,819} & {\footnotesize 11,847} & {\footnotesize 6,813}\tabularnewline
 &  &  &  &  &  & \tabularnewline
\hline 
\multicolumn{7}{l}{{\footnotesize \textbf{Panel C: $\beta^{Agentic}$ and $\beta^{AI}$ in the same regression}}}\tabularnewline
 &  &  &  &  &  & \tabularnewline
{\footnotesize $\beta^{Agentic}$} & {\footnotesize $-2.804${*}{*}{*}} & {\footnotesize $-4.940${*}{*}{*}} & {\footnotesize $-4.113${*}{*}} & {\footnotesize $-1.746${*}} & {\footnotesize $-3.687${*}{*}{*}} & {\footnotesize $-2.348$}\tabularnewline
 & {\footnotesize $(-3.070)$} & {\footnotesize $(-4.149)$} & {\footnotesize $(-2.278)$} & {\footnotesize $(-1.889)$} & {\footnotesize $(-3.243)$} & {\footnotesize $(-1.302)$}\tabularnewline
{\footnotesize $\beta^{AI}$} & {\footnotesize $0.999$} & {\footnotesize $0.597$} & {\footnotesize $-0.732$} & {\footnotesize $0.507$} & {\footnotesize $0.426$} & {\footnotesize $-1.772$}\tabularnewline
 & {\footnotesize $(1.210)$} & {\footnotesize $(0.549)$} & {\footnotesize $(-0.475)$} & {\footnotesize $(0.603)$} & {\footnotesize $(0.383)$} & {\footnotesize $(-0.995)$}\tabularnewline
 &  &  &  &  &  & \tabularnewline
{\footnotesize Adj $R^{2}$} & {\footnotesize $0.064$} & {\footnotesize $0.026$} & {\footnotesize $0.006$} & {\footnotesize $0.077$} & {\footnotesize $0.064$} & {\footnotesize $0.063$}\tabularnewline
{\footnotesize Obs} & {\footnotesize 17,090} & {\footnotesize 12,038} & {\footnotesize 6,920} & {\footnotesize 16,819} & {\footnotesize 11,847} & {\footnotesize 6,813}\tabularnewline
 &  &  &  &  &  & \tabularnewline
\hline
\end{tabular}
}{\footnotesize\par}
\end{table}

\clearpage
\begin{table}[!htbp]
\stepcounter{anch}
{\footnotesize\caption{{\small\textbf{Posting Growth after the Release on the Agentic and AI Betas -- Single-Month Growth\label{tab:S1_single}}}}
\medskip{}
}{\footnotesize\par}

{\footnotesize This table reports regression results for posting growth from the three months before the 29 September 2025 release of Claude Code 2.0 to a single month after the release. At each horizon $h$ from 1 to 9 months, the dependent variable is $100 \times [\ln(1 + x) - \ln(1 + y)]$, where $x$ is the number of postings in month $h$ after the release and $y$ is average monthly postings over the three months before the release. The agentic and AI betas are dated 26 September 2025 and standardized across firms. Panel A includes the agentic beta alone. Column (1) includes no controls and column (2) adds Fama-French 49 industry fixed effects and the nine firm characteristic controls. Panel B includes both betas in the same regression. Columns (1) and (2) report the two coefficients without controls, and columns (3) and (4) report them with the same industry fixed effects and firm characteristic controls. The number of observations and the adjusted $R^{2}$ of each regression are reported. t-statistics are reported in parentheses. {*}, {*}{*}, and {*}{*}{*} denote statistical significance at the 10\%, 5\%, and 1\% levels, respectively.\par}

{\footnotesize\medskip{}
}{\footnotesize\par}
\centering{}{\scriptsize{}%
\renewcommand{\arraystretch}{0.85}%
\begin{tabular}{lcccccc}
\hline 
\multicolumn{7}{l}{{\scriptsize \textbf{Panel A: $\beta^{Agentic}$ alone}}}\tabularnewline
 & \multicolumn{2}{c}{{\scriptsize (1)}} & \multicolumn{2}{c}{{\scriptsize (2)}} & \multicolumn{2}{c}{{\scriptsize Adj $R^{2}$}}\tabularnewline
{\scriptsize Months after the release} & \multicolumn{2}{c}{{\scriptsize No controls}} & \multicolumn{2}{c}{{\scriptsize With controls}} & {\scriptsize (1)} & {\scriptsize (2)}\tabularnewline
\hline 
 &  &  &  &  &  & \tabularnewline
{\scriptsize 1} & \multicolumn{2}{c}{{\scriptsize $-0.870$}} & \multicolumn{2}{c}{{\scriptsize $-2.167$}} & {\scriptsize $0.000$} & {\scriptsize $0.002$}\tabularnewline
 & \multicolumn{2}{c}{{\scriptsize $(-0.535)$}} & \multicolumn{2}{c}{{\scriptsize $(-1.011)$}} &  & \tabularnewline
{\scriptsize 2} & \multicolumn{2}{c}{{\scriptsize $-4.627${*}{*}{*}}} & \multicolumn{2}{c}{{\scriptsize $-4.930${*}{*}}} & {\scriptsize $0.005$} & {\scriptsize $0.014$}\tabularnewline
 & \multicolumn{2}{c}{{\scriptsize $(-2.602)$}} & \multicolumn{2}{c}{{\scriptsize $(-2.279)$}} &  & \tabularnewline
{\scriptsize 3} & \multicolumn{2}{c}{{\scriptsize $-2.132$}} & \multicolumn{2}{c}{{\scriptsize $-2.113$}} & {\scriptsize $0.000$} & {\scriptsize $0.005$}\tabularnewline
 & \multicolumn{2}{c}{{\scriptsize $(-1.010)$}} & \multicolumn{2}{c}{{\scriptsize $(-0.763)$}} &  & \tabularnewline
{\scriptsize 4} & \multicolumn{2}{c}{{\scriptsize $-6.866${*}{*}{*}}} & \multicolumn{2}{c}{{\scriptsize $-7.551${*}{*}{*}}} & {\scriptsize $0.010$} & {\scriptsize $0.035$}\tabularnewline
 & \multicolumn{2}{c}{{\scriptsize $(-3.465)$}} & \multicolumn{2}{c}{{\scriptsize $(-3.072)$}} &  & \tabularnewline
{\scriptsize 5} & \multicolumn{2}{c}{{\scriptsize $-6.642${*}{*}{*}}} & \multicolumn{2}{c}{{\scriptsize $-7.126${*}{*}{*}}} & {\scriptsize $0.008$} & {\scriptsize $0.052$}\tabularnewline
 & \multicolumn{2}{c}{{\scriptsize $(-3.345)$}} & \multicolumn{2}{c}{{\scriptsize $(-2.874)$}} &  & \tabularnewline
{\scriptsize 6} & \multicolumn{2}{c}{{\scriptsize $-6.375${*}{*}{*}}} & \multicolumn{2}{c}{{\scriptsize $-5.943${*}{*}}} & {\scriptsize $0.008$} & {\scriptsize $0.039$}\tabularnewline
 & \multicolumn{2}{c}{{\scriptsize $(-3.299)$}} & \multicolumn{2}{c}{{\scriptsize $(-2.576)$}} &  & \tabularnewline
{\scriptsize 7} & \multicolumn{2}{c}{{\scriptsize $-5.321${*}{*}{*}}} & \multicolumn{2}{c}{{\scriptsize $-3.048$}} & {\scriptsize $0.005$} & {\scriptsize $0.071$}\tabularnewline
 & \multicolumn{2}{c}{{\scriptsize $(-2.667)$}} & \multicolumn{2}{c}{{\scriptsize $(-1.369)$}} &  & \tabularnewline
{\scriptsize 8} & \multicolumn{2}{c}{{\scriptsize $-5.097${*}{*}}} & \multicolumn{2}{c}{{\scriptsize $-4.954${*}{*}}} & {\scriptsize $0.004$} & {\scriptsize $0.052$}\tabularnewline
 & \multicolumn{2}{c}{{\scriptsize $(-2.561)$}} & \multicolumn{2}{c}{{\scriptsize $(-2.125)$}} &  & \tabularnewline
{\scriptsize 9} & \multicolumn{2}{c}{{\scriptsize $-6.044${*}{*}{*}}} & \multicolumn{2}{c}{{\scriptsize $-5.357${*}{*}}} & {\scriptsize $0.006$} & {\scriptsize $0.065$}\tabularnewline
 & \multicolumn{2}{c}{{\scriptsize $(-3.045)$}} & \multicolumn{2}{c}{{\scriptsize $(-2.480)$}} &  & \tabularnewline
 &  &  &  &  &  & \tabularnewline
{\scriptsize Obs} & \multicolumn{2}{c}{{\scriptsize 1,725}} & \multicolumn{2}{c}{{\scriptsize 1,394}} &  & \tabularnewline
 &  &  &  &  &  & \tabularnewline
\hline 
\multicolumn{7}{l}{{\scriptsize \textbf{Panel B: $\beta^{Agentic}$ and $\beta^{AI}$ in the same regression}}}\tabularnewline
 & {\scriptsize (1)} & {\scriptsize (2)} & {\scriptsize (3)} & {\scriptsize (4)} & \multicolumn{2}{c}{{\scriptsize Adj $R^{2}$}}\tabularnewline
 & \multicolumn{2}{c}{{\scriptsize No controls}} & \multicolumn{2}{c}{{\scriptsize With controls}} &  & \tabularnewline
{\scriptsize Months after the release} & {\scriptsize $\beta^{Agentic}$} & {\scriptsize $\beta^{AI}$} & {\scriptsize $\beta^{Agentic}$} & {\scriptsize $\beta^{AI}$} & {\scriptsize (1)--(2)} & {\scriptsize (3)--(4)}\tabularnewline
\hline 
 &  &  &  &  &  & \tabularnewline
{\scriptsize 1} & {\scriptsize $-0.600$} & {\scriptsize $-0.504$} & {\scriptsize $-0.683$} & {\scriptsize $-2.664$} & {\scriptsize $-0.001$} & {\scriptsize $0.003$}\tabularnewline
 & {\scriptsize $(-0.308)$} & {\scriptsize $(-0.325)$} & {\scriptsize $(-0.281)$} & {\scriptsize $(-1.304)$} &  & \tabularnewline
{\scriptsize 2} & {\scriptsize $-5.743${*}{*}{*}} & {\scriptsize $2.086$} & {\scriptsize $-4.640${*}} & {\scriptsize $-0.521$} & {\scriptsize $0.005$} & {\scriptsize $0.013$}\tabularnewline
 & {\scriptsize $(-2.618)$} & {\scriptsize $(1.125)$} & {\scriptsize $(-1.811)$} & {\scriptsize $(-0.236)$} &  & \tabularnewline
{\scriptsize 3} & {\scriptsize $-0.836$} & {\scriptsize $-2.424$} & {\scriptsize $0.835$} & {\scriptsize $-5.291${*}{*}} & {\scriptsize $0.001$} & {\scriptsize $0.008$}\tabularnewline
 & {\scriptsize $(-0.335)$} & {\scriptsize $(-1.201)$} & {\scriptsize $(0.263)$} & {\scriptsize $(-1.997)$} &  & \tabularnewline
{\scriptsize 4} & {\scriptsize $-7.464${*}{*}{*}} & {\scriptsize $1.117$} & {\scriptsize $-5.745${*}{*}} & {\scriptsize $-3.241$} & {\scriptsize $0.010$} & {\scriptsize $0.036$}\tabularnewline
 & {\scriptsize $(-3.329)$} & {\scriptsize $(0.622)$} & {\scriptsize $(-2.149)$} & {\scriptsize $(-1.442)$} &  & \tabularnewline
{\scriptsize 5} & {\scriptsize $-6.840${*}{*}{*}} & {\scriptsize $0.371$} & {\scriptsize $-6.288${*}{*}} & {\scriptsize $-1.505$} & {\scriptsize $0.008$} & {\scriptsize $0.051$}\tabularnewline
 & {\scriptsize $(-2.971)$} & {\scriptsize $(0.184)$} & {\scriptsize $(-2.242)$} & {\scriptsize $(-0.596)$} &  & \tabularnewline
{\scriptsize 6} & {\scriptsize $-5.825${*}{*}} & {\scriptsize $-1.029$} & {\scriptsize $-4.542${*}} & {\scriptsize $-2.515$} & {\scriptsize $0.007$} & {\scriptsize $0.039$}\tabularnewline
 & {\scriptsize $(-2.514)$} & {\scriptsize $(-0.508)$} & {\scriptsize $(-1.732)$} & {\scriptsize $(-0.975)$} &  & \tabularnewline
{\scriptsize 7} & {\scriptsize $-5.241${*}{*}} & {\scriptsize $-0.149$} & {\scriptsize $-3.348$} & {\scriptsize $0.538$} & {\scriptsize $0.005$} & {\scriptsize $0.070$}\tabularnewline
 & {\scriptsize $(-2.224)$} & {\scriptsize $(-0.074)$} & {\scriptsize $(-1.339)$} & {\scriptsize $(0.219)$} &  & \tabularnewline
{\scriptsize 8} & {\scriptsize $-4.931${*}{*}} & {\scriptsize $-0.309$} & {\scriptsize $-4.210$} & {\scriptsize $-1.336$} & {\scriptsize $0.004$} & {\scriptsize $0.051$}\tabularnewline
 & {\scriptsize $(-1.994)$} & {\scriptsize $(-0.150)$} & {\scriptsize $(-1.495)$} & {\scriptsize $(-0.538)$} &  & \tabularnewline
{\scriptsize 9} & {\scriptsize $-5.386${*}{*}} & {\scriptsize $-1.231$} & {\scriptsize $-4.335$} & {\scriptsize $-1.834$} & {\scriptsize $0.006$} & {\scriptsize $0.065$}\tabularnewline
 & {\scriptsize $(-2.211)$} & {\scriptsize $(-0.621)$} & {\scriptsize $(-1.577)$} & {\scriptsize $(-0.720)$} &  & \tabularnewline
 &  &  &  &  &  & \tabularnewline
{\scriptsize Obs} & {\scriptsize 1,725} & {\scriptsize 1,725} & {\scriptsize 1,394} & {\scriptsize 1,394} &  & \tabularnewline
 &  &  &  &  &  & \tabularnewline
\hline
\end{tabular}
}{\scriptsize\par}
\end{table}

\clearpage
\begin{table}[!htbp]
\stepcounter{anch}
{\footnotesize\caption{{\small\textbf{Future Posting Growth on the Open-Weight and Closed-Weight Agentic Betas\label{tab:S11}}}}
\medskip{}
}{\footnotesize\par}

{\footnotesize This table reports panel regression results for future posting growth on the trailing open-weight agentic beta (Panel A) and the trailing closed-weight agentic beta (Panel B). Each beta uses weekly log growth in OpenRouter agentic tokens for its model group and is standardized within each month. Open-weight models account for 65 percent of agentic tokens over the weeks used to estimate the trailing betas and 9 percent over the weeks used to estimate the pre-release betas. The correlation between the two weekly token growth rates is 0.04. The dependent variable is growth in monthly job postings from the three months ending in month $t$ to the following 3, 6, or 9 months, in percent. Columns (1) to (3) include time fixed effects only. Columns (4) to (6) add Fama-French 49 industry fixed effects and the nine firm characteristic controls. The number of observations and the adjusted $R^{2}$ are reported. t-statistics are reported in parentheses. Standard errors are clustered by firm. {*}, {*}{*}, and {*}{*}{*} denote statistical significance at the 10\%, 5\%, and 1\% levels, respectively.\par}

{\footnotesize\medskip{}
}{\footnotesize\par}
\centering{}{\footnotesize{}%
\renewcommand{\arraystretch}{1.10}%
\begin{tabular}{lcccccc}
\hline 
 & {\footnotesize (1)} & {\footnotesize (2)} & {\footnotesize (3)} & {\footnotesize (4)} & {\footnotesize (5)} & {\footnotesize (6)}\tabularnewline
 & \multicolumn{3}{c}{{\footnotesize Time fixed effects}} & \multicolumn{3}{c}{{\footnotesize With controls}}\tabularnewline
{\footnotesize Horizon (months)} & {\footnotesize 3} & {\footnotesize 6} & {\footnotesize 9} & {\footnotesize 3} & {\footnotesize 6} & {\footnotesize 9}\tabularnewline
\hline 
\multicolumn{7}{l}{{\footnotesize \textbf{Panel A: Open-weight $\beta^{Agentic}$}}}\tabularnewline
 &  &  &  &  &  & \tabularnewline
{\footnotesize Open-weight $\beta^{Agentic}$} & {\footnotesize $0.034$} & {\footnotesize $-0.426$} & {\footnotesize $0.492$} & {\footnotesize $-0.082$} & {\footnotesize $-0.993$} & {\footnotesize $-0.993$}\tabularnewline
 & {\footnotesize $(0.058)$} & {\footnotesize $(-0.510)$} & {\footnotesize $(0.436)$} & {\footnotesize $(-0.134)$} & {\footnotesize $(-1.187)$} & {\footnotesize $(-0.900)$}\tabularnewline
 &  &  &  &  &  & \tabularnewline
{\footnotesize Adj $R^{2}$} & {\footnotesize $0.058$} & {\footnotesize $0.062$} & {\footnotesize $0.009$} & {\footnotesize $0.069$} & {\footnotesize $0.095$} & {\footnotesize $0.062$}\tabularnewline
{\footnotesize Obs} & {\footnotesize 17,090} & {\footnotesize 12,038} & {\footnotesize 6,920} & {\footnotesize 16,819} & {\footnotesize 11,847} & {\footnotesize 6,813}\tabularnewline
 &  &  &  &  &  & \tabularnewline
\hline 
\multicolumn{7}{l}{{\footnotesize \textbf{Panel B: Closed-weight $\beta^{Agentic}$}}}\tabularnewline
 &  &  &  &  &  & \tabularnewline
{\footnotesize Closed-weight $\beta^{Agentic}$} & {\footnotesize $-1.521${*}{*}{*}} & {\footnotesize $-2.744${*}{*}{*}} & {\footnotesize $-4.206${*}{*}{*}} & {\footnotesize $-0.929$} & {\footnotesize $-1.956${*}{*}{*}} & {\footnotesize $-3.079${*}{*}{*}}\tabularnewline
 & {\footnotesize $(-2.704)$} & {\footnotesize $(-3.549)$} & {\footnotesize $(-3.954)$} & {\footnotesize $(-1.552)$} & {\footnotesize $(-2.590)$} & {\footnotesize $(-2.938)$}\tabularnewline
 &  &  &  &  &  & \tabularnewline
{\footnotesize Adj $R^{2}$} & {\footnotesize $0.059$} & {\footnotesize $0.065$} & {\footnotesize $0.017$} & {\footnotesize $0.069$} & {\footnotesize $0.096$} & {\footnotesize $0.065$}\tabularnewline
{\footnotesize Obs} & {\footnotesize 17,090} & {\footnotesize 12,038} & {\footnotesize 6,920} & {\footnotesize 16,819} & {\footnotesize 11,847} & {\footnotesize 6,813}\tabularnewline
 &  &  &  &  &  & \tabularnewline
\hline
\end{tabular}
}{\footnotesize\par}
\end{table}

\clearpage
\begin{table}[!htbp]
\stepcounter{anch}
{\footnotesize\caption{{\small\textbf{Six-Month Posting Growth by Job Type -- Trailing Agentic Beta\label{tab:cuts-ag}}}}
\medskip{}
}{\footnotesize\par}

{\footnotesize This table reports panel regression results for six-month posting growth by job type. The independent variable is the trailing agentic beta, standardized within each month. For each job type, X is the point estimate for growth in postings of that type, and Not X is the point estimate for growth in the comparison group. Relative is the X point estimate minus the Not X point estimate. Columns (1) to (3) include time fixed effects only. Columns (4) to (6) add Fama-French 49 industry fixed effects and the nine firm characteristic controls. The number of observations is reported. t-statistics are reported in parentheses. Standard errors are clustered by firm. {*}, {*}{*}, and {*}{*}{*} denote statistical significance at the 10\%, 5\%, and 1\% levels, respectively.\par}

{\footnotesize\medskip{}
}{\footnotesize\par}
\centering{}{\footnotesize{}%
\renewcommand{\arraystretch}{1.10}%
\begin{tabular}{lcccccc}
\hline 
 & {\footnotesize (1)} & {\footnotesize (2)} & {\footnotesize (3)} & {\footnotesize (4)} & {\footnotesize (5)} & {\footnotesize (6)}\tabularnewline
 & \multicolumn{3}{c}{{\footnotesize Time fixed effects}} & \multicolumn{3}{c}{{\footnotesize With controls}}\tabularnewline
{\footnotesize Posting type} & {\footnotesize X} & {\footnotesize Not X} & {\footnotesize Relative} & {\footnotesize X} & {\footnotesize Not X} & {\footnotesize Relative}\tabularnewline
\hline 
 &  &  &  &  &  & \tabularnewline
{\footnotesize All postings} & {\footnotesize $-3.033${*}{*}{*}} &  &  & {\footnotesize $-2.114${*}{*}{*}} &  & \tabularnewline
 & {\footnotesize $(-4.031)$} &  &  & {\footnotesize $(-2.916)$} &  & \tabularnewline
{\footnotesize Junior titles} & {\footnotesize $-2.965${*}{*}{*}} & {\footnotesize $-2.919${*}{*}{*}} & {\footnotesize $-0.046$} & {\footnotesize $-2.217${*}{*}{*}} & {\footnotesize $-2.086${*}{*}{*}} & {\footnotesize $-0.132$}\tabularnewline
 & {\footnotesize $(-4.006)$} & {\footnotesize $(-3.867)$} & {\footnotesize $(-0.060)$} & {\footnotesize $(-2.873)$} & {\footnotesize $(-2.889)$} & {\footnotesize $(-0.175)$}\tabularnewline
{\footnotesize Senior titles} & {\footnotesize $-0.891$} & {\footnotesize $-3.118${*}{*}{*}} & {\footnotesize $2.227${*}{*}{*}} & {\footnotesize $-0.570$} & {\footnotesize $-2.133${*}{*}{*}} & {\footnotesize $1.562${*}{*}}\tabularnewline
 & {\footnotesize $(-1.380)$} & {\footnotesize $(-4.071)$} & {\footnotesize $(3.561)$} & {\footnotesize $(-0.889)$} & {\footnotesize $(-2.948)$} & {\footnotesize $(2.498)$}\tabularnewline
{\footnotesize Asks 0--1 years} & {\footnotesize $-2.735${*}{*}{*}} & {\footnotesize $-2.664${*}{*}{*}} & {\footnotesize $-0.071$} & {\footnotesize $-2.408${*}{*}{*}} & {\footnotesize $-1.740${*}{*}} & {\footnotesize $-0.667$}\tabularnewline
 & {\footnotesize $(-3.933)$} & {\footnotesize $(-3.356)$} & {\footnotesize $(-0.082)$} & {\footnotesize $(-3.294)$} & {\footnotesize $(-2.299)$} & {\footnotesize $(-0.770)$}\tabularnewline
{\footnotesize Asks 6+ years} & {\footnotesize $-1.766${*}{*}} & {\footnotesize $-3.143${*}{*}{*}} & {\footnotesize $1.377${*}} & {\footnotesize $-0.968$} & {\footnotesize $-2.352${*}{*}{*}} & {\footnotesize $1.383${*}}\tabularnewline
 & {\footnotesize $(-2.340)$} & {\footnotesize $(-4.162)$} & {\footnotesize $(1.909)$} & {\footnotesize $(-1.321)$} & {\footnotesize $(-3.194)$} & {\footnotesize $(1.823)$}\tabularnewline
{\footnotesize Expects AI-tool use} & {\footnotesize $-0.278$} & {\footnotesize $-3.536${*}{*}{*}} & {\footnotesize $3.258${*}{*}} & {\footnotesize $-1.141$} & {\footnotesize $-2.349${*}{*}{*}} & {\footnotesize $1.209$}\tabularnewline
 & {\footnotesize $(-0.263)$} & {\footnotesize $(-4.444)$} & {\footnotesize $(2.535)$} & {\footnotesize $(-1.108)$} & {\footnotesize $(-3.065)$} & {\footnotesize $(0.981)$}\tabularnewline
{\footnotesize Remote or hybrid} & {\footnotesize $-2.344${*}{*}} & {\footnotesize $-3.056${*}{*}{*}} & {\footnotesize $0.713$} & {\footnotesize $-1.686${*}} & {\footnotesize $-2.168${*}{*}{*}} & {\footnotesize $0.483$}\tabularnewline
 & {\footnotesize $(-2.461)$} & {\footnotesize $(-3.862)$} & {\footnotesize $(0.663)$} & {\footnotesize $(-1.729)$} & {\footnotesize $(-2.865)$} & {\footnotesize $(0.437)$}\tabularnewline
{\footnotesize Top half, Eloundou} & {\footnotesize $-2.875${*}{*}{*}} & {\footnotesize $-2.521${*}{*}{*}} & {\footnotesize $-0.354$} & {\footnotesize $-1.903${*}{*}} & {\footnotesize $-2.045${*}{*}{*}} & {\footnotesize $0.141$}\tabularnewline
 & {\footnotesize $(-3.751)$} & {\footnotesize $(-3.283)$} & {\footnotesize $(-0.465)$} & {\footnotesize $(-2.524)$} & {\footnotesize $(-2.630)$} & {\footnotesize $(0.177)$}\tabularnewline
{\footnotesize Top half, Webb AI} & {\footnotesize $-2.772${*}{*}{*}} & {\footnotesize $-1.956${*}{*}{*}} & {\footnotesize $-0.816$} & {\footnotesize $-1.875${*}{*}} & {\footnotesize $-1.614${*}{*}} & {\footnotesize $-0.262$}\tabularnewline
 & {\footnotesize $(-3.598)$} & {\footnotesize $(-2.699)$} & {\footnotesize $(-1.224)$} & {\footnotesize $(-2.513)$} & {\footnotesize $(-2.139)$} & {\footnotesize $(-0.366)$}\tabularnewline
{\footnotesize Top half, Anthropic exposure} & {\footnotesize $-3.064${*}{*}{*}} & {\footnotesize $-2.044${*}{*}} & {\footnotesize $-1.020$} & {\footnotesize $-2.121${*}{*}{*}} & {\footnotesize $-1.369${*}} & {\footnotesize $-0.752$}\tabularnewline
 & {\footnotesize $(-3.992)$} & {\footnotesize $(-2.466)$} & {\footnotesize $(-1.172)$} & {\footnotesize $(-2.833)$} & {\footnotesize $(-1.664)$} & {\footnotesize $(-0.844)$}\tabularnewline
 &  &  &  &  &  & \tabularnewline
{\footnotesize Obs} & {\footnotesize 12,038} & {\footnotesize 12,038} & {\footnotesize 12,038} & {\footnotesize 11,847} & {\footnotesize 11,847} & {\footnotesize 11,847}\tabularnewline
 &  &  &  &  &  & \tabularnewline
\hline
\end{tabular}
}{\footnotesize\par}
\end{table}

\clearpage
\begin{table}[!htbp]
\stepcounter{anch}
{\footnotesize\caption{{\small\textbf{Future Posting Growth on Firm AI Overlap\label{tab:O1}}}}
\medskip{}
}{\footnotesize\par}

{\footnotesize This table reports panel regression results for future posting growth on firm AI overlap measures. Panels A, B, and C use the \citet{Eloundou2024} GPT exposure beta, the \citet{webb2020impact} AI exposure percentile, and Anthropic Economic Index observed exposure, respectively. The dependent variable is growth in monthly job postings from the three months ending in month $t$ to the following 3, 6, or 9 months, in percent. Each AI overlap measure is standardized within each month. Columns (1) to (3) include time fixed effects only. Columns (4) to (6) add Fama-French 49 industry fixed effects and the nine firm characteristic controls. The number of observations and the adjusted $R^{2}$ are reported. t-statistics are reported in parentheses. Standard errors are clustered by firm. {*}, {*}{*}, and {*}{*}{*} denote statistical significance at the 10\%, 5\%, and 1\% levels, respectively.\par}

{\footnotesize\medskip{}
}{\footnotesize\par}
\centering{}{\footnotesize{}%
\renewcommand{\arraystretch}{1.10}%
\begin{tabular}{lcccccc}
\hline 
 & {\footnotesize (1)} & {\footnotesize (2)} & {\footnotesize (3)} & {\footnotesize (4)} & {\footnotesize (5)} & {\footnotesize (6)}\tabularnewline
 & \multicolumn{3}{c}{{\footnotesize Time fixed effects}} & \multicolumn{3}{c}{{\footnotesize With controls}}\tabularnewline
{\footnotesize Horizon (months)} & {\footnotesize 3} & {\footnotesize 6} & {\footnotesize 9} & {\footnotesize 3} & {\footnotesize 6} & {\footnotesize 9}\tabularnewline
\hline 
\multicolumn{7}{l}{{\footnotesize \textbf{Panel A: Eloundou GPT exposure}}}\tabularnewline
 &  &  &  &  &  & \tabularnewline
{\footnotesize Firm AI overlap} & {\footnotesize $1.190${*}{*}} & {\footnotesize $1.457${*}} & {\footnotesize $1.064$} & {\footnotesize $0.816$} & {\footnotesize $0.656$} & {\footnotesize $-0.509$}\tabularnewline
 & {\footnotesize $(2.324)$} & {\footnotesize $(1.799)$} & {\footnotesize $(1.061)$} & {\footnotesize $(1.145)$} & {\footnotesize $(0.612)$} & {\footnotesize $(-0.371)$}\tabularnewline
 &  &  &  &  &  & \tabularnewline
{\footnotesize Adj $R^{2}$} & {\footnotesize $0.059$} & {\footnotesize $0.064$} & {\footnotesize $0.010$} & {\footnotesize $0.070$} & {\footnotesize $0.095$} & {\footnotesize $0.062$}\tabularnewline
{\footnotesize Obs} & {\footnotesize 16,872} & {\footnotesize 11,892} & {\footnotesize 6,845} & {\footnotesize 16,606} & {\footnotesize 11,704} & {\footnotesize 6,738}\tabularnewline
 &  &  &  &  &  & \tabularnewline
\hline 
\multicolumn{7}{l}{{\footnotesize \textbf{Panel B: Webb AI exposure}}}\tabularnewline
 &  &  &  &  &  & \tabularnewline
{\footnotesize Firm AI overlap} & {\footnotesize $3.084${*}{*}{*}} & {\footnotesize $3.897${*}{*}{*}} & {\footnotesize $4.235${*}{*}{*}} & {\footnotesize $1.200$} & {\footnotesize $0.635$} & {\footnotesize $-0.694$}\tabularnewline
 & {\footnotesize $(5.457)$} & {\footnotesize $(4.833)$} & {\footnotesize $(4.043)$} & {\footnotesize $(1.614)$} & {\footnotesize $(0.622)$} & {\footnotesize $(-0.532)$}\tabularnewline
 &  &  &  &  &  & \tabularnewline
{\footnotesize Adj $R^{2}$} & {\footnotesize $0.063$} & {\footnotesize $0.070$} & {\footnotesize $0.017$} & {\footnotesize $0.071$} & {\footnotesize $0.096$} & {\footnotesize $0.063$}\tabularnewline
{\footnotesize Obs} & {\footnotesize 16,865} & {\footnotesize 11,889} & {\footnotesize 6,844} & {\footnotesize 16,599} & {\footnotesize 11,701} & {\footnotesize 6,737}\tabularnewline
 &  &  &  &  &  & \tabularnewline
\hline 
\multicolumn{7}{l}{{\footnotesize \textbf{Panel C: Anthropic Economic Index observed exposure}}}\tabularnewline
 &  &  &  &  &  & \tabularnewline
{\footnotesize Firm AI overlap} & {\footnotesize $-0.907${*}} & {\footnotesize $-1.158$} & {\footnotesize $-2.350${*}{*}} & {\footnotesize $-1.000$} & {\footnotesize $-1.269$} & {\footnotesize $-2.606${*}{*}}\tabularnewline
 & {\footnotesize $(-1.720)$} & {\footnotesize $(-1.438)$} & {\footnotesize $(-2.175)$} & {\footnotesize $(-1.412)$} & {\footnotesize $(-1.262)$} & {\footnotesize $(-1.976)$}\tabularnewline
 &  &  &  &  &  & \tabularnewline
{\footnotesize Adj $R^{2}$} & {\footnotesize $0.059$} & {\footnotesize $0.064$} & {\footnotesize $0.012$} & {\footnotesize $0.070$} & {\footnotesize $0.096$} & {\footnotesize $0.063$}\tabularnewline
{\footnotesize Obs} & {\footnotesize 16,872} & {\footnotesize 11,892} & {\footnotesize 6,845} & {\footnotesize 16,606} & {\footnotesize 11,704} & {\footnotesize 6,738}\tabularnewline
 &  &  &  &  &  & \tabularnewline
\hline
\end{tabular}
}{\footnotesize\par}
\end{table}

\clearpage
\begin{table}[!htbp]
\stepcounter{anch}
{\footnotesize\caption{{\small\textbf{Unconditional Posting Growth by Pre-Release Agentic-Beta Quartile\label{tab:agentic_beta_quartiles}}}}
\medskip{}
}{\footnotesize\par}

{\footnotesize This table reports the average firm-level posting growth by quartile of the agentic beta dated 26 September 2025. Q1 contains the lowest betas and Q4 the highest. At horizon $h$, growth is 100 times the log change in one plus average monthly postings from July--September 2025 to the first $h$ months after the release. Entries are in log-growth points, approximately percent growth. Q4 $-$ Q1 is mean posting growth in Q4 minus mean posting growth in Q1. The t-statistics in parentheses test each mean and the difference in means against zero. The number of firms is reported below. {*}, {*}{*}, and {*}{*}{*} denote significance at the 10\%, 5\%, and 1\% levels, respectively.\par}
{\footnotesize\medskip{}}
\centering{}{\footnotesize{}%
\renewcommand{\arraystretch}{1.10}%
\begin{tabular}{lccccc}
\hline
 & {\footnotesize (1)} & {\footnotesize (2)} & {\footnotesize (3)} & {\footnotesize (4)} & {\footnotesize (5)}\tabularnewline
{\footnotesize Months after release} & {\footnotesize Q1} & {\footnotesize Q2} & {\footnotesize Q3} & {\footnotesize Q4} & {\footnotesize Q4 $-$ Q1}\tabularnewline
 & {\footnotesize Lowest beta} &  &  & {\footnotesize Highest beta} & \tabularnewline
\hline
{\footnotesize 1} & {\footnotesize $-6.55${*}{*}} & {\footnotesize $-3.71$} & {\footnotesize $-3.83$} & {\footnotesize $-7.01${*}{*}} & {\footnotesize $-0.47$}\tabularnewline
 & {\footnotesize $(-2.385)$} & {\footnotesize $(-1.475)$} & {\footnotesize $(-1.644)$} & {\footnotesize $(-2.392)$} & {\footnotesize $(-0.116)$}\tabularnewline
{\footnotesize 3} & {\footnotesize $-13.84${*}{*}{*}} & {\footnotesize $-14.59${*}{*}{*}} & {\footnotesize $-15.30${*}{*}{*}} & {\footnotesize $-16.62${*}{*}{*}} & {\footnotesize $-2.78$}\tabularnewline
 & {\footnotesize $(-5.392)$} & {\footnotesize $(-6.467)$} & {\footnotesize $(-6.908)$} & {\footnotesize $(-6.932)$} & {\footnotesize $(-0.792)$}\tabularnewline
{\footnotesize 6} & {\footnotesize $1.39$} & {\footnotesize $-4.51${*}{*}} & {\footnotesize $-5.11${*}{*}} & {\footnotesize $-6.31${*}{*}{*}} & {\footnotesize $-7.70${*}{*}}\tabularnewline
 & {\footnotesize $(0.531)$} & {\footnotesize $(-1.983)$} & {\footnotesize $(-2.494)$} & {\footnotesize $(-2.777)$} & {\footnotesize $(-2.224)$}\tabularnewline
{\footnotesize 9} & {\footnotesize $7.10${*}{*}{*}} & {\footnotesize $-0.53$} & {\footnotesize $-1.58$} & {\footnotesize $-1.33$} & {\footnotesize $-8.43${*}{*}}\tabularnewline
 & {\footnotesize $(2.652)$} & {\footnotesize $(-0.232)$} & {\footnotesize $(-0.768)$} & {\footnotesize $(-0.566)$} & {\footnotesize $(-2.368)$}\tabularnewline
\hline
{\footnotesize Firms} & {\footnotesize 432} & {\footnotesize 431} & {\footnotesize 431} & {\footnotesize 431} & {\footnotesize 863}\tabularnewline
\hline
\end{tabular}
}{\footnotesize\par}
\end{table}

\end{document}